\documentclass[CRPHYS,Unicode,manuscript]{cedram}

\usepackage{graphicx}% Include figure files
\usepackage{dcolumn}% Align table columns on decimal point
\usepackage{amsmath,amsthm}
\usepackage{amsfonts}
\usepackage{amssymb}
\usepackage{bbold}
\usepackage{amssymb}
\usepackage{epstopdf}
\usepackage{times}
\usepackage{mathrsfs}
\usepackage{color}
\usepackage{gensymb}
\usepackage{times}
\usepackage{subfigure}
\usepackage[]{graphicx}
\usepackage{amssymb}
\usepackage{amsmath}
\usepackage{bm}
\DeclareGraphicsExtensions{.PNG,.jpg,.pdf,.gif,.eps}
\usepackage{amsmath,amssymb,amsthm,amsfonts,eucal}
\usepackage[english]{babel}
\usepackage{eucal}

\usepackage{amsmath}
\usepackage{amssymb}
\usepackage{amsfonts}
\usepackage{color}
\usepackage{mathrsfs}
\usepackage{bm}
\usepackage[utf8]{inputenc}
\def\Det{{\tt det}\hspace{1mm}}
\def\tr{{\tt tr}\hspace{1mm}}

\makeatother

 2 
 \font\fiverm=cmss9

\def\i{\item}

\def\RR{\vbox{\hbox to 8.9pt {I\hskip-2.1pt R\hfill}}}
\def\NN{\vbox{\hbox to 9pt {I\hskip-2.1pt N\hfill}}}
\def\CC{\vbox{\hbox to 9pt { {\fiverm l}\hskip-4.5pt \bf C\hfill}}}

\title{Discontinuous yielding of pristine micro-crystals}

\sameaddress %If authors have the same address

\author[\firstname{O.U.} \lastname{Salman}\IsCorresp]{\firstname{O.} \middlename{U.}
  \lastname{Salman}\IsCorresp} 
\address{CNRS, LSPM UPR3407,  Université Sorbonne Paris Nord, Villetaneuse, 93430, France} 
\email[O.U.S]{umut.salman@lspm.cnrs.fr}
\author[\firstname{R.} \lastname{Baggio}]{\firstname{R.} \lastname{Baggio}}
\address{CNRS, LSPM UPR3407, Université Sorbonne Paris Nord, Villetaneuse, 93430, France} 
\email[R.B.]{robi.bagg.colpi@gmail.com}
\author[\firstname{B.} \lastname{Bacroix}]{\firstname{B.} \lastname{Bacroix}}
\email[B.B.]{brigitte.bacroix@lspm.cnrs.fr}
\address{CNRS, LSPM UPR3407,  Université Sorbonne Paris Nord, Villetaneuse, 93430, France} 
\endsameaddress %closes group of authors with same addressz
\sameaddress
\author[\firstname{G.} \lastname{Zanzotto}]{\firstname{G.}  \lastname{Zanzotto}}
\email[G.Z.]{zanzotto@dmsa.unipd.it}
\address{DPG, Université di Padova, Via Venezia 8, 35131 Padova, Italy}
\endsameaddress

\sameaddress %If authors have the same address
\author[\firstname{N.} \lastname{Gorbushin}]{\firstname{N.} \lastname{Gorbushin}}
\address{PMMH, CNRS-UMR 7636, ESPCI PSL, F-75005 Paris, France}
\email[N.G.]{nikolai.gorbushin@espci.fr}
\author[\firstname{L.}  \lastname{Truskinovsky}]{\firstname{L.} \lastname{Truskinovsky}\IsCorresp}
\email[L.T.]{lev.truskinovsky@espci.fr}
\address{PMMH, CNRS-UMR 7636, ESPCI PSL, F-75005 Paris, France}
\endsameaddress %closes group of authors with same address



\begin{abstract} 
We study the mechanical response of a dislocation-free 2D crystal under homogenous shear using a new mesoscopic approach to crystal plasticity, a Landau-type theory, accounting for the global invariance of the energy in the space of strain tensors while operating with 
 an infinite number of equivalent energy wells. 
 The advantage of this approach is that it eliminates arbitrariness in dealing with topological transitions involved, for instance, in  nucleation and annihilation of dislocations.  We use discontinuous yielding of pristine micro-crystals as a benchmark problem for the new theory and show that the nature of the catastrophic instability, which in this setting inevitably follows the standard affine response, depends not only on lattice symmetry but also on the orientation of the crystal in the loading device. The ensuing dislocation avalanche involves cooperative dislocation nucleation, resulting in the formation of complex microstructures controlled by a nontrivial self-induced coupling between different plastic mechanisms.  \end{abstract}

\begin{document}

% Use the \maketitle command after the abstract
\maketitle
%% Beginning of text
%% Abridged versions
%% 1. English one if the paper is in French
% \selectlanguage{english}
% \section*{Abridged English version}
% <your text here>
% \selectlanguage{french}
% to go back to main language.

%% 2. French one if the paper is in English
% \selectlanguage{french}
% \section*{Version française abrégée}
%Nous serons brefs.
\selectlanguage{english}
%to go back to main language.

% Example of section
\section{Introduction}

With the advance of nanotechnology and broad fabrication of nano-scale structures, the focus in the study of plastic deformation has shifted to atomic dimensions.  The emerging  science of nano-materials  deals, for instance,   with machine parts printed by chemical vapor deposition or made of  nano-grained metals.  At these manufacturing  scales, external and internal (microstructural) lengths become comparable, and the  dislocation-based description of plasticity comes to the forefront in providing design guidelines for  miniaturized mechanical devices \cite{Lee2009-as,Li2016-cz,Lu2016-ag}.  It was found, for instance,  that sub-micron size objects, serving as components of such systems, are characterized by remarkably high strength. This opened  a way  to a broad range of \emph{novel} engineering applications, including nano-metric machining and hierarchical steels, e.g. \cite{Schaedler2011-gn}. 

Considerable research efforts have been focused on the study of dislocation plasticity in sub-micron crystals in an attempt to assess the extent of the failure of traditional inelastic constitutive models concerning phenomena at these scales  \cite{Mordehai2011-to,Maas2012-ib,Han2015-db,Papanikolaou2017-ld,Maas2018-qu}. As a result of these efforts, it has become clear that the deformation mechanisms, which we habitually associate with plastic flows, change dramatically once the sample size is reduced below the micrometer range. It was found that the strength of such crystals  reaches theoretical (ideal) limit  \cite{Nix1998-xr,Uchic2004-ax,Greer2005-ak,Dimiduk2006-fz} and  that the plastic flow  proceeds through stress drops reminiscent of brittle fracture~\cite{bei2008effects,chrobak2011deconfinement,wang2012pristine,Cui2017-xn}. The attendant intermittency compromises the reliable functioning of ultra-small machinery and jeopardizes our ability to ensure the predictable performance of MEMS and other similar systems~\cite{Csikor2007-ua,Benzerga2009-ny,Uchic2009-jl,Argon2013-fv,Zhang2017-cg}.

Of primary interest to our study will be the phenomenon of \emph{discontinuous yielding} in sub-micron (initially) dislocation free volumes. According to the classical continuum paradigm, the elastic deformation in strained crystals must be followed by either an abrupt brittle-like failure or gradual plastic deformation. Instead, loading of sub-micron crystals revealed a mixed behavior. Micro-crystals, exhibiting smooth classical yield at macro-scales, were shown to fail catastrophically, with conventional work hardening replaced by an abrupt system spanning plastic event ~\cite{Greer2011-av}. %In particular, strained micro-pillars, with only a few initial dislocations, were shown to exhibit (almost) reversible elastic response till they reach high critical stress and then

The precipitous brittle yield, revealing a high level of dislocation correlations and manifesting plastic collapse, was first discovered in metal whiskers ~\cite{Brenner1956-rx,Brenner1958-dr}. Brittleness of this type is now routinely observed in nanoparticles, which “break plastically” while generating a large number of dislocations~\cite{Sharma2018-iw,Mordehai2018-qm}; similar global plastic instabilities have also been recorded during nanoindentation  \cite{Lilleodden2006-jq, Corcoran1997-vt}. The implied system size plastic avalanches result from a large number of highly cooperative individual dislocation nucleation and dislocation glide events, taking place almost simultaneously~\cite{Dimiduk2006-fz,Csikor2007-ua,Maas2013-tq,Lee2020-nx}.

% The spatial and temporal cooperativity relies on long-range spatial and temporal correlations known to characterize generic mature non-equilibrium systems 

%, including size-dependent strength at theoretical limits, and apparent brittleness, 
The distinctive  features of crystal plasticity at sub-micron scales can be attributed to the high degree of structural perfection of such samples, which are usually almost pristine~\cite{He2016-bi,Merabet2018-cv}. In particular, the catastrophic yield depends critically on the absence of dislocations before straining. Despite the global, system-spanning scale of experienced plastic deformation, such  samples produce nearly pristine postmortem microstructures. Almost all dislocations mediating the explosive stress drop manage to escape to the boundaries, sometimes even producing pristine-to-pristine transition~\cite{wang2012pristine,Chisholm2012-ki}. It was also observed that as the deformation volume of a material decreases, the effect of crystal orientation on the operative deformation mechanisms increases; in particular, the mechanical response of sub-micron defect-free nanopillars is different when they are deformed along high symmetry orientations and low symmetry orientations~\cite{Ziegenhain2010-tu,Bagheripoor2020-qm}. The compression tests on confined micropillars also revealed  higher (and therefore closer to theoretical limit)  stresses of massive dislocation nucleation than in the presence of free surfaces~\cite{Bagheripoor2020-qm,Bagheripoor2020-gs}. 

In this paper, addressing  the mechanical response of   dislocation-free  2D crystals under homogenous shear, we attempt to rationalize the above  
observations  using the  mesoscopic model of crystal plasticity \cite{Salman2011-ij,Salman2012-oa,Baggio2019-rs}.  
%We  study  using the new  mesoscopic approach to crystal plasticity, a Landau-type theory,  accounting for the global invariance of the energy in the space of strain tensors while  operating with 
% an infinite number of equivalent energy wells. 
%% We use  athermal dynamics, which in case of quasi-static loading reduces to parametric energy minimization. 
% The advantage of this approach is that it eliminates arbitrariness in dealing with  topological transitions involved, for instance,  in   nucleation and annihilation of dislocations.  
  We use  discontinuous yielding of pristine micro-crystals as a benchmark problem for this new theory and  study in detail how  the nature of the catastrophic instability depends  on  the    lattice symmetry  and the  orientation of the crystal in the loading device. The initial plasticity in pristine nano-scale volumes is of particular theoretical importance as the process dominated by cooperative dislocation nucleation vs more conventional flow mechanisms involving   multiplication and glide  of pre-existing dislocations.

The rest of the paper is organized as follows. In Section 2, we further  motivate our study reviewing the problem from the viewpoint of materials science of sub-micron crystals. We then briefly discuss the existing computational tools.  The  new mesoscopic model is   introduced in Section 3, where we recall the geometrically nonlinear kinematics of crystal lattices and discuss the construction of   strain-energy density respecting the finite strain symmetry of the crystal lattice. In Section 4, we present the macroscopic picture of the discontinuous yield revealed by our numerical experiments and, to justify these results,  compute  the theoretical (ideal) shear strength  for perfect crystals with different   point groups. In Section 5, we  study post-bifurcation behavior of pristine  crystals  and   compare the  representations of post avalanche dislocation  patterns.  Along with their  representation in the physical space, we also visualize such patterns in the configurational space of  lattice strains.  In  Section 6, we briefly discuss the mechanical  response of the simulated  crystals beyond the catastrophic brittle  event. Finally, in Section 7 we   present our conclusions.
 
\section{Some background}

Plastic deformation in crystalline solids involves discontinuous changes in the configuration of nearest neighbors as some stress/strain thresholds are exceeded upon external loading. The associated relative movement of atoms can be interpreted as the creation and propagation of topological lattice defects. The most prominent among them are dislocations, moving along crystallographic planes and leaving behind quantized lattice slip   \cite{Read1955-ps,Cottrell2002-dh,kubin2013dislocations}. Understanding the collective motion of dislocations is the key to control the ductile  failure of crystals as these linear defects are  crucially  involved in  shear band formation, fatigue, and even fracture ~\cite{Differt1986-qi,Antolovich2014-ki,Weiss2019-yl}. 

Dislocation motion is inherently complex due to  long-range elastic interactions  of dislocation lines and strongly nonlinear, threshold type  short-range interactions  of dislocation cores ~\cite{Madec2002-hk,Sethna2017-tm}. Facilitated by lattice trapping, transient and sessile  dislocation patterns  cover a broad range of scales from microscopic (junctions) to macroscopic (grains) ~\cite{Gomez-Garcia2006-ur,Chen2010-sq,Li2011-nf}. 

Despite the presence of microscopic  heterogeneities,  plastic flows appear at the scale of bulk materials as  smooth phenomena amenable to continuous description~\cite{Takeuchi1975-vw}. Therefore,  macroscopic crystal plasticity is usually formulated within classical continuum mechanics and involves macroscale  constitutive relations. The inelastic component of  strain tensor  is parametrized by a finite number of order parameters representing amplitudes of pre-designed plastic mechanisms; in rate independent limit  each of them is assumed to be  governed  by dry  friction  dynamics~\cite{Han2005-uh,Gurtin2009-bl,Roters2010-cw,Stein2018-gr,Forest2019,Salman2021-sb}. The fact that   fluctuations are effectively averaged  out \footnote{Recently continuum  models were proposed with internal fluctuations  accounted for through stochastic constitutive equations~\cite{Weiss2015-eh,Zaiser2019-qm}.},  
opens for continuum plasticity (CP) access to macroscopic time and length scales and allows one to model realistic 3D structures with complex geometries while accounting phenomenologically for such complex effects as hardening, rate-dependence and even polycrystallinity ~\cite{Franciosi1985-cu,Zhang2007-hb,Roters2010-cw,Forest2019-au,Marano2019-yo}. 

However, it has been recently realized that plastic flows in ultra-small samples are beyond the reach of such theories. A description of the cross-over from a `mild' plastic flow in bulk materials to a `wild' scale-free intermittent plastic response at the sub-micron scale requires significant paradigm change \cite{Weiss2015-eh}. In particular, the current CP theory fails to resolve intermittent  stress-drops (or strain-jumps)   at submicron scales \cite{Uchic2004-ax,Dimiduk2006-fz,Zhang2017-cg} and  is also unable to explain catastrophic events  accompanying plastic flows in nanoparticles \cite{wang2012pristine}, nanowires \cite{Lu2011-cu} and nano-pillars \cite{bei2008effects,Issa2015-mq,Hu2018-yn,Zhang2020-ax}. None of these phenomena can be rationalized without a direct reference to dislocation motion.

If in  bulk materials dislocation  motion is largely uncorrelated and can be indeed averaged out, plastic flows in sub-micron crystal involve  highly cooperative dislocational rearrangements. The most striking effect of such cooperativity is provided by intermittent  system size dislocation avalanches, which defy self-averaging and challenge any attempt of  continuous  description. More generally, the scale-free CP fails at sample sizes comparable to the emerging microstructural scale of defect patterning, simply because the latter is assumed to be  zero in the classical  theory.  The  attempts to regularize the CP and link  the   internal length scale with the presence of `geometrically necessary' strain gradients  have been so far  only partially successful as the size effect was also observed in the absence of strain gradients, e.g. \cite{Bittencourt2019-tg}.
 
An alternative approach to  the rationalization of the size effect was to shift the attention  from the role of  gradients in dislocation arrangements to the scarcity of dislocation sources in small crystals. It was noticed that when  sufficiently small crystals were strained to the bulk yield stresses, the rate of dislocation escape to the surfaces grew to become larger than the rate of dislocation multiplication, and therefore, plasticity could  proceed only  by the nucleation of new dislocations at considerably higher stresses~\cite{Shan2008-nh}. The implied 'dislocation starvation' was therefore linked to the fact   the 'breeding' distance for dislocation multiplication becomes  larger than the system size \cite{Rao2008-df,Weinberger2008-qw,Greer2011-av,Bagheripoor2018-nt,Shan2008-nh,Mordehai2011-ys,Shan2008-nh}. 

Following this logic,   the `brittle' response of pristine ultra-small samples after they reached the level of theoretical (ideal) strength,   can be explained by   massive homogeneous nucleation of  dislocations \cite{Plans2007-cx,Miller2008-rr,Garg2016-kz}. In other words, in such `starved' samples   yielding can be expected to proceed as   a dislocation nucleation avalanche which can reach the size of the system \cite{Kiener2011-hc,Oh2009-uf,Benzerga2009-ny,Lee2011-oi,Ng2009-cc}. The implied  cooperative response becomes possible because the defect-free environment allows the nucleated dislocations to dynamically self-organize, taking full advantage of the un-screened  long-range elastic interactions \cite{Miguel2001-fi,Koslowski2004-sa,Uchic2009-rj}. The emerging  mechanical instability ultimately  originates from  strain softening whose origin   will be  revealed as  a part of this study.

 We note that  sharp  peak stresses  at yielding  have been also observed in some  \emph{macroscopic}   samples where the large yield drop is usually followed by a Luders-like flow instability,  e.g. \cite{Li2002-tw, Ruan2019-bx}. 
% Various  theories ,  advanced to explain this phenomenon,    emphasize  low mobile dislocation density in the material before deformation. 
In such cases  the initial mobile dislocation density   is usually  low  either because  dislocations  were annihilated during the annealing process   or  because  they were  blocked by  solute atoms;  in both cases  the discontinuous yielding is  associated with  either collective depinning   from Cottrell atmosphere  or explosive   nucleation   from  nano-sized grain boundary sources. 
%. The grain boundary source is usually less important unless for the crystals with nano-sized grains.
%It may be more appropriate to say that the discontinuous yielding is associated with collective depinning of dislocations from Cottrell atmosphere. The grain boundary source is usually less important unless for the crystals with nano-sized grains. 
% In both case the rapid stress drop during the discontinuous yielding  is associated with  the   formation of the large quantities of mobile  dislocations in a collective avalanche-like process.   
  A theoretical explanation of discontinuous yielding in the prototypical context of transformational plasticity was proposed in \cite{Truskinovsky2004-xp} where the stress peaks were linked  to the difference between (homogeneous) nucleation and propagation thresholds for internal instabilities.

%\subsection{Computational approaches}

The development of the  computational  tools   accounting for the dislocational nature  of plastic flows has  become a priority  because the inadequacy of the conventional methods  based on CP were  placing severe restrictions    on the possibility to model plastically deforming ultra-small structural elements  ~\cite{Zheng2010-ag,Zhang2017-cg,Wang2018-fk,Parakh2020-mk}.  In particular, it was realized that to simulate \emph{discontinuous yielding} in sub-micron crystals  which   involves  massive nucleation of dislocations,  the   modeling approach cannot ignore lattice effects and must  necessarily  account for phenomena at the scale of dislocation  cores.  Capturing the attendant pattern formation   is   also challenging because the nucleated dislocations  continue to  interact strongly at  many different scales
%,   from  long-range interactions due to the elastic far-fields all the way to short-range reactions involving deformations of the defect cores, exemplified  by annihilation and multiplication phenomena
~\cite{Xia2015-vo,Clouet2015-zk,Salvalaglio2019-kt,Van_der_Giessen2020-gl,Bertin2020-hv}.

%With the modern trend  of extreme minituarization of structural elements  and the   increased tendency to manufacture composite materials with sub micron crystal grains the development of computational  tools compatible with 'dislocation driven'  plastic flows has became unavoidable. 

%It would be  desirable to incorporate  the discrete features of  small-scale plasticity into  the  continuum crystral plasticity (CP)   framework, which   reduces computational cost  drastically and usually allows for   straightforward analytical and numerical treatment. However, capturing dislocation self-organization  within a continuous description is  challenging because dislocations   interact strongly at  many different scales,   from  long-range interactions due to the elastic far-fields all the way to short-range reactions involving deformations of the defect cores, exemplified  by annihilation and multiplication phenomena~\cite{Xia2015-vo,Clouet2015-zk,Salvalaglio2019-kt,Van_der_Giessen2020-gl,Bertin2020-hv}.

Numerous computational  alternatives to CP   have been developed  targeting different time and length-scales. They range     from  molecular dynamics (MD) \cite{Niiyama2015-gv,Zepeda-Ruiz2017-gu} and   similarly microscopic  phase-field crystal (PFC) method \cite{Salvalaglio2020-eb,Chan2010-ha},  to the more coarse grained   phase-field dislocation dynamics    (PFDD) \cite{Finel2000-um,Koslowski2002-dn,Hunter2011-vp},   discrete dislocation dynamics (DDD) \cite{Vattre2014-be,Ispanovity2014-ra,El-Awady2016-ea,Bertin2019-bb}, and   continuum dislocation dynamics (CDD) \cite{Varadhan2006-xb,Hochrainer2014-sk,El-Azab2020-fu}. The multi-scale   quasi-continuum  model (QC)  attempts to bridge all the scales while accounting for  each of them fully comprehensibly \cite{Shenoy1999-gh,Miller2002-pa}. 
%However, the closest to the approach we choose  is  the  phase-field dislocation dynamics method  (PFDD) \cite{Hunter2011-vp,Koslowski2002-dn}. Each of these computational platforms  was created to address  particular time and length-scales and usually higher-scale models (more coarse-grained) rely on  input from lower-scale models (less coarse-grained). To justify our choice, we now briefly recall the advantages and the shortcomings of each of the proposed approaches.

MD approaches, including  Molecular Statics (MS), and Density Functional
Theory (DFT), accurately represent   micro-mechanisms of plastic response while  relying  minimally on phenomenology~\cite{Cia2012-pj}. 
MD simulations have been particularly   instrumental in the study of the homogeneous and heterogeneous dislocation nucleation~\cite{Miller2008-rr,Zepeda-Ruiz2017-gu,Parakh2020-mk}. 
%They were also used to study  the role of short range interactions in  plastic intermittency~\cite{Niiyama2015-gv},  the irreversible behavior of colloidal polycrystals~\cite{Moretti2011-xw}, the mobility of edge dislocations~\cite{Queyreau2011-ar} and the peculiarities of grain boundary migration~\cite{Baruffi2019-ii}, to mention just a few of the relevant effects. 
The only  relative shortcoming of the  atomistic simulations  is that they are  still computationally rather expensive   in most applications, even at the small time and length scales of interest;  also  the problem of mapping to the macroscopic description in terms of the measurable quantities like stresses and strains  is  apparently  not yet  fully resolved ~\cite{Zimmerman2001-cz,Zimmerman2009-my,Zepeda-Ruiz2020-cl,Lim2018-bj}. Partial temporal averaging of   atomistic molecular dynamics has emerged in the form  a coarse-grained continuum  PFC theory    \cite{Elder2002-qt}. The PFC approach   was applied to the description of   an ensemble of few interacting dislocation dipoles and was able to  capture some cooperative features of crystal plasticity ~\cite{Chan2010-ha,Skaugen2018-rf,Salvalaglio2020-eb}. However,  such a detailed description of atomic lattices still remains rather demanding in terms of computer time, at least  in the case of  developed plastic  flows with realistic  number of interacting dislocations.

Discrete dislocation dynamics (DDD) approach was created to  overcome the short time and length scales  of atomistic methods. This model treats adequately    long range interaction of dislocations without  resolving   the  fine structure of the dislocation cores.    To account for short range interactions, specific `local rules' must be added governing, for instance,  intersections and locks and  some particular  lattice scale  effects  have been indeed successfully  modeled  with the help of such additional phenomenological constructs, typically motivated by atomistic simulations~\cite{Kubin1992-bs,Devincre1992-mu,Cazacu2013-no,Po2014-qu,Wang2014-sb, Geslin2017-fb,Kohnert2021-ts}.  Being reinforced in this way,  the  DDD approach has emerged as an extremely useful approach for modeling  evolution of many interacting dislocations~\cite{Cai2006-fe,Dmitrieva2010-qn}. However, it  still  remains challenging  to account in the DDD framework for large  deformations, crystal symmetry, lattice rotation, as well as the emergence of non-dislocational defects.  Coarse graining of DDD,  which opens the way towards modeling of  dislocation patterning, has been attempted in the framework of the  mesoscopic  CDD methods where dislocation microstructures are modeled by continuum  dislocation density fields \cite{Starkey2020-qb,Acharya2006-nm,Sandfeld_undated-nm}. While various  phenomenological closure relations have been proposed to model the evolution of the  
dislocation density, the systematic  development of this approach is hindered by the fact that   \emph{rigorous} statistical averaging in the ensemble of strongly interacting dynamic defects is still  a big challenge~\cite{Valdenaire2016-yp}.

In search of a micro-macro compromise a  QC  approach was   proposed  in \cite{Tadmor1996-qi} and  then significantly developed in  \cite{Shenoy1999-gh,Miller2002-pa,Dobson2007-ax,Sorkin2014-jy,Kochmann2016-bv}. It is based on the observation that a fully atomistic resolution is  necessary  only in small spatial regions, while in most of the modeling domain the deformation fields  can be  represented  by  the classical continuum theory. A   necessity of patching the continuum and discrete subdomains poses, however,   a complex problem ~\cite{Miller2002-pa,Tadmor2011-mo}. The QC method has been successfully used in many applications, including the study of nano-indentation, deformation of grain-boundaries and crack tip evolution~\cite{Rodney1999-em,Knap2003-pp,Yu2017-mg,Jin2018-ty}.

Of particular interest to  our study is the local version of the QC method representing a meso-scale compromise between  continuum and atomistic approaches~\cite{Tadmor1996-qi,Sorkin2014-jy}; a closely related approach  is  the interatomic potential FEM (IPFEM) \cite{Van_Vliet2003-yg,Zhu2004-wr}.   In these effectively single scale methods  the spatial domain is subdivided into discrete elements, as in a finite element method (FEM), and  the  deformation in each element is taken to be piece-wise affine.  The  energy density  of an element, which then depends only on the displacement gradient at this point,  is  evaluated using  interatomic potentials and the   Cauchy-Born rule \cite{Ericksen2008-kx,Weinan2007-hd,Steinmann2006-qw,Podio-Guidugli2010-hx}. These approaches were shown to capture adequately the basic structure of  dislocation cores,  even though  they can misrepresent  some truly atomistic features like  deformations varying rapidly within the cut-off distance of the interatomic potential. 

A  mesoscale approach to crystal plasticity, which also allows one to treat dislocations in a fully continuum framework,  and which is particularly  relevant for our own development, is the  phase-field dislocation dynamics (PFDD)~\cite{Koslowski2002-dn,Rodney2003-wy,Beyerlein2016-av,Ruffini2017-fs}. This method has evolved from  the original  Landau  approach to the modeling of phase transitions~\cite{Chen2002-cn,salman_thesis,Finel2010-rd,Salman2012-zg,Shchyglo2012-nz,Salman2019-cg}.   In PFDD lattice  slips are described  by scalar order parameters and the energy wells  represent quantized lattice invariant shears ~\cite{Jin2001-fw,Rodney2003-wy,Zheng2018-zn};   transition layers,  separating    regions with different amount of shear,    represent the locations of dislocation lines.  The Landau energy functional couples the  \emph{tensorial} linear elastic    energy with the\emph{ scalar}  lattice energy whose periodic structure  is usually  informed by  atomic scale simulations based on the Cauchy-Born rule.  The PFDD method  enables simulations of much larger crystal sizes and much longer time scales  than  atomistic simulations~\cite{Hu2001-cp,Chen2002-cn,Louchez2017-ui,Qiu2019-ue}. While  the extensions  of PFDD to finite strains have  recently appeared \cite{Biscari2016-qy,Javanbakht2016-dr}, the remaining challenge is that the   structure of lattice invariant shears is  resolved  by scalar order parameters only  approximately  \cite{Xu2019-ex,Li2017-dz,Salvalaglio2020-eb}. 

% One  challenge for this method is  the rationalization for the choice  of the internal length scale controlling the size of the dislocation cores.  Since in PFDD another  challenge is to maximally adequately reproduce the underlying tensorial periodicity of the  energy in this restricted setting.
 
% It  has been originally  used to study equilibrium configurations involving only few  individual dislocations  ~\cite{Hu2001-cp,Chen2002-cn}.
%However in   recent years it  has been   successfully  applied to  the  study of dislocation dynamics and kinetics, to the modeling of  grain boundaries,  to the explanation of the texture formation  in thin films  and for  the prediction of the fine structure of  dislocation networks, among other equally important problems  \cite{Louchez2017-ui,Qiu2019-ue}. Extensions of PFDD to finite strains have been developed as well \cite{Biscari2016-qy,Javanbakht2016-dr}. 

\section{Mesoscopic  tensorial  model (MTM)}
 
We have seen that despite many important advances  in various specific problems, the sufficiently versatile  computational method, allowing for  natural coupling of different   plastic `mechanisms'  while  addressing  realistic space and time scales \footnote{Allowing one, for instance, to resolve the system size dislocation nucleation avalanche for  micron size  crystals.}, is still missing.  The  implied challenge is to   build  a synthetic mesoscopic tensorial  approach dealing with large   strains, while accounting correctly for  both  anisotropy and   discreteness of  the simplest Bravais  lattices.
 
Different attempts along these lines can be found in the literature.  The corresponding    scalar approaches \cite{Landu1994-wt,Carpio2003-yu,Salman2011-ij} can be viewed as  generalizations of the  minimalistic  1D  Frenkel-Kontorova  model \cite{Frenkel1939-tk,Peierls2002-xo,Nabarro2002-js} for the 2D case when only one slip system is activated.  Despite their simplicity such models have been successful in describing  dislocation cores \cite{Kovalev1993-se},  in  simulate dislocation nucleation  \cite{lomdahl1986dislocation,srolovitz1986dislocation,Plans2007-cx,Bonilla2007-jk,Geslin2014-ad} and  even in modeling of plastic intermittency \cite{Salman2012-oa,Zhang2020-ax}.
The  tensorial   models   with    \emph{linearized} kinematics were proposed  in \cite{bulatov1994stochastic, Minami2007-ew,Onuki2003-ln,Carpio2005-fl}.  They produced  a more realistic  picture of plastic flows but  still could not account adequately  for   lattice-invariant  shears  associated with finite plastic  slip~\cite{kaxiras1994energetics}.  

The first attempts to incorporate  the effects   of geometrical nonlinearity were made in the context of a model of reconstructive phase transitions   allowing for partial plastic accommodation  \cite{Conti2004-sv}.  A finite strain model  focused    directly on the modeling of  plastic deformation was first developed for the case of  highly anisotropic  lattices of  HCP type with a single slip system \cite{Salman2011-ij}.  Both of these papers can be viewed as different  realizations of the  original program of   Ericksen  who proposed that in nonlinear elasticity the energy periodicity   should be made compatible with geometrically nonlinear kinematics of Bravais lattices \cite{Ericksen1970-rx, Ericksen1973-yt,Ericksen1977-pj,Ericksen1980-km},  see  also the  subsequent  important  developments of the mathematical formalism in \cite{Parry1976-zt,Folkins1991-em,Parry1998-sv,pitteri2002continuum}.  
 
Behind the coarse-grained  approach of Ericksen was the  general assumption  that meso-scale material elements are exposed to an effective  energy landscape which is globally periodic due to the presence of   lattice invariant shears.
 From the perspective of such Landau-type continuum  theory with an infinite number of  equivalent energy wells,  plastically deformed crystal emerges as   a multi-phase mixture of    equivalent   `phases'.   Plastic yield  can be then    interpreted as  an escape from the    reference  energy well  and   plastic `mechanisms'  can be linked      to  low-barrier valleys in the  energy landscape.  Friction type dissipation controlling dynamics  in continuum crystal plasticity  emerges in such theory as a result of a homogenized description  of an overdamped athermal dynamics in a rugged energy landscape \cite{Puglisi2005-lg,Mielke2011-ck}. 

The 2D scalar model  of crystal  plasticity  presented in \cite{Salman2011-ij,Salman2012-oa}  was effectively a single slip system  version  of such theory.  Building upon \cite{Conti2004-sv},  its  2D  tensorial version was presented in \cite{Baggio2019-rs}. Its main advantage is that it   incorporates   adequately the full   $GL(2,\mathbb Z)$  symmetry of the Bravais lattices~\cite{Ericksen1970-rx,Ericksen1980-km}. The invariance described by the \emph{infinite} group $GL(2,\mathbb Z)$ is different from the  one associated with the classical \emph{finite}  point group  limited to orthogonal transformations.  The global group $ GL(2,\mathbb{Z})$ is much broader,  accounting not only  for orthogonal transformations but also for non-orthogonal transformations representing  lattice invariant shears.                   
%The symmetries  described by the groups $P(\mathbf{e}_I)$ and $G(\mathbf{e}_I)$   are compatible, as the first can be obtained from the second, provided that the global group  is restricted to a suitable neighborhood, known as the \emph{Ericksen-Pitteri neighborhood} (EPN) \cite{pitteri1984reconciliation}. Intuitively, the  EPN  is the sub domain where the system deforms "elastically".  
Recall first that in 2D  a   Bravais lattice  can be viewed as an  infinite integer translation of two linearly independent vectors or,  more formally, it is a set of vectors   $\mathbf{x}\in\mathbb{R}^2$ such that $ \mathbf{x}=
v_K{\mathbf{e}}_K,$ with $ v_K\in{\mathbb Z}.$
%$\{\mathbf{e}_K\}$:
%$
%  % {\mathcal{L}}({\mathbf{e}}_I)=
%  \left\{
%   \mathbf{x}\in\mathbb{R}^2,\,\mathbf{x}=
%v_K{\mathbf{e}}_K,\, v_K\in{\mathbb Z}\, \right\}\,.
%$
%Bravais lattices are also referred to as simple lattices  
%We can associate with a basis $\{\mathbf{e}_K\}$ a \emph{metric tensor} $\textbf{C}$ with components 
%$
%   C_{KL} =    \mathbf{e}_K \cdot \mathbf{e}_L  
%%\quad   (1\le I, J \le 2)\quad.
%$
%The tensor $\textbf{C}$ is always symmetric and positive definite.
For a given Bravais lattice, there are infinite choices for the basis $\{\mathbf{e}_K\}$.  The associate global symmetry group is the maximal subgroup of the group  of invertible linear mappings  leaving the  lattice invariant. In particular, for  2D Bravais lattices it is  $GL(2,\mathbb{Z}) = \left\{
%{\bf M},\, 
M_{KL}\in{\mathbb Z},\, {\det(\bf M)}=\pm1 \right\},
$ where $K,L=1,2$.
 %\emph{global symmetry group}.   
%$G({\mathbf{e}}_I)$ of lattice $\mathcal{L}({\mathbf{e}}_I)$.
% It is 
%\begin{align}
%\label{global}
%\begin{split}
%G({\mathbf{e}}_I):=\left\{ {\bf H}\in Aut :\quad\mathcal{L}({\bf He}_I)=\mathcal{L}({\bf e}_I) %\right\}\\                  
%=\left\{ {\bf H}\in Aut :\quad{\bf He}_I=m_{JI}{\bf e}_J,\quad {\bf m}\in GL(2,\mathbb{Z}) %\right\}\,.
%\end{split}
 %
 %In other words, the invariance within this group means that we may choose for the same lattice an infinite number of different bases related through 
%$
%{\bf \bar {e}}_K  =  M_{KL}  {\bf e}_L $.
%The corresponding symmetries  for  metric tensors  $\textbf{C}$  are  given by
%${\bf \bar C} = {\bf M}^T{\bf C} {\bf M}$.
%To summarize, every time that we perform a change of basis we are operating within the group $GL(2,{\mathbb Z})$. Similarly, every time that the lattice is deformed into a symmetrically equivalent structure, such deformation can be also described as the action of $GL(2,{\mathbb Z})$ on the initial configuration. Equations (\ref{eq:invba}) and (\ref{eq:mtCm}) describe the action of the global symmetry group $GL(2,{\mathbb Z})$ on the configurational space $\mathcal{B}$ of basis vectors (4D space of linearly-independent 2D vectors) and on the space ${\mathcal Q}^+_2$  of lattices metrics (the 3D space of positive definite symmetric second order tensors $\in \mathbb{R}^2$). In what follows we will mostly refer to the invariance with respect to configurational space $\mathcal{Q}^+_2$, which naturally accounts for frame indifference.  
In other words, the  global symmetry group of a 2D Bravais lattice   is  represented  by   $2\times 2$ invertible matrices with integer entries and determinant $\pm1$.     
 
In the   kinematically   nonlinear  Landau type theory  with $GL(2,\mathbb{Z})$ symmetry the role of the order parameter is  played by the  metric tensor (defined below), and    the  bottoms of the energy wells correspond to lattice invariant deformations. 
%The  energy density,  respecting  locally the symmetry imposed by a  point  group,  is also invariant under  the global infinite symmetry group.  
  Since the    ground state in the continuum  theory of this type   is  necessarily degenerate (hydrostatic)  \cite{Ericksen1973-yt,Fonseca1987-pd},  the  regularization is necessary, as in the case of PFDD or any other Landau-type theory. In \cite{Baggio2019-rs}   such regularizing internal length scale was associated with  the size of the  mesh generating discrete  elastic \emph{elements}. In other words, the mesh scale was treated as a \emph{physical} parameter  which defines the meso-scale.  Given  the large magnitude of the `transformation strain',   different  `phases'  ended up being    localized at the scale of such mesoscopic \emph{elements}  and the domain boundaries appeared macroscopically  as linear defects  mimicking  dislocations. 
 
In what follows we refer to the approach proposed in \cite{Baggio2019-rs}  as the  mesoscopic tensorial model  (MTM).  The main advantage of this approach is that it is formulated in terms of macroscopically measurable quantities (stress and strain)  while being able to distinguish between different symmetry induced  configurations of dislocation cores. It can therefore account  adequately  for both long- and short-range interactions between dislocations.  Most importantly, it allows for topological transitions associated with dislocation nucleation and annihilation
 even though the corresponding `reactions' appear as blurred  on the scale of  regularization. Last but not least, in the MTM  approach the interaction of dislocations   with various obstacles,  including pinning  by impurities and  depinning from nucleation sources, can be handled  without introducing  ad-hoc relations. 

%It is important to notice that such  approach   incorporates both  long- and short-range  dislocation interactions 

% which can be also viewed as a scalar reduction of the local QC theory  and the PFDD  approach based on 
%
%A tensorial version of such a 2D model  based on nonlinear elasticity and aimed first at the description of  reconstructive phase transitions  was    proposed in \cite{Conti2004-sv}.  It  incorporated   adequately the full tensorial symmetry of the crystal lattice. 
%The  main assumption was  that meso-scale material elements are exposed to the  periodic   energy landscape. 

%The model \cite{Salman2011-ij,Salman2012-oa} was  shown to be  reducible  to a computationally effective integer-valued discrete
%automaton and capable of  dealing  with millions of mesoscopic elements and tens of thousands of dislocations. Despite its geometrical  simplicity this scalar model could  account  for both short- and long-range elastic interactions and describe both  dislocation nucleation and immobilization without introducing  any  phenomenological assumptions. Most notably, it  allowed one  to accumulate sufficient statistics  and adequately capture the  inherently intermittent nature of   plastic flows in highly anisotropic HCP crystals. 

In this paper  the MTM is  used to  conceptualize  the brittle-like response of nominally ductile sub-micron crystals.  Through a series of  numerical experiments we uncover  the origin of  the correlated dislocational response of  such crystals with different symmetry  at  loading  levels approaching  the theoretical (ideal) strength.   In particular, we reveal  the mechanism of  the   observed orientation dependence of the   degree of `brittleness' and rationalize the  system size dependence of the associated collective  nucleation event.

\subsection{$GL(2,\mathbb Z)$  energy density}

For simplicity,  we   deal  in this paper   with the simplest two dimensional Bravais lattices:  square (with lower, square symmetry) and triangular (with higher, hexagonal symmetry). In addition to its theoretical interest, the study of plasticity in 2D crystals is technologically relevant because such crystals  have been  recently found important in many applications \cite{Bhimanapati2015-bw,Chen2016-bd,Hoang2016-hm,Zhang2017-mb,Akinwande2017-mf,Chen2018-gt,Kryuchkov2018-ud,Van_Hoang2019-ob,Ma2020-mi}. 

The main ingredient  of the  MTM approach   is the   objective elastic energy density which  respects
%
%landscape.   The ruggedness reflects the \emph{global} symmetry of the original crystal described by the (infinite and discrete) symmetry group
%$GL(d,Z)$,   constituted by the $d\times d$ invertible matrices with integer entries and determinant $\pm1$; here  $d$ is the spatial dimension  ~\cite{Ericksen1970-rx,Ericksen1980-km,Conti2004-sv,Ericksen2008-kx}.  In the resulting   Landau type theory with an infinite number of equivalent energy wells  the role of the order parameter is  played by the metric tensor (characterizing local deformation)  and the bottoms of the energy wells correspond to lattice-invariant deformations.
the  $GL(2,\mathbb Z)$ symmetry.  To construct such energy, we need to first introduce   the  deformation  of  a  continuum  body $\bold y=\bold y(\bold x)$,  where $\bold y$ and  $\bold x$ are the position vectors in  the current  (Eulerian) configuration  and the  reference (Lagrangian) configuration, respectively. If  ${\bf F}  $ is the deformation gradient  $\nabla \bold y  = \partial \bold y/\partial \bold x$, the  frame indifferent elastic  energy density  $\Phi (\bold C)$ can  depend  on $\bold F$ only through the metric tensor $\bold C = \bold F^T\bold F$.  To account adequately for all deformations that map a Bravais lattice onto itself, we require that the strain energy density   satisfies
$\Phi (\bold C) = \Phi (\bold M^T\bold C\bold M)$
      for any matrix $\bold M$  in $GL(2,\mathbb Z)$. 
      This invariance  follows   from the fact that the same    lattice  can be generated by the  two sets of lattice vectors ${\bf \bar e}_K$ and ${\bf e}_K$     if and only if  $  {\bf\bar e}_K  =  M_{KL}{\bf e}_L$. 
       In the presence of such symmetry, the space of metric tensors $\bold C$ with components 
      $ C_{KL}={\bf e}_K.{\bf e}_L$
        partitions into periodicity domains.  Therefore, when the energy is known in one of the  domains, it can be automatically extended to all  other equivalent domains by the use of global tensorial periodicity. 
%       In other words, if  we know the energy in one of such domains we  can use periodicity 
%       %, for instance, the Lagrange reduction mapping \cite{Engel_undated-vp} 
%       to find it  in any other domain. 

\begin{figure}[h!]
\includegraphics[scale=.23 ]{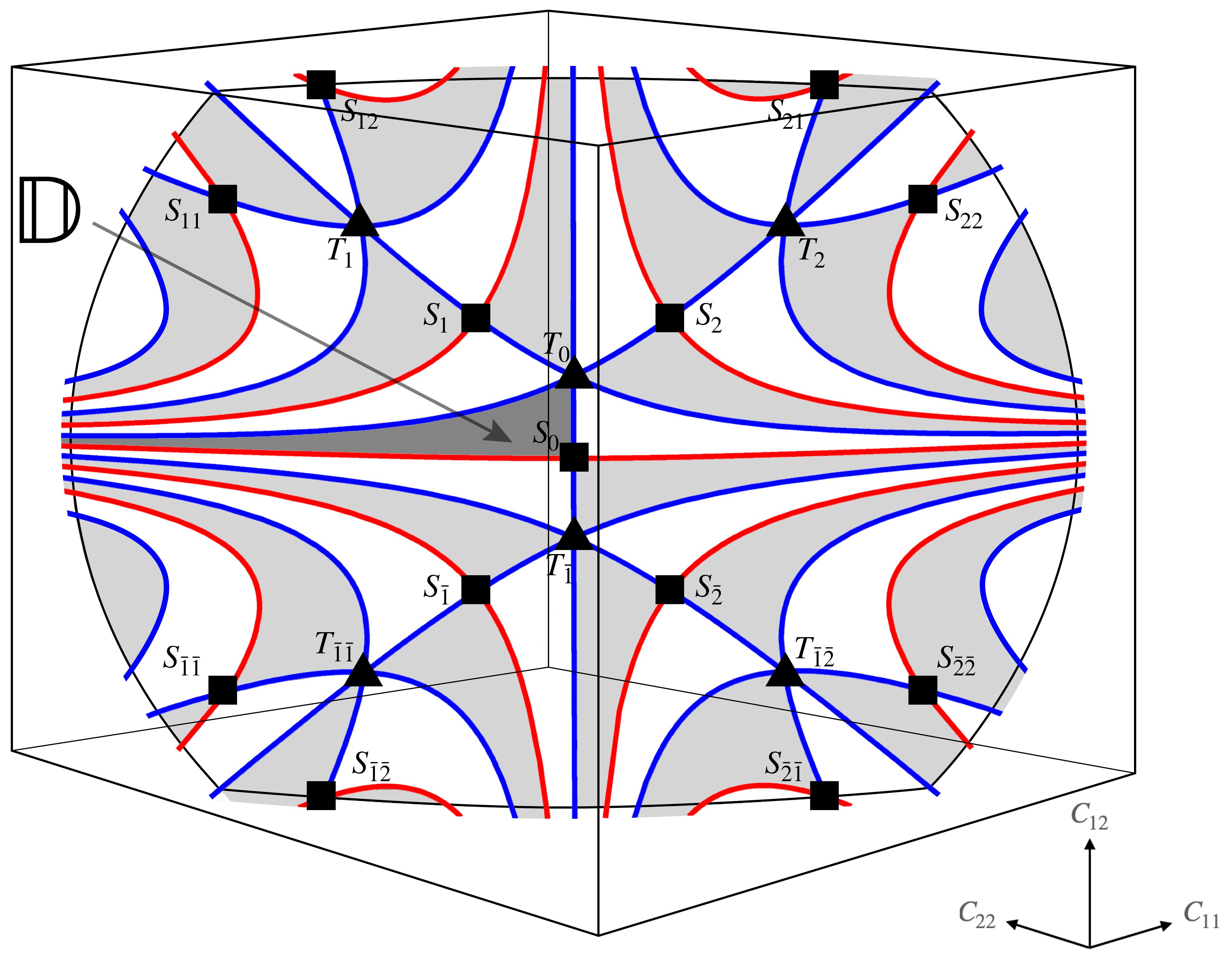}
\caption{\scriptsize {The structure of the  $GL(2,\mathbb Z)$ periodicity domains  in  the space of metric tensors ${\bf C}$. The presented  portion of the infinite  surface $\det{\bf C}=1$  represents  simple lattices with  $\det{\bold C}>0$.  The projected section of the original hyperbolic surface is defined by the inequalities 
%which identifies an hyperboloid in the $C_{11}$,$C_{12}$,$C_{22}$ space.
% We used the portion of the infinite surface $C_{11}C_{22}-C_{12} = 1$ whose the coordinates belong to the interval
  ${0<C_{11} <6}, {0<C_{22}<6}, \text{ and } {-3<C_{12}<3}.$}}
\label{fig:detc1}
\end{figure}

The  space of metric tensors in the 2D case  is three dimensional and in Fig. \ref{fig:detc1} we  show  a two dimensional  section of this  space defined by the condition $\det \bold C = 1$.  For better visibility equivalent  tensorial  periodicity  subdomains are shown in  two alternating colors,  gray and white.  Each of these subdomains contains necessarily one and only one copy of a particular Bravais lattice as shown in Fig. \ref{fig:detc1}.  In this figure, equivalent zero-parametric families of square lattice configurations are illustrated with solid squares $\mathrm{S_i,S_{ij},...S_{\bar i},S_{\bar{i}\text{ }\bar{j} },.}..$  while solid triangles $\mathrm{T_i,T_{ij},...T_{\bar i},T_{\bar{i}\text{ } \bar{j}},..}.$ corresponding to similar families of equivalent triangular lattices.  Red lines correspond to one-parametric families of rectangular lattices and blue lines to similar families of  rhombic lattices. Finally,  the open 2D domains correspond to two-parametric families of equivalent oblique (monoclinic) lattices, see also \cite{Cayron2019-vs,Gao2020-qy,Gao2020-bx,Gao2020-zy}. Note that in this representation we do not see the  `volumetric' part of the  metric tensor responsible, for instance,    for void   formation ~\cite{Marconi2005-cp}.

The darkened area  in Fig. \ref{fig:detc1},  denoted  by  $\mathbb D$,   is known as the ‘fundamental’ domain  \cite{Engel1986-tm}.   Among all the equivalent metrics describing  symmetry-related lattices, this domain contains the  "minimal" ones, known as   ‘reduced forms of Lagrange’.  
%because  all other  periodicity domains  are   its   copies.  
%As we have already seen, points in the interior of $\mathbb D$ correspond  to metrics (and thus lattices) with trivial symmetry, while points on its boundary  correspond to metrics possessing nontrivial symmetries. 
%The chosen fundamental domain 
An arbitrary basis $\bold e_i$  producing an arbitrary metric $ \bold C $ has its   unique `Lagrange reduced'  copy $\tilde{\bf e}_i$ with the corresponding metric  $ \tilde{\bold C} $ in  $\mathbb D$. 
 
The Lagrange reduction is obtained using the following rule:   among all symmetry related copies  of the original basis, the reduced one  is formed by  the shortest non-colinear  vectors  with an acute angle between them, see for instance  \cite{Conti2004-sv}. Therefore, the   domain  $\mathbb D$  can be  defined as:
\begin{equation}
 \label{eq:reduced}
\mathbb D = \bigg\{0<C_{11}\leq C_{22},\text{   }
0<C_{12}\leq \frac{C_{11}}{2}\bigg\}.
\end{equation}
Inside this domain,  the  `Lagrange reduced'   square  lattice,  with  basis vectors aligned with the close-packed directions  ${\bf e}_1=(1,0)$, ${\bf e}_2=(0,1)$,  corresponds to point  $S_0$ in Fig. \ref{fig:detc1}. Instead, the  `Lagrange reduced'  triangular lattice with its  basis  vectors   ${\bf h}_1=\gamma(1,0)$  and ${\bf h}_2=\gamma(1/2,\sqrt{3}/2))$, where $\gamma =  (4/3)^ {1/4}$,  corresponds to the point  $T_0$ in Fig. \ref{fig:detc1}.
 
In view of the  $GL(2,\mathbb Z)$  periodicity,  it is sufficient  to know the elastic energy for $ \tilde{\bold C} \in \mathbb D$.  
%
%function  $\Phi^0 ( \tilde{\bf C})$ for $ \tilde{\bold C} \in \mathbb D$ which, in particular,  implies   that $\det \bold C = 1$ ({\bf this is not necessary, the volumetric part is not influced by the Lagrange reduction}).   
One can then use the symmetry to write   $\Phi(\bold C) =\Phi^0 ( \tilde{\bf C}), $ where  $\tilde{\bf C}=\bold M^T\bold C\bold M$  is  the Lagrange reduced metric,  and   the function $\Phi^0$  is defined only inside $\mathbb D$. 
%to an  arbitrary periodicity domain characterized by the nontrivial matrix $\bold M$. 
The    condition that the function $\Phi^0$  is twice-differentiable can be satisfied if  the deviatoric part of  $\Phi^0$ is a sixth order polynomial \cite{Parry1976-zt, Parry1998-sv}.  With  hexagonal symmetry chosen as global, one  can write   the minimal  expression for such polynomial \cite{Conti2004-sv} 
 %the general Lagrange reduced function}  $\Phi_d$,  
\begin{equation}
\Phi^0_{d}(\tilde{\bf C}/(\det  {\bf C})^{1/2}) =\beta\psi_1(\tilde{\bf C}/(\det  {\bf C})^{1/2})+ \,\psi_2(\tilde{\bf C}/(\det  {\bf C})^{1/2}),
\label{eq:sigCZ}
\end{equation}
where, $
\psi_1={I_1}^4\,I_2 - 41\,{I_2}^3/99 +
7\,I_1\,I_2\,I_3/66 + {I_3}^2/1056, $ and $
\psi_2 = 4\,{I_2}^3/11  + {I_1}^3\,I_3 -  8\,I_1\,I_2\,I_3/11  +  17\,{I_3}^2/528. 
$
and  we used  the following hexagonal invariants  
$
I_1 =  \frac{1}{3} (\tilde{C}_{11} + \tilde{C}_{22} - \tilde{C}_{12})$, 
$I_2= \frac{1}{4} (\tilde{C}_{11} - \tilde{C}_{22})^2 + \frac{1}{12}(\tilde{C}_{11} + \tilde{C}_{22} -
4 \tilde{C}_{12})^2$, and 
$I_3 =  (\tilde{C}_{11} - \tilde{C}_{22})^2 (\tilde{C}_{11} + \tilde{C}_{22} - 4 \tilde{C}_{12}) - \frac{1}{9} (\tilde{C}_{11} + \tilde{C}_{22} -
4 \tilde{C}_{12})^3$. 
The elastic energy density \eqref{eq:sigCZ}  depends on a  single    parameter $\beta$ that may be used to enforce  a particular  symmetry on the reference state. For instance,  the choice   $\beta=-1/4$  guarantees that  the  global  energy minimizers  correspond to  square  lattices while the choice $\beta=4$ shifts the  bias towards the triangular lattice.  In what follows, the total elastic energy will be then taken in the form $ \Phi^0( \tilde {\bf C}) =  \Phi^0_d(\tilde{\bf C}/(\det  {\bf C})^{1/2})+\Phi_v(\det  {\bf C})$.  The   additive volumetric part of the energy density, which   plays only minor role in this study by affecting the structure of dislocation cores and controlling the formation of voids, %are effected; Bauschinger effect even in fully symettric loading paths such as $ \phi=0$}; 
%needs to be mentioned briefly, a more natural choice is a Lennard-Jones like form}, 
 will  be chosen in the  form  $\Phi_v(s)=\mu  (s-\log(s))$. This choice, instead of a more realistic Lennard-Jones type potential, will allow us to  avoid  various  volumetric instabilities  while still excluding infinite  compression. Moreover,  to ensure  that the strain field  remains  close to the surface $\det \bold C = 1$ in our numerical experiments we choose large value of the bulk modulus by setting $\mu=25$.
 
We remark that the particular potential  \eqref{eq:sigCZ} was chosen for illustrative purposes only and for most applications in crystal plasticity the piece-wise quadratic potentials with $GL(2,\mathbb Z)$ symmetry would be sufficient, see for instance \cite{Salman2012-oa,Zhang2020-ax}. Outside crystal plasticity, the general approach of MTM can be used as well with potentials  satisfying less restrictive symmetry constraints, see for instance, \cite{Friesecke2002-hb}.
 
\begin{figure}[tbp]
\centering
\includegraphics[scale=.3]{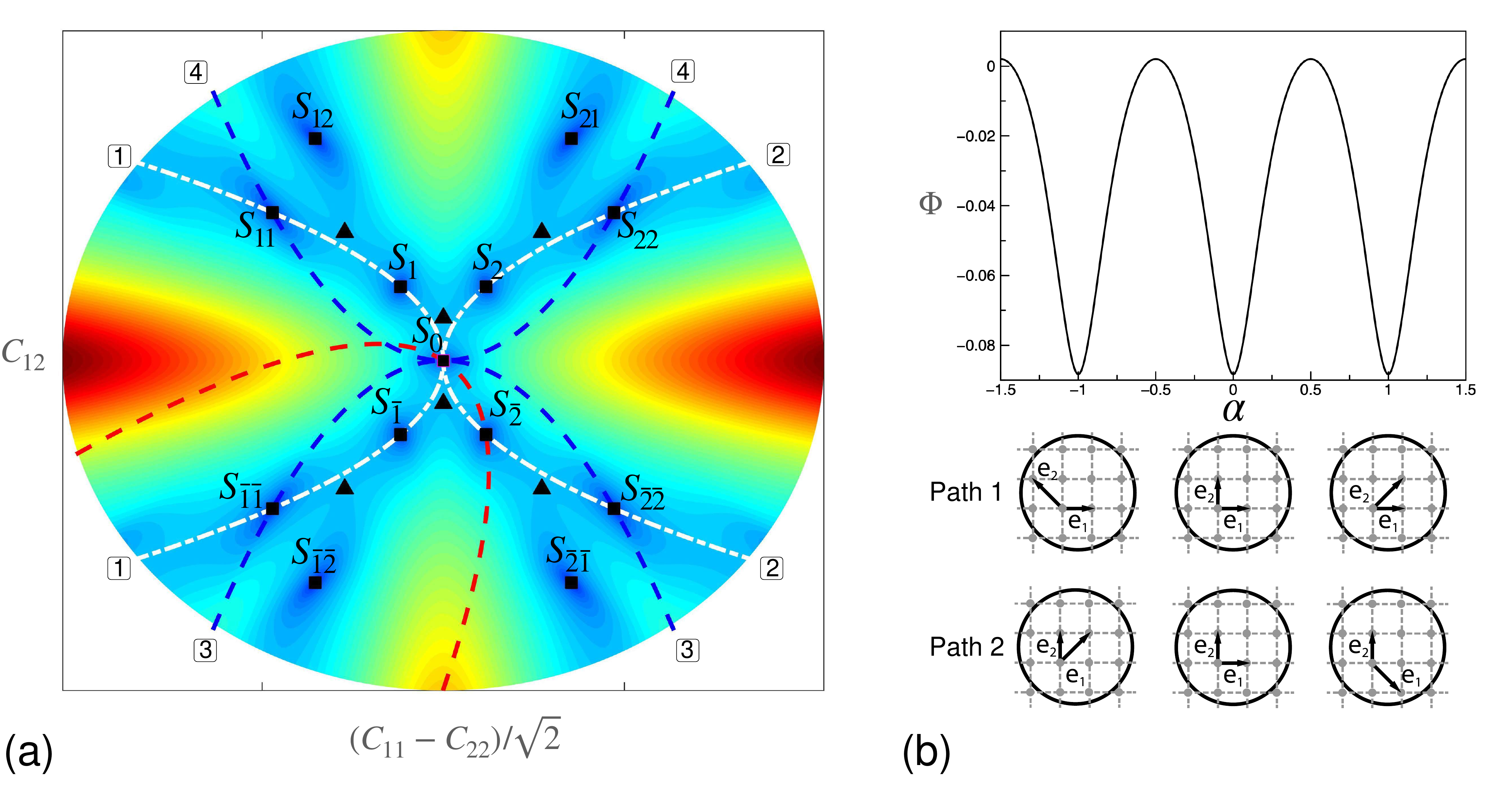}
\caption{\scriptsize {(a) The level sets of  the strain-energy density   (\ref{eq:sigCZ})  with $\beta=-1/4$. Colors indicate the energy level: blue, low; red, high. Dashed lines correspond to the  simple shear loading paths $\bold F(\alpha,\phi)$ defined  in  \eqref{paths_sq}.
%The white lines correspond to the two symmetric shears along dense planes of a square crystal $\bar{\bf F}(\alpha,0), \bar{\bf F}(\alpha,\pi/2)$. 
(b) The periodic strain-energy profile along the two symmetric paths $1-1$ and $2-2$ and and the current state of lattice vectors.}
 \label{fig:detc1ensq}}
\end{figure}

\subsection{Energy landscape}

\begin{figure}[tbp!]
\centering
\includegraphics[scale=.31]{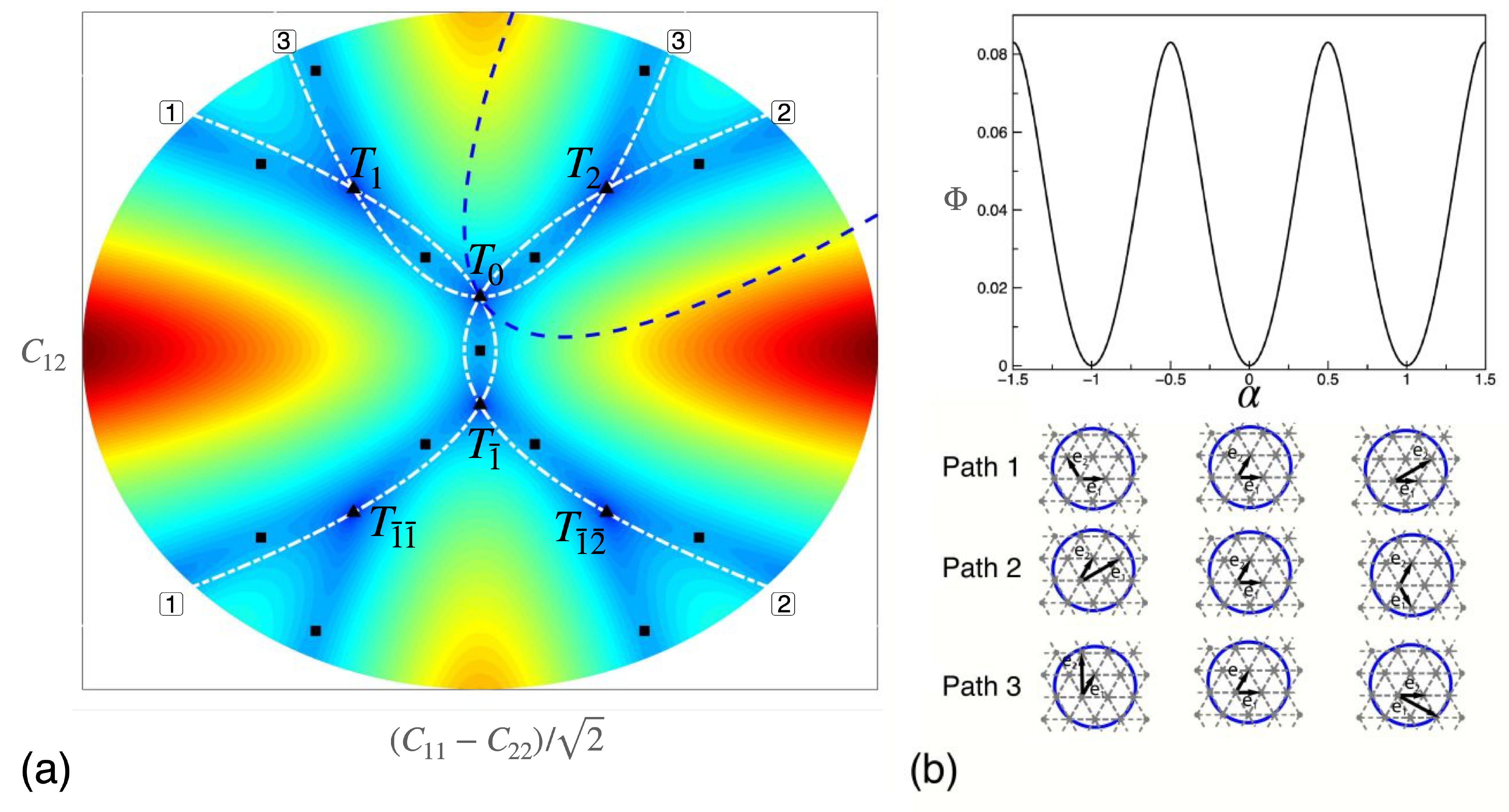}
\caption{\scriptsize {(a) The level sets of the  strain-energy  (\ref{eq:sigCZ})  with $\beta=4$. Colors indicate the energy level: blue, low; red, high. Dashed lines correspond to the  simple shear loading paths $\bold F(\alpha,\phi)$ defined  in  \eqref{pathshex}.
%The whites lines correspond to the three symmetric shears along dense planes of a triangular crystal $\bar{\bf F}(\alpha,0), \bar{\bf F}(\alpha,\pi/3), \bar{\bf F}(\alpha,2\pi/3)$. 
(b) The periodic strain-energy profile along the three symmetric paths $1-1$, $2-2$, and $3-3$ and the current state of lattice vectors.
%In each loading direction, the original Bravais lattice is recovered for the integer values of $\alpha$. 
}}
 \label{fig:detc1enhex}
\end{figure}

To visualize   the   local energy landscape  we  use the  plane with  coordinates  $((C_{11}-C_{22})/\sqrt{2},C_{12})$.  In Fig.~\ref{fig:detc1ensq}(a) we  illustrate  the potential (\ref{eq:sigCZ})  with   $\beta=-1/4$ describing an energy landscape whose absolute minima are the equivalent square lattices.  
To model  plastic flows  in triangular lattices  with  higher, hexagonal symmetry,  we  choose in \eqref{eq:sigCZ} the parameter value $\beta=4$ and the resulting elastic potential is illustrated  in Fig. \ref{fig:detc1enhex}(a). 
%  while    the   integer valued  matrices $\bold M$  corresponding to  such   crystallographically  indistinguishable  `phases'  are different.  
 
  In Fig.~\ref{fig:detc1ensq}(a) and Fig. \ref{fig:detc1enhex}(a)  one can   clearly distinguish between the  `soft' directions  located inside the deep valleys of the energy landscape (marked in blue), and the  `hard' directions pointing away from the valleys and forcing the  system   to climb  high energy barriers separating the valleys (yellow and red regions).  Inside each low energy  valley the energy is periodic with  the bottoms of the   energy wells corresponding to equivalent lattices. Thus, in the case of square lattices (Fig.~\ref{fig:detc1ensq}(a)  the  equivalent minima are denoted by   $\mathrm{S_i,S_{ij},...S_{\bar i},S_{\bar{i}\text{ }\bar{j} },.}..$ and in the case of triangular lattices (Fig. \ref{fig:detc1enhex}(a)), by  $\mathrm{T_i,T_{ij},...T_{\bar i},T_{\bar{i}\text{ } \bar{j}},..}.$   
  
Note that while the energy is locally convex   near the bottom of each  energy well and can be approximated by the standard Hookean paraboloid, the global energy landscape is highly nonconvex and can be rather characterized  as rugged. In particular this means that outside the immediate vicinity of the energy bottoms the system finds itself in spinodal regions where the constitutive behavior \emph{softens} opening the possibility for  various mechanical  instabilities. Note that the local maxima of the one parametric energy landscape along the low energy valleys turn out to be located  very close to the configurations corresponding to saddles describing perfect  lattices with alternative  symmetries: triangular lattices, in the case  when we shear a square lattice (see triangles  near the local maxima along the white dashed paths in Fig.~\ref{fig:detc1ensq}(a)) and square lattices, in the case  when we shear a triangular lattice (see squares near the local maxima along the white dashed paths in Fig.~\ref{fig:detc1enhex}(a)).

CP theory obviously appreciates  the  implied complexity of the energy landscape but not in the same way as MTM. Thus, the  low energy valleys are modeled as  zero energy plastic `mechanisms'. The periodicity of the energy inside the valleys is neglected but instead the phenomenological friction type lays are introduced to govern the `working' of such mechanisms.  Elasticity is usually assumed to be linear and is only accounted for around the reference energy well,  which is then extended globally. This does not create  problems in CP because at large strains plastic deformation along the valleys always wins against  elasticity which becomes at such strains prohibitively expensive energetically. To describe in the CP framework the coupling between different plastic `mechanisms', involved for instance in latent hardening \cite{Ortiz1999-pp}, additional phenomenological relations have to be formulated. In MTM those will be automatically accounted for due to the presence of various saddle points  (mountain paths) that are apparent in    Fig.~\ref{fig:detc1ensq}(a) and Fig. \ref{fig:detc1enhex}(a).
 
To illustrate  a  more detailed structure of the energy landscape along the particular low energy valleys  we now consider  two parametric families of square lattices  transformed to one another  by  the volume preserving shear deformations of the form
\begin{equation}
\label{paths_sq}
\bar{\bf F}(\alpha,\phi) =\mathbb{1} +\alpha{\bf a} (\phi)\otimes{\bf a} ^\perp(\phi),
\end{equation}
where $\alpha$ is the amplitude of shear, ${\bf a}(\phi) ={\bf R(\phi)}{\bf e}_1$, ${\bf R}(\phi)$  is a matrix describing  a   counterclockwise rotation by the  angle $\phi$, ${\bf e}_1$ is a unit vector of the Cartesian coordinate system  directed along the  close-packed  $x$ direction and ${\bf a} ^\perp$ is a  vector orthogonal to  ${\bf a} $. 

For instance, if we fix one of the parameters at $\phi=0$ and consider a one parametric family  involving the  metric tensors $\bold C(\alpha)$, we obtain the path $1$-$1$,  shown  in Fig.~\ref{fig:detc1ensq}(a)  by a dashed white line facing left.  It starts at  the reference state $S_0$, corresponding to the unstressed square lattice,  and at integer values of parameter $\alpha$  generates   an infinite sequence of equivalent unstressed replicas  of the same square lattice.  For instance,  the `closest'  lattices configurations  $S_1$ and $S_{\bar1}$ can be  reached from  $S_0$ by the elementary shear deformations  $\bar{\bf F}(\pm1,0)$. 

If we choose instead $\phi =\pi/2$, we obtain another one parametric family of shears described by the  path $2$-$2$, shown in Fig.~\ref{fig:detc1ensq}(a) by a dashed white line facing right. It also starts at the reference state $S_0$ and produces  at integer values of parameter $\alpha$ an infinite sequence of equivalent unstressed configurations of the square lattice. Once again,    the `closest'  lattices   $S_2$ and $S_{\bar2}$ can be reached  reached through the shear deformations $\bar{\bf F}(\mp1,\pi/2)$. 

The   energy landscape along  these symmetric paths is the same, see   Fig.~\ref{fig:detc1ensq}(b). Note that around the symmetric energy minima $\alpha=0$ and  $\alpha=\pm 1$, visible in such a graph, the energy is quadratic but then it loses convexity and the corresponding mechanical response starts to soften. Such softening is ultimately behind the break up of elastic response and the emergence of yield. If the system is quasi-statically driven through such an energy landscape, it undergoes a succession of snap-back instabilities which, under overdamped dynamics,  merge  into  the  dissipative  plastic  response which is postulated phenomenologically in the classical CP approach, see \cite{Puglisi2005-lg} for details.

Note  that a \emph{composition} of  shear deformations from the paths  $1$-$1$ and  $2$-$2$ can be interpreted as an activation of a double-slip. For instance,  the mapping   $\bar{\bf F}(-1,\pi/2)\bar{\bf F}(1,0)$ brings the system from the reference lattice configuration   $S_0$ to  the equivalent lattice configuration   $S_{12}$ after it  slips in two perpendicular directions. In fact,  all the  bottoms of the energy wells $\mathrm{ S_{ij},... S_{\bar{i}\text{ }\bar{j} },.}..$,  corresponding to equivalent  square lattices,   can be reached similarly by combining    quantized simple shears    \eqref{paths_sq} with $\phi=0$ or $\phi=\pi/2$. Moreover, as we have already mentioned, a simple shear path leading away from one of the  energy wells passes near the saddles corresponding to ideal lattices with different symmetries. It is then natural to conclude that these saddles serve as a switching points activating double slip and inciting other composite shears.  Based on these observations, one can argue that  multiple alternative lattice symmetries, that are virtually invisible in the `single symmetry' classical CP approach,  may  be  contributing  fundamentally to the   complexity  of plastic flows in crystals. 

Periodicity of the energy along the particular  \emph{tensorial} directions, illustrated in  in   Fig.~\ref{fig:detc1ensq}(b),  is usually directly postulated in PFDD,   with each  of these directions bringing in a \emph{scalar} order parameter of its own. To describe  a \emph{composition} of  the corresponding shear deformations, the  PFDD  order parameters need to be coupled phenomenologically. Instead, in the MTM,  such coupling is automatic due to the global symmetry requirements.  Moreover the MTM approach  allows one to assess the height and the structure of the barriers  separating the corresponding valleys and controlling the activation of  double-slip.  Note also, that a combination of scalars can represent a globally defined tensorial variable (metric tensor) only approximately.
%Note that in the continuum CP, the implied periodicity of the energy landscape and the corresponding quantization of the slip  are neglected and the energy along the  simple shear type tensorial valleys like \eqref{paths_sq} is assumed to be flat. 
   
Turning back to   Fig.~\ref{fig:detc1ensq}(a), we observe that the   two other symmetric paths  $3$-$3$ and  $4$-$4$, shown   by  dashed blue lines,    correspond to simple shears applied to  the rotated crystal   with either $\phi=\pi/4$
or  $\phi=3\pi/4$.  While  these symmetric shears are no longer    aligned with the dense planes  of our  square lattice, the corresponding   loading paths are still  special since they bring the system from the reference  energy well $S_0$ directly to the second nearest   energy wells, represented in  Fig.~\ref{fig:detc1ensq}(a)  by  the small squares  $S_{11}$, $S_{22}$, $S_{\bar 1 \bar1}$ and  $S_{\bar 2 \bar 2}$.  Along such paths the energy landscape is also periodic but the period doubles and the separating barriers become much higher as these paths are largely located outside the low energy valleys.  While in CP such paths are not viewed as plastic `mechanisms' that can be activated, they are  still accounted for in the MTM  which deals with the whole tensorial landscape and therefore encompasses all possible paths.

Finally, the red dashed line in Fig.~\ref{fig:detc1ensq}(a)  illustrates  a non-symmetric loading path with  $\phi =\arctan(1/2)$, also originating in  $S_0$.  This path  exhibits the characteristic loading-unloading asymmetry  of \emph{generic} simple shear deformations.  In particular, along this path  the equivalent square energy well   $S_{\bar 2}$  can be  reached through  `unloading'  at $\bar{\bf F}(-1,\arctan(1/2))$ but none of the  nearest square wells are reached upon  `loading' with    $\alpha$   positive.  

The analysis above suggests the strong dependence of the plastic response  on the orientation of the crystalline sample inside a hard loading device,  even if it always applies the same simple shear deformation. The fact that there are soft and hard orientations has been long known in crystal plasticity. MTM approach presents a more  nuanced picture, showing that the soft paths may come not only with different size  of the periodically placed  barriers  but also with different periodicity, while along  the hard paths one may expect  extremely high barriers and generic loading-unloading asymmetry. The crucial advantage of the MTM  is that all this complexity does not have to be postulated and emerges instead from the global symmetry of the energy landscape.

 The obtained general picture is confirmed if we turn from square to triangular lattices, however, as  the analysis below shows, the important differences between the energy landscape in   lower symmetry (square) and higher symmetry (triangular) lattices exists.
 
 In the case of triangular lattices we can again illustrate the available plastic `mechanisms'   by  considering   again the one parametric  simple shear paths.  The single slip  plasticity in triangular  lattices can be modeled by the shears 
\begin{equation}
\label{pathshex}
\bar{\bf F}(\alpha,\phi) =\mathbb{1} +\alpha{\bf b}_1(\phi)\otimes{\bf b}^2(\phi),
\end{equation}
 where  ${\bf b}_A(\phi) = {\bf R(\phi)}{\bf h}_A$ with $A=1,2$.  The (non-unit) lattice vectors   $ {\bf h}_A$   have been introduced earlier and   the vectors  ${\bf b}^A$  are   their \emph{duals}  defined by the conditions ${\bf b}_A.{\bf b}^B =  \delta_A^B$.  

Using \eqref{pathshex},   we can generate    three symmetric shear paths  originating in the unstressed lattice  $T_0$ and  describing plastic slips along the  three   close-packed directions.  They correspond to crystal orientations  $\phi=0,\pi/3,2\pi/3$, see the white dashed lines in  Fig. \ref{fig:detc1enhex}(a).  Here again,  for the integer values of  $\alpha$  we obtain  infinite families  of equivalent  replicas of the original triangular  lattice. In particular,  the   lattice configurations   $T_{1}$ and $T_{\bar 1}$,  located on the path $1$-$1$,  can be  reached by the shear deformations $\bar{\bf F}(\pm1, 0)$ starting from the lattice configuration  $T_{0}$ (see the dashed white line facing left)  Upon further   `unloading'   $\bar{\bf F}(-2, 0)$, another  triangular lattice  $T_{\bar 1\bar 1}$ can be reached, while the  lattice configuration $T_{1 1}$ is reachable by the matching additional  `loading'  $\bar{\bf F}( 2, 0)$ (not visible in Fig. \ref{fig:detc1enhex}(a)).
  
The symmetry related  paths  $2$-$2$  and $3$-$3$, also shown in Fig.~\ref{fig:detc1enhex}(a)  by dashed white lines, correspond to  the shear deformations  $\bar{\bf F}(\alpha, \pi/3)$  and $\bar{\bf F}(\alpha, 2\pi/3)$. Both of them also originate in the point  $T_{0}$ and produce at integer values of  $\alpha$   infinite families  of equivalent  replicas of the original triangular  lattice.  Thus, along the path $2$-$2$, one  can reach  the  equivalent  lattices    $T_{2}$, $T_{\bar 1}$, and  $T_{\bar 1\bar 2}$, all visible in Fig.~\ref{fig:detc1enhex}(a). Similarly,   the equivalent   lattice configurations   $T_{1}$ and  $T_{2}$ are reachable along  the path  $3$-$3$.  
 
 For each of these paths, which all correspond to the known plastic `mechanisms',  the one dimensional periodic energy landscape is exactly the same, see  Fig.~\ref{fig:detc1enhex}(b). The maxima of such energy landscapes are located close to the saddle points of the global energy landscape corresponding to square lattice configurations. Because they follow the low energy valleys of the global energy landscape (shown in blue in Fig.~\ref{fig:detc1enhex}(a)), these simple shear paths  correspond to  `soft' loading directions for a triangular lattice.  Observe  that such valleys  are  splitting  into two in each of the energy wells which is clearly an easy  mechanism for the  activation of multi-slip in such higher symmetry lattices.

A     non generic   path with $\phi=\pi/2$ is also shown  in Fig.~\ref{fig:detc1enhex}(a) (see the blue dashed line). In this case, the crystal is not driven along one of the  low energy valleys,  however it  does   pass through   the  bottoms of the energy wells (with period 3!). If we  load the  crystal  in shear along  this `hard' direction,  we can  expected to reach the `yellow' and even `red' zones in Fig.~\ref{fig:detc1enhex}(a) and therefore acquire considerable elastic energy before the ultimate breakdown of an elastic state. Curiously enough, as we show below, this does not happen.

Since we  use in this paper  a  particular expression for the energy  \eqref{eq:sigCZ},  one may think that, given that  the results  depend on this choice,  we cannot obtain  generic picture   for square and tetragonal crystals. However, the proposed  energy  has several  generic features  such as the exact location of the energy wells and the fact that they are  quadratic   close to   lattice invariant shears.  Moreover,  the configuration of the low energy valleys and the connecting saddle points is also universal in the sense that it is  dictated exclusively by the  $GL(2,\mathbb Z)$ symmetry. Of course, the   potential-specific features of the energy landscape, like  the curvatures of the wells  and the height of the energy barriers, remain.  In fact, such \emph{quantitative} ambiguity is the element which MTM shares with all other Landau type theories. To assess this ambiguity,  we have conducted  a preliminary comparison of our results with the outcome of the MTM,   where the energy was constructed directly from an atomic model. While the fine structure of the dislocation cores was affected, we did not record any qualitative differences as far as the global post yield dislocation pattern  is concerned \cite{R_Baggio}. 

Note also that the level of   smoothness  of the Landau potential  can be  improved if we replace the six order polynomials used in \eqref{eq:sigCZ},  by the higher order polynomials.  Various exponential approximations were discussed in  \cite{Folkins1991-em} and an example of an analytic function with $GL(2,\mathbb Z)$ symmetry  was  presented in~\cite{Baggio2019-rs}.  However, in general, such potentials are analyzed numerically and the possibility of a relatively simple analytic representation  is not an issue.    Therefore, physically  relevant potentials   can be constructed  directly  inside the fundamental domain $\mathbb D$  by analyzing  homogeneous deformations of atoms and applying the Cauchy-Born rule. To this end one can use  interatomic potentials, embedded atom methods and  even quantum mechanics. Such numerically constructed energy potentials can  be then continued by symmetry  using the Lagrange reduction mentioned above or other similar approaches \cite{Engel1986-tm, Grosse-Kunstleve2004-oi,Andrews2019-qp}.

\subsection{Computational approach}

We now turn to the numerical implementation of the MTM approach. Its underlying  idea  is that   rate-independent crystal  plasticity  can be  modeled within the framework of  the appropriately discretized  nonlinear elasticity    combined with   athermal,  overdamped  dynamics.  If the loading is sufficiently slow, the exact nature of such  dynamics is not essential  as long as it  effectively performs incremental energy minimization \cite{Puglisi2005-lg,Mielke2011-ck}.  As the loading evolves the system remains on the same branch of local minima of the energy till the latter ceases to exist and then it switches to another equilibrium branch. Such switching is a dynamic process (avalanche) which can be considered instantaneous at the time scale of the loading but it still contributes to  dissipation. In this representation,   plastic flow  emerges as  a  set of equilibrium solutions of  nonlinear  elasticity with dissipation taking place exclusively during the  abrupt  branch switching events \cite{Puglisi2005-lg}. 

We recall that  solution of an  elastic  problem implies  local  minimization of  the energy $$W=\int_{\Omega}\Phi(\nabla\bold y ) d\bold x, $$ which is prescribed  on  a  reference domain  $\Omega$ with unit volume. We assume that the system is  loaded  by  an affine displacement field prescribed  on $\partial \Omega$ (hard device).  The conditions of mechanical equilibrium  are usually formulated in terms  of the first Piola-Kirchhoff stress tensor ${\bf P}= \partial\Phi/{\partial\bf F} $. In the index free form they can be written as  
$$
\nabla\cdot\mathbf{P}=0.  
$$
Using the Eulerian   $i,j=1,2$ and the Lagrangian  $K,L =1,2$ indexes and assuming summation on repeated indexes, we can rewrite  these equations  in the form
$$
A_{iKjL} y_{j,KL}=0,
$$
where  
 \begin{equation}A_{iKjL}= \frac{\partial^2\Phi^0{({\bf\tilde C})}}{\partial F_{iK}\partial F_{jL}}.
 \label{eq:ATred}
 \end{equation}
Here  $\tilde{\bf C}=\bold M^T\bold C\bold M$, where the integer-valued matrix $\bold M$  with determinant equal to one is computed for each value of $\bold C$ using the Lagrange reduction algorithm.  The matrix ${\bf M}$ depends on the current state of deformation $\bf F$ in a piece-wise constant manner and  therefore can be considered as constant in \eqref{eq:ATred} which can be then rewritten as 
 \begin{equation} \label{compact}
A_{iKjL}=\text{M}_{MN} \text{M}_{PQ} \left[ 
\frac{\partial C_{MP}}{\partial F_{jL}} \frac{\partial C_{AB}}{\partial F_{iK}}\text{M}_{AC}\text{M}_{BD}  \text{H}_{MQCD}+\frac{\partial ^2C_{MP}}{\partial F_{iK}\partial F_{jL}}   \Sigma_{NQ} 
 \right].
\end{equation}
%where  we   introduced the tensors  ${\bf \Sigma} = 
%  \partial\Phi^0/\partial \widetilde{\bf C} $ and  ${\bf H}= 
% \partial^{2}\Phi^{0}/\partial\widetilde{\bf C}\partial\widetilde{\bf C}$. Their components are explicitly  given as follows:
Here  we   introduced the tensors  
\begin{equation}
\label{eqn:Phi}
{\pmb \Sigma}= \scalebox{.75}{$
\left[\begin{array}{cc}
\frac{\partial\Phi^{0}}{\partial\widetilde{C}_{11}} & \frac{1}{2}\frac{\partial\Phi^{0}}{\partial\widetilde{C}_{12}}\\
\frac{1}{2}\frac{\partial\Phi^{0}}{\partial\widetilde{C}_{12}} & \frac{\partial\Phi^{0}}{\partial\widetilde{C}_{22}}
\end{array}\right]$},
\end{equation}
and
  \begin{alignat*}{1}
  \bf H=\label{eqn:ftenscom}
  \scalebox{.75}{$
\left[\begin{array}{cc}
\left[\begin{array}{cc}
\frac{\partial^{2}\Phi^{0}}{\partial\widetilde{C}_{11}^{2}} & \frac{1}{2}\frac{\partial^2\Phi^{0}}{\partial\widetilde{C}_{12}\partial\widetilde{C}_{11}}\\
\frac{1}{2}\frac{\partial^2\Phi^{0}}{\partial\widetilde{C}_{12}\partial\widetilde{C}_{11}} & \frac{\partial^{2}\Phi^{0}}{\partial\widetilde{C}_{22}\partial\widetilde{C}_{11}}
\end{array}\right] & \frac{1}{2}\left[\begin{array}{cc}
\frac{\partial^2\Phi^{0}}{\partial\widetilde{C}_{11}\partial\widetilde{C}_{12}} & \frac{1}{2}\frac{\partial^2\Phi^{0}}{\partial\widetilde{C}_{12}^{2}}\\
\frac{1}{2}\frac{\partial^2\Phi^{0}}{\partial\widetilde{C}_{12}^{2}} & \frac{\partial^2\Phi^{0}}{\partial\widetilde{C}_{22}\partial\widetilde{C}_{12}}
\end{array}\right]\\
\frac{1}{2}\left[\begin{array}{cc}
\frac{\partial^2\Phi^{0}}{\partial\widetilde{C}_{11}\partial\widetilde{C}_{12}} & \frac{1}{2}\frac{\partial^2\Phi^{0}}{\partial\widetilde{C}_{12}^{2}}\\
\frac{1}{2}\frac{\partial^2\Phi^{0}}{\partial\widetilde{C}_{12}^{2}} & \frac{\partial^2\Phi^{0}}{\partial\widetilde{C}_{22}\partial\widetilde{C}_{12}}
\end{array}\right] & \left[\begin{array}{cc}
\frac{\partial^{2}\Phi^{0}}{\partial\widetilde{C}_{11}\partial\widetilde{C}_{22}} & \frac{1}{2}\frac{\partial^2\Phi^{0}}{\partial\widetilde{C}_{12}\partial\widetilde{C}_{11}}\\
\frac{1}{2}\frac{\partial^2\Phi^{0}}{\partial\widetilde{C}_{12}\partial\widetilde{C}_{11}} & \frac{\partial^{2}\Phi^{0}}{\partial\widetilde{C}_{22}^{2}}
\end{array}\right]
\end{array}\right]$},
 \end{alignat*} 
 which are assumed  as known   inside  the fundamental domain $\mathbb D$.  We also recall here for convenience that $  \partial C_{KL}/\partial F_{ij} = \delta_{Kj}F_{iL} + \delta_{Lj}F_{iK}
 $, and  $ \partial ^2C_{KL}/\partial F_{iM}\partial F_{jN}  = (\delta_{KM}  \delta_{LN}+ \delta_{KN} \delta_{LM})\delta_{ij}.$ 

As we have already mentioned, the continuum   elasticity problem, formulated above, is highly degenerate which is the property the MTM shares with other similar Landau type theories. Usually such theories are regularized  through the introduction of an  internal length scale. In contrast to the conventional Ginzburg-Landau approaches like PFDD, relying for regularization on higher gradients of the order parameters,  in    MTM the  regularization  is achieved by spatial discretization.  More precisely,   deformation is assumed to be piecewise linear and the    elastic  response is attributed   to   discrete material elements  whose size is viewed as a  physical parameter of the model  \cite{Conti2004-sv,Salman2011-ij,Baggio2019-rs}.

 We therefore  need to reduce the space of admissible deformations   to compatible piece-wise-affine mappings.  To this end we   build  a  network whose discrete nodes are labeled by  integer valued coordinates $a =1,..., N^2$. We assume that each   element of the network is a deformable triangle  and write  the displacement field   in the form ${\bf u} ({ \bf x})= {\bf u}^a {\mathcal N}^a(\bf x )$, where ${\mathcal N}^a(\bf x)$ are  the  compactly supported  shape functions,  ${\bf u}^a$ are the amplitudes of nodal displacements and summation over repeated indexes effectively extends over elements containing or  bounding point $\bf x$. 
%;  the summation  over all nodes $a$ is implied. 
%
%have integer valued coordinates.
%We then uniformly mesh this domain based on the nodes in their reference positions using the three node linear triangles shape functions $N^a({\bf x}_e)$, whose nodes of each triangular element $e$ correspond to the positions 
%
% ${\bf r}^a$ chosen such that they correspond to the stress-free reference state  of the underlying crystal symmetry. 
%
The  mesoscopic deformation gradient is   then    ${\bf F}(\bf x)  = \mathbb{1}  +\nabla {\bf u}(\bf x)$, 
%  The  columns of the matrix ${}_{e}{\bf F}^a$ represent  the components of the deformed basis lattice vectors ${}_{e}{\bf f}_{K}$, with $K= 1,2$ and the local value of the metric tensor is ${}_{e}C_{KL}={}_{e}{\bf f}_{K}.{}_{e}{\bf f}_{L}$. 
and the 
%discretized strain-energy can be written in the form $W=\int _{\Omega}  \Phi( \mathbb{1}  +  {\bf u}^a \nabla N^a (\bf x))  d \bf x$.  The
 equilibrium equations can be written in the form
\begin{equation}
\label{numerics}
\frac{\partial W}{\partial {\bf u}^a }=\int _{\Omega} {\bf P}({\bf F}) \nabla {\mathcal N}^a ({\bf x}) d  {\bf x}  =0,
\end{equation}
where  ${\bf P} = 2 {\bf F } {\bf M } {\pmb \Sigma } {\bf M}  ^T.$  This problem   can be solved by quasi-Newton  method followed by the  so called  NR `refinement'  when  the initial guess is too far from the solution for Newton–Raphson method  to converge initially~\cite{Tadmor1996-qi}. 

% In both cases we will require the first and second variation of the total potential energy   with respect to the nodal degrees of freedom referred to    in FEM   terms as the  global out-of-balance force residual and the global stiffness matrix. 

More specifically, to solve  \eqref{numerics} for ${\bf u}^a$ we first   use  the L-BFGS algorithm~\cite{Bochkanov2013-lk} which      builds a positive definite linear approximation of \eqref{numerics}  allowing one  to make a quasi-Newton step lowering  $W$.  Such  iterations continue till  the increment  in total energy $W$ becomes sufficiently small. The obtained approximate solution is then   used as an initial guess  ${\bf w}^a$ to solve, using LU factorization~\cite{Sanderson2016-ht,itensor},  the  equations for the correction  $\text{d}{\bf w}^a$
%around such an initial guess for the displacement field ${\bf w}^a +  \text{d}{\bf w}^a$ yields a system of linear equations:
\begin{equation}
\label{newton}
 K^{ab}_{ij}dw_j^b+R_i^a =0,
\end{equation}
where  
\begin{equation}K^{ab}_{ij}= A_{iKjL}({\bf F}) \frac{\partial {\mathcal N}^a}{\partial x_K}\frac{\partial {\mathcal N}^b}{\partial x_L},\,\,\ R^a_i=  P_{iK}( {\bf F}) \frac{\partial {\mathcal N}^a}{\partial x_K}.
\end{equation}  The displacement field can be updated in this way till   the value of  the integral in  \eqref{numerics} is sufficiently small and  then  the loading parameter can be  advanced again. In the case  when  the applied deformation is,  a homogeneous shear $\bar{\bf F}(\alpha,\phi)$  with  fixed orientation $\phi$, the loading parameter   is the shear amplitude   $\alpha$.  By changing this parameter in increments of $10^{-4}$,  we advance   the  displacement field   ${\bf u}(\alpha,\phi)= (\bar{\bf F}(\alpha,\phi)-\mathbb{1}){\bf x}$ for all nodes $a$ on the boundary of the body $\partial \Omega$. 

The outcome  of such  numerical experiments  depends on the value of the internal length scale   entering the discrete problem  through the finite element size $h$. Each element contains $n^2$ interacting atoms, where $n=h/a$ and $a$ is a fixed  atomic scale. For such element deformed in simple shear  we can  use one of the known  ab initio methods to compute its energy, which  becomes close  to periodic  as we  increase the  value of $h$.  The  value of this internal length scale  used in MTM depends on the smallest range of  energy periodicity  required  in the problem; in many crystals  the  periodicity  at the level of the few first energy wells  can be  captured   already for $h \sim 10a$.
%
%  As a consequence of the inherent locality of the ensuing mesoscopic description, some aspects of a truly atomistic description will be necessarily lost.
%In other words we assume that the periodic structure of the elastic energy of an element emerges as a result of homogenizing out the scale smaller than $h$.
%

Since  the linear size of the macroscopic sample is $L=1$, the small dimensionless parameter of the MTM is $h/L=1/N$ where $N^2$ is the number of the nodes. Thus, if $h$ is  in $nm$ size range, the simulations with $N \sim 10^3$ would describe a micrometer size crystal. 
%In numerical experiments we assume that the (normalized) $L=1$ and $h=1/N$. 
When $N$ is small, dislocation cores emerge as \emph{blurred} because the scales smaller than $1/N$ are homogenized out. While such cut-offs may compromise the short-range interaction of dislocations, long-range interactions at distances larger than $1/N$ will be still captured correctly.
 
\section{Numerical experiments}

To understand the paradoxical crack-free brittle behavior of very small, initially dislocation-free crystals, we used MTM and conducted a series of numerical experiments with square and triangular crystals. The samples were subjected to simple shear, and we performed experiments with several different orientations of the same  samples in the same hard loading device. We used square cut samples in all our experiments and could vary the system size up to $N=1024$.  

\subsection{Macroscopic response}

\begin{figure}[h!]
\centering
\includegraphics[scale=.4]{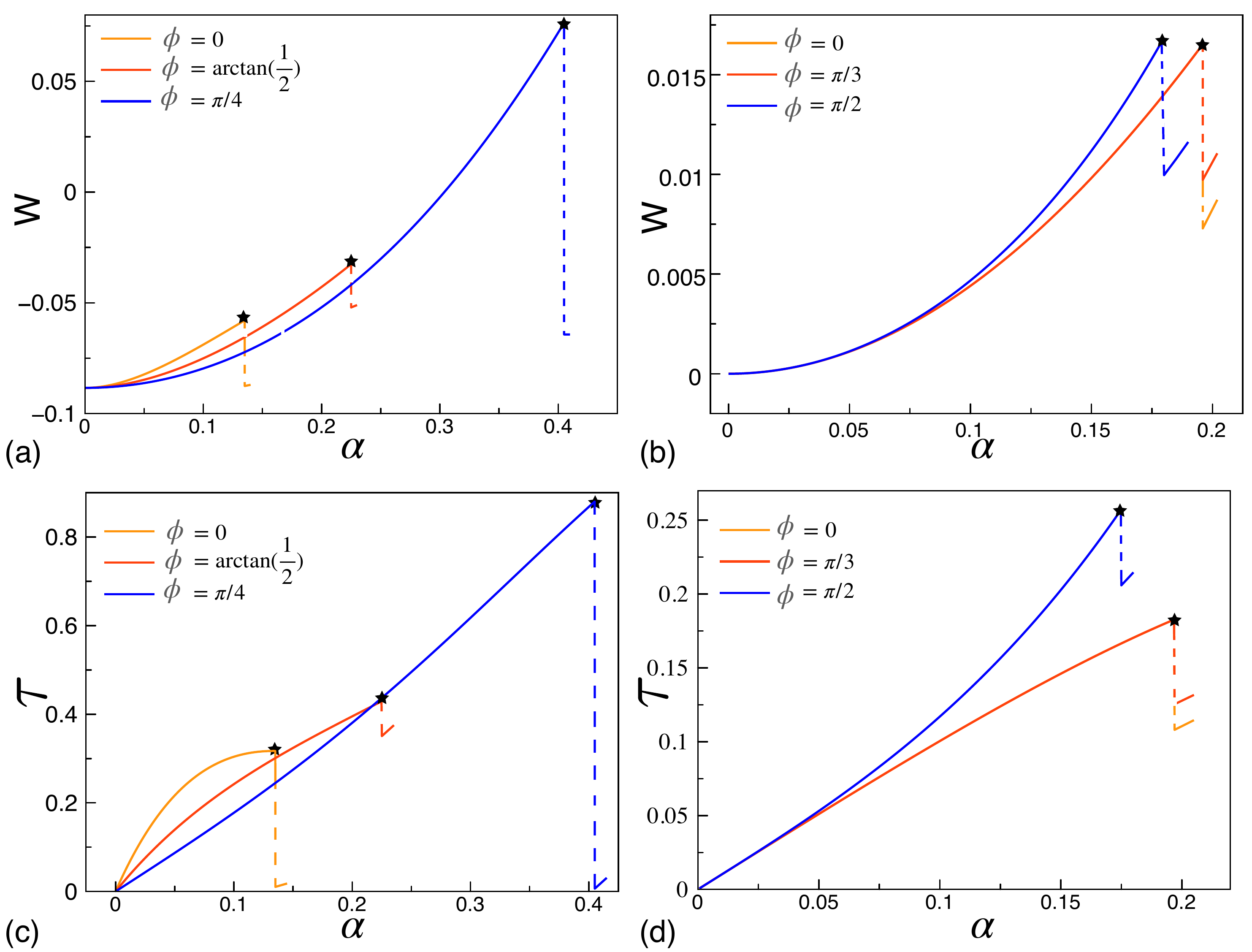}
\caption{\scriptsize {Macroscopic mechanical response of square (a-c) and triangular  crystals (b-d) subjected to simple shear loading: (a,b) the  equilibrium elastic energy $\mathrm{W}(\alpha)$ (c,d);   the equilibrium stress defined as $\tau(\alpha)=\text{d}\mathrm{W}/\text{d}\alpha=\int_{\Omega}{\bf P}:(\text{d}\bar {\bf F}/\text{d}\alpha)dx$.
% $\sigma(\alpha)=\text{d}\mathrm{W}/\text{d}\alpha$ 
%  (c-d) as a function of the loading strength $\alpha$ along the all considered paths. In each case, a sharp  drop in the energy/stress is observed originated  from the dislocation nucleation event.
  }
 \label{fig:energydrops}}
\end{figure}

The obtained  macroscopic responses for  crystals  described by the piece-wise smooth   potential \eqref{eq:sigCZ} are summarized in Fig.~\ref{fig:energydrops}(a) and Fig.~\ref{fig:energydrops}(b). We show separately  the energy-strain and the stress-strain   responses  for  square and triangular crystals loaded in  different orientations. The energy was obtained by incremental local minimization starting from the reference state. To compute the stress-strain curve,  we define the resolved-shear stress in the direction of loading as $\tau(\alpha)=\text{d}\mathrm{W}/\text{d}\alpha=\int_{\Omega}{\bf P}:(\text{d}\bar {\bf F}/\text{d}\alpha)dx$.

In each of our numerical experiments, the initial phase of the deformation history is (nonlinear) elastic. During this stage, all elements have associated metrics that remain inside the extended fundamental neighborhood of the initial phase   (Pitteri neighborhood \cite{pitteri2002continuum}), comprised of four (in case of square symmetry) or six (in case of triangular symmetry) symmetric replicas of the fundamental domains. While the response of the crystal inside such domain is nominally elastic, our experiments show that the affine deformation becomes unstable before the boundary of this neighborhood is reached. The instability takes place at a critical value of the loading parameter $\alpha=\alpha^*_c$ which depends not only on crystal symmetry and crystal orientation, see Fig.~\ref{fig:energydrops}, but also on the value of $N$ (see below). The breakdown of affine elastic regime takes the form of a catastrophic drop in both stress and energy. While these drops are still associated with plastic deformation,  they are highly reminiscent of brittle fracture.

We now discuss how the macroscopic mechanical behavior during such discontinuous yield depends on crystal symmetry and how it varies when crystals of the same symmetry are differently orientated in the loading machine.

We start with a pristine square crystal that we load along the principal slip direction with $\phi=0$. In such an `easy' glide, the instability is preceded by the purely elastic softening, and the yield takes place near the maximal load. However, instead of conventional continuous yielding, the crystal experiences discontinuous yield as it abruptly loses almost all accumulated energy with stress dropping almost to zero. This means that the crystal manages to expel almost all nucleated dislocations away from its bulk, either by annihilating or sending them towards the boundaries. If the boundaries were unconstrained, the transition would be pristine-to-pristine, but in the hard-device loading conditions, at least some of the nucleated dislocations end up forming low energy piles-up near the boundaries (see more about this in the next Section). 

A similar almost pristine-to-pristine discontinuous-yield occurs in the case of a less symmetric but still non-generic shear with $\phi=\pi/4$ but only after both the elastic energy and the stress reach a much higher value. As we have already seen, along this loading path, the periodicity strain doubles, and the barriers become much higher than in the case of shear loading directed along the low energy valley.  Instead, the crystal is first driven away from such valley and, before yielding, manages to accumulate a considerable amount of energy without an apparent softening. It is then rather remarkable that the eventual breakdown of the elastic state leads to almost complete relaxation of considerable elastic stress. The explanation probably lies in the fact that the crystal's orientation is still special vis a vis the applied load. 

Loading of the crystal in a non-generic tensorial direction $\phi=\arctan (1/2)$ leads to a moderate softening in the elastic range, which again ends with a discontinuous yield. The resulting dislocation avalanche relaxes the stored elastic energy only minimally.  This suggests that the crystal is left with considerable amount of dislocations which may not be  strongly correlated  because the stress relaxes only minimally.

In triangular lattices with hexagonal crystallographic symmetry, the macroscopic mechanical response is different. For two loading paths with $\phi=0$ and $\phi=\pi/3$, passing inside the low energy valleys,  the overall softening nonlinear elastic response terminates with a yielding avalanche which takes place at exactly the same level of energy and the same value of the loading parameter. The pristine state is not recovered after the discontinuous yield because higher symmetry gives rise to geometric frustration, and the stored elastic energy cannot be fully resolved through dislocation self-organization. However, the resulting stress drop is considerable, which suggests that long-range correlations have been created. The slight difference in the terminal post-avalanche state for the paths with $\phi=0$ and $\phi=\pi/3$ is a finite size effect suggesting that the corresponding instability modes interact differently with the incommensurate square shape of the sample.

Along the non-generic `hard' loading path with $\phi=\pi/2$, the crystal shows hardening nonlinear elastic response, however, the yielding stress is only slightly higher than in the case of the `soft' paths considered above. The discontinuous yield leads again to relatively small energy relaxation, suggesting that a high symmetry environment favors the considerable accumulation of dislocations. However, the stress drop is smaller than along more symmetric `soft' paths, which hints towards a higher level of frustration and weaker long-range correlations in the post-buckling equilibrium state.

To summarize, our numerical experiments suggest that nominally ductile crystals, which are expected to  yield continuously in the bulk form (large $N$, pre-existing defects), can still undergo a brittle-to-ductile (BD) transition when the value of $N$ and the initial dislocation density both drop below certain thresholds. Such brittleness has been indeed  realized to be a characteristic feature of sub-micron samples with high initial purity (absence of solutes, precipitates, and dislocations).  The implied discontinuous yield results from massive homogeneous nucleation of dislocations in the form of a highly cooperative avalanche. As we have shown,  the MTM can successfully simulate such avalanches and captures not only the  catastrophic stress drop but also, in some cases,  the recovery of almost pristine post-avalanche states. Even more importantly, MTM allows one to study the \emph{subtle} quantitative dependence of the parameters of such discontinuous yield on crystal symmetry and sample orientation.  Some features of the observed behavior may depend on our particular  choice of the  Landau  potential, like for instance, the  crossover from elastic softening to elastic hardening depending on the orientation of the sample; the question whether such  behavior is instead a generic property of all $GL(2,\mathbb Z)$ symmetric potentials will be addressed elsewhere.

%In particular, depending on orientation,  crystals with the  same symmetry may exhibit either hardening or softening behavior, exhibit vastly different strength  and also the  difference in  the structure of the terminal  system size avalanche can lead to different level of stress drop.  Below we show that the appropriately defined theoretical strength limit is also almost achieved. 

\subsection{Ideal shear strength}
 
To rationalize the observed behavior,  we  now  study  analytically the  linear  stability  of  an affine elastic response  in  homogeneously deformed  crystals.  From  the    classical continuum elasticity  standpoint,  the  instability would mean  that a homogeneous configuration is no longer   a weak local minimum of the elastic energy. In  a hard device,  such instability results  from the local loss  of   rank-1 convexity of the elastic energy density which also means  the  loss of strong ellipticity of the  equilibrium equations ~\cite{ogden_nl,Grabovsky2014-fb}. The instability in  the regularized  model is necessarily  delayed which is one of the  sources of the `smaller is stronger' size effect in sub-micron crystals. 

Consider a homogeneous configuration  of an elastic crystal with deformation gradient ${\bf F}$ rigidly imposed on its boundary. Suppose that this state becomes linearly unstable at the critical value of the imposed strain ${\bf \bar F}^c$.  To find this threshold,   we need to linearize   the  equilibrium equations around the homogeneous  state.   If we  write them  in terms of   small incremental displacements  ${\bf u}$ superimposed on the homogeneous state ${\bf F}$ (used as the  reference state),  we obtain the system
$$
{ \mathcal  A}_{piqj}({\bf F})  u_{j,pq}=0,
$$
where 
$
{\mathcal A}_{piqj} ({\bf F})= F_{pK}F_{qL}A_{KiLj} 
$
is the (Eulerian) tensor of  incremental moduli. 
%It is known that the solution of the continuum equilibrium problem   with displacements controlled at the whole boundary is stable with respect to small perturbations as long as the system of equations is  strongly elliptic . Along the loading path, this condition will be satisfied  till a  certain critical value of the imposed strain ${\mathbf F}^c$ is reached. At this value of the loading parameter  the   homogeneous solution will cease to be stable.
%
%manifested by the non-positive definiteness of the acoustic tensor
The unstable mode can be represented as a combination of Fourier components  ${\bf u} =\pmb{\eta}\exp{(i k  {{\pmb n}{\bf x}})}$, where $\pmb{n}$ is the unit normal selecting the modulation direction,  $\pmb{\eta}$ is the amplitude, and $k$ is the wave number.   In the Fourier space,   this equation can be re-written as $$[{\pmb{\mathcal Q}}({\bf F},\pmb{n})\pmb{\eta}]\pmb{\eta} =0,
$$
where ${\mathcal Q_{jl}({\bf F}, \pmb{n}})={\mathcal A}_{ijml}n_in_m ({\bf F})$ is the Eulerian acoustic tensor~\cite{Merodio,Kumar2015-xo}. 
% The strong ellipticity condition reads:
%$$
%Q_{im}(\pmb{n})\eta_{i}\eta_{m}>0,
%$$
%where  $\pmb{\eta}$ is  an arbitrary unit vector.
 Stability of the homogeneous state  ${\bf F}$ is lost when there exists a 
non-trivial $\pmb{n}$ such that   $$\Det {{ \pmb{\mathcal Q}}}({\bf F}, \pmb{n})=0.$$ This condition can be used to identify   the  unstable values of  $\bf C$  \cite{Hill1962-fb,Rice1976-ta, ogden1997non}.   Generalizing  the classical definition of Frenkel \cite{Anderson2017-qz}, we can   associate the    ideal  shear strength   with such $\bf C$  and compute them by performing  a sequence of monotone loading tests. The ensuing critical orientation $\pmb{n}$ and the associated polarization $\pmb{\eta}$   can be interpreted as the characteristics of the incipient  defects, see,  for instance,   \cite{ Van_Vliet2003-yg, Zhu2004-wr,Zhong2008-pj,Miller2008-rr,Bigoni2012-by, Garg2016-kz}.

\begin{figure}[h!]
\centering
\includegraphics[scale=.35]{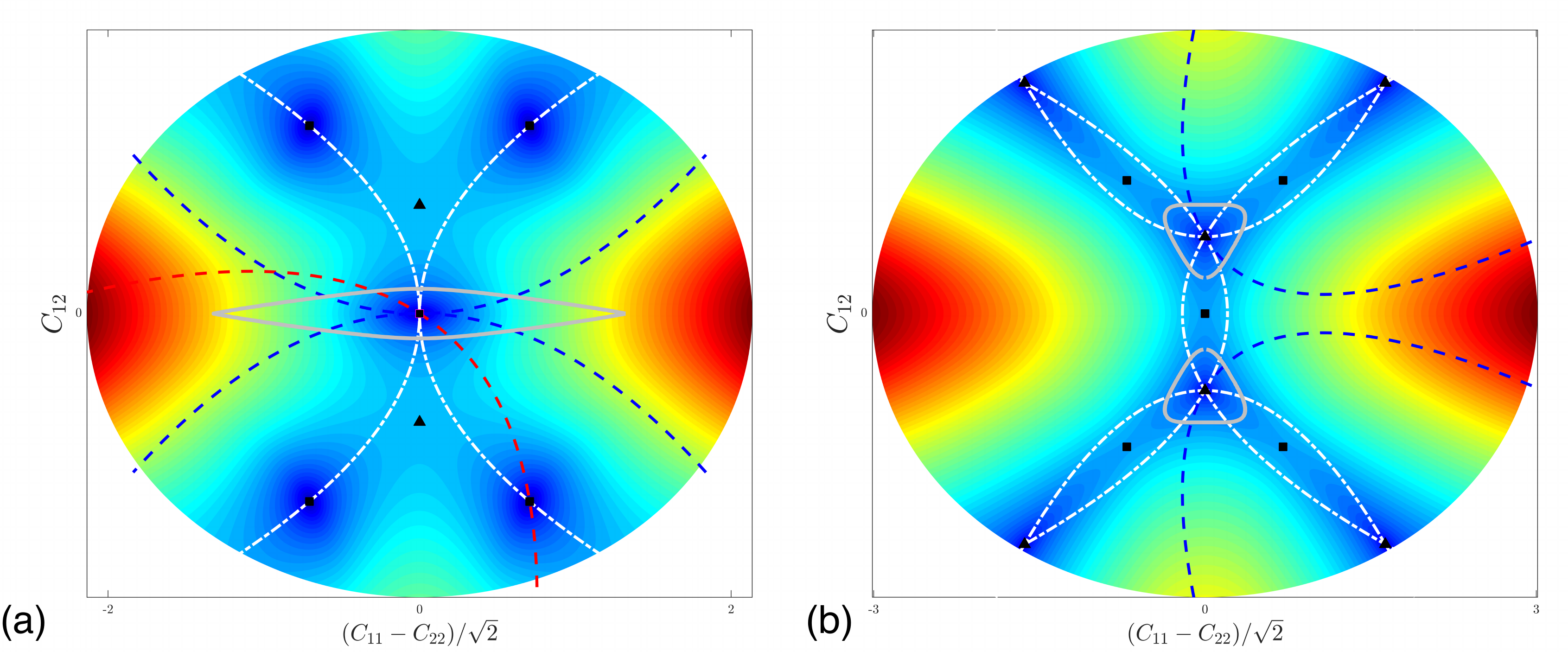}
\caption{\scriptsize {Surfaces of ideal shear strength for square (a) and triangular (b)  crystals. Silver contours delimit the domains of linear stability for   homogeneous states.   The strain-energy density is taken from  (\ref{eq:sigCZ}).
%: (a) square crystal,  (b) triangular crystal. 
}
 \label{fig:detc1en_ys}}
\end{figure} 
 
To identify the sub-domain in the space of reduced metric tensors  (inside the extended fundamental domain)  where  the  homogeneous  deformations are stable, we conducted a series of numerical experiments with differently oriented lattices   loaded by  shear deformations of the form  \eqref{paths_sq} and \eqref{pathshex}. Interpolating over a family of   one parametric paths we obtained  the surfaces of ideal strength for both square and triangular lattices, see  the light grey contours   in Fig. ~\ref{fig:detc1en_ys}(a) and  Fig. ~\ref{fig:detc1en_ys}(b), respectively. These contours represent   upper bounds for the sets which can be interpreted as  continuum yield surfaces.

% limits   by determining  the first value of the loading parameter $\alpha = \alpha_c$ for which the acoustic tensor degenerates.  As we  spanned the configurational space of  metric tensors with many simple shear trajectories,   collected the corresponding  bifurcation points and performed interpolation, we obtained  the  surface of theoretical strength  which can be interpreted as the continuum yield surface. Such surfaces for square and triangular crystals are shown by  
% 
 
Observe that  such surfaces may have rather different shapes: more elongated   in the case of less symmetric square lattices and more round in the case of more symmetric triangular lattices. To maintain the symmetry of the configurational space, we showed in  the case of triangular lattice two symmetric yield surfaces.  Similar stability boundaries  can be, of course,  constructed around each of the equivalent replicas of the unstressed lattices ~\cite{Baggio2019-rs}.

The strong asymmetry of such   yield surface in the case of square lattices, see Fig. ~\ref{fig:detc1en_ys}(a),  shows that if an anisotropic  crystal is loaded in `soft' and `hard' tensorial directions it can  exhibit very different strength. Thus, the instability can take place either inside the low energy valley or get delayed till after the crystal has first stored   considerable amount of elastic energy. Instead,  in high symmetry crystals the apparent yield surface is almost isotropic, see Fig. ~\ref{fig:detc1en_ys}(b),  and since in this case all tensorial loading directions are almost equally `soft', the expected theoretical strength  is  low independently of the  orientation of the applied shear.

In our numerical experiments,  the   parameters  of discontinuous yield were practically indistinguishable from the computations based on the idea of ideal shear strength, see  small stars on the stress-strain and energy-strain graphs in Fig.~\ref{fig:energydrops}. This means that  inside the elastic range the system with  $N=1024$ is adequately described by the continuum limit. This also justifies the experimental  observations  that pristine sub-micron  crystals  yield close to  the limits of theoretical strength which therefore cannot be improved without changing the nature of the crystal.

\begin{figure}[h!]
\centering
\includegraphics[scale=.2]{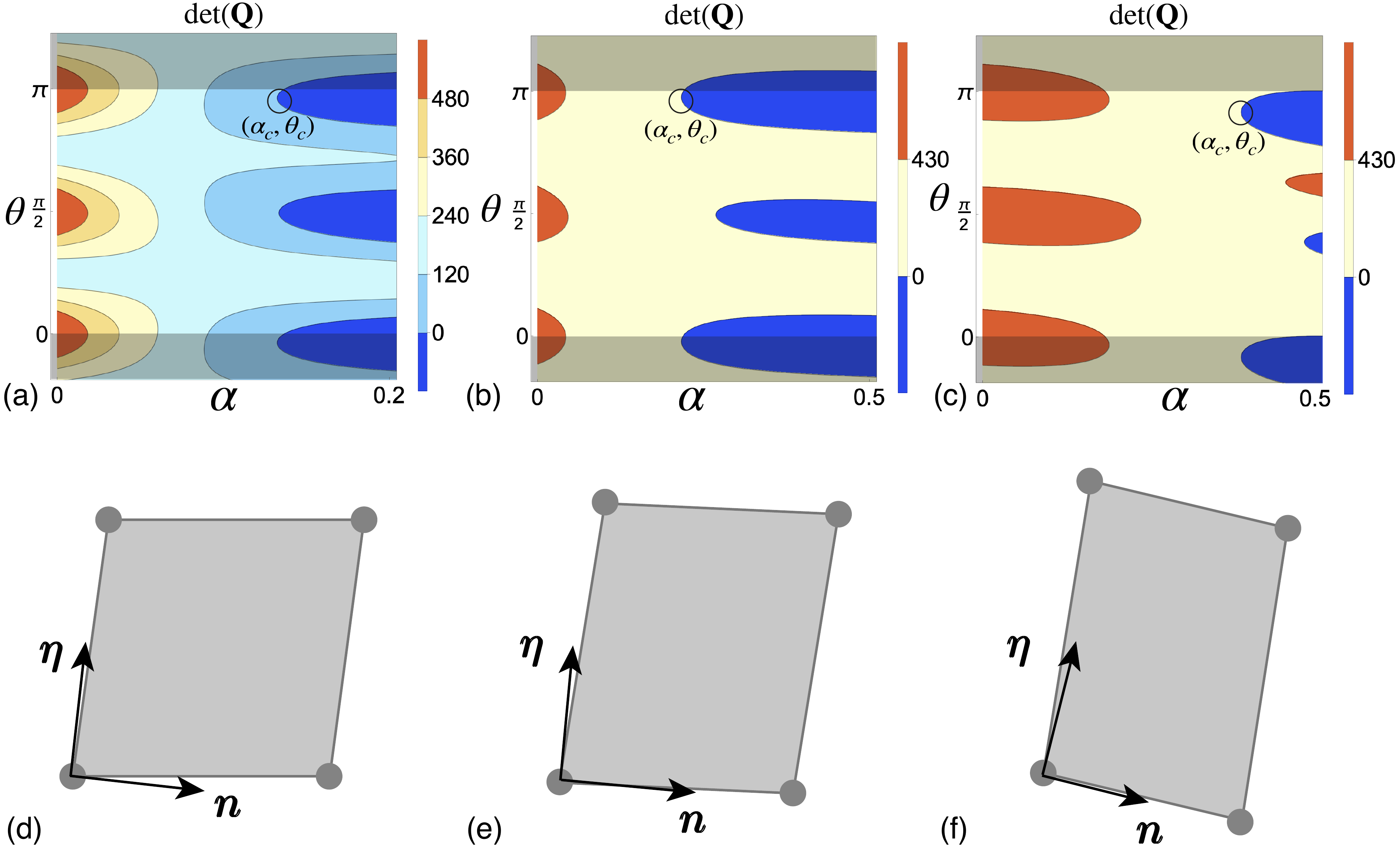}
\caption{\scriptsize {Instability limits for square lattices: (a-c) level sets for the determinant of the Eulerian acoustic tensor, (d-f) parameters of the unstable modes vis a vis the deformed lattice vectors.
%function versus  the angle $\phi$ for  crystal orientations:
 (a) $\phi=0$, (b) $\phi=\arctan{1/2}$ and (c) $\phi=\pi/4$. The  parameters   ($\alpha_c$,$\theta_c$) are calculated from the condition $\det(\bold{\mathcal Q})=0$.}
\label{fig:at_sq}}
\end{figure}

It is instructive to look in more detail at the  instability modes  for  crystals  reaching  different  points $\alpha = \alpha_c(\phi)$  along  the surface of ideal strength. The knowledge  of the corresponding vectors $\pmb{n}_c$ and $\pmb{\eta}_c$ allows one to identify the primary  instability  modes which, at least for sufficiently large $N$,  guide  the  eventual  nucleation of dislocational dipoles.  Since the value of the bulk modulus $\mu$  in the strain-energy density was chosen to be sufficiently large, our crystals are  effectively  incompressible  and  therefore we can always expect that approximately   $\pmb{n}_c\perp\pmb{\eta}_c$ and  the instability mode is close to a simple shear.  
%
%These directions can be also linked with the orientation of the activated slip planes.
%. Even though the bifurcated solutions are unstable for sufficiently large $N$, we can 
\begin{figure}[h!]
\centering
\includegraphics[scale=.23]
{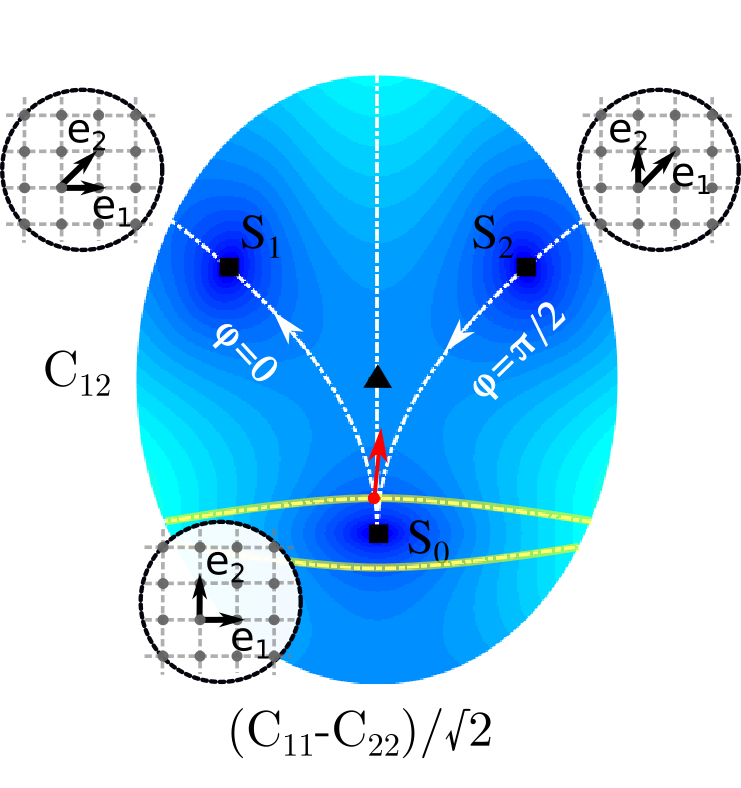}
%{/home/rbaggio/Documents/Bozze/answers_CR/images/lineariz.png}\\
\caption{\scriptsize {The orientation  (in the configurational space of metric tensors) of the first unstable mode  along  the simple shear loading path  $\phi=0$ originating in  $S_0$ (red arrow). Yellow curve is the theoretical strength threshold. The white dashed path passing through the black triangle (traingular lattice $T_0$) corresponds to   pure shear.   Shades of blue show the level sets of the  potential. White arrows point in the direction of growing simple shear amplitude $\alpha$.}
\label{fig:lineariz}}
\end{figure} 

Consider first  square lattices, and assume that   the macroscopic loading paths  are again  given by  \eqref{paths_sq} with  $\phi=0,\arctan( 1/2), \pi/4$, see   Fig. \ref{fig:detc1ensq}.  In  Fig. \ref{fig:at_sq}(a-c) we show the evolution of the determinant of the Eulerian acoustic tensor along each of these path as a function of the parameter $\theta$ defining the orientation of the unit vector   $\pmb{n} = (\cos \theta,\sin \theta)$. The external boundaries of the dark blue regions define the  stability limits and the smallest value of the loading parameter $\alpha$  on  one of such boundaries, defines the maximal  homogeneous strain achievable along the corresponding loading path. Note that the whole pattern is periodic in $\theta$  and only one period is shown in Fig. \ref{fig:at_sq}(a-c).

We recall that the  `soft'  path  with $\phi=0$ corresponds to a simple shear along one of the dense planes of the square lattice.  According to Fig. \ref{fig:at_sq}(a), at the critical value of the loading parameter $\alpha_c \approx  0.134$,  the associated $\theta_c\approx 3.035\text{rad}$. Since the unit  vectors  $  \pmb{n}_c $ and $ \pmb{\eta}_c \approx \pmb{n}^\perp$ are almost parallel  to the current (deformed)  lattice vectors, see Fig. \ref{fig:at_sq}(d),  the unstable mode   essentially activates a   single slip in the `vertical' direction. This is illustrated in Fig. \ref{fig:lineariz} where we show the orientation  of the   unstable mode.   Note  that  the red arrow in Fig. \ref{fig:lineariz} points to the right of the pure shear path.  This, together with  the global topography of the energy surface suggests that the instability will develop in the direction of the energy well $S_2$ rather than $S_1$.

\begin{figure}[h!]
\centering
\includegraphics[scale=.23]{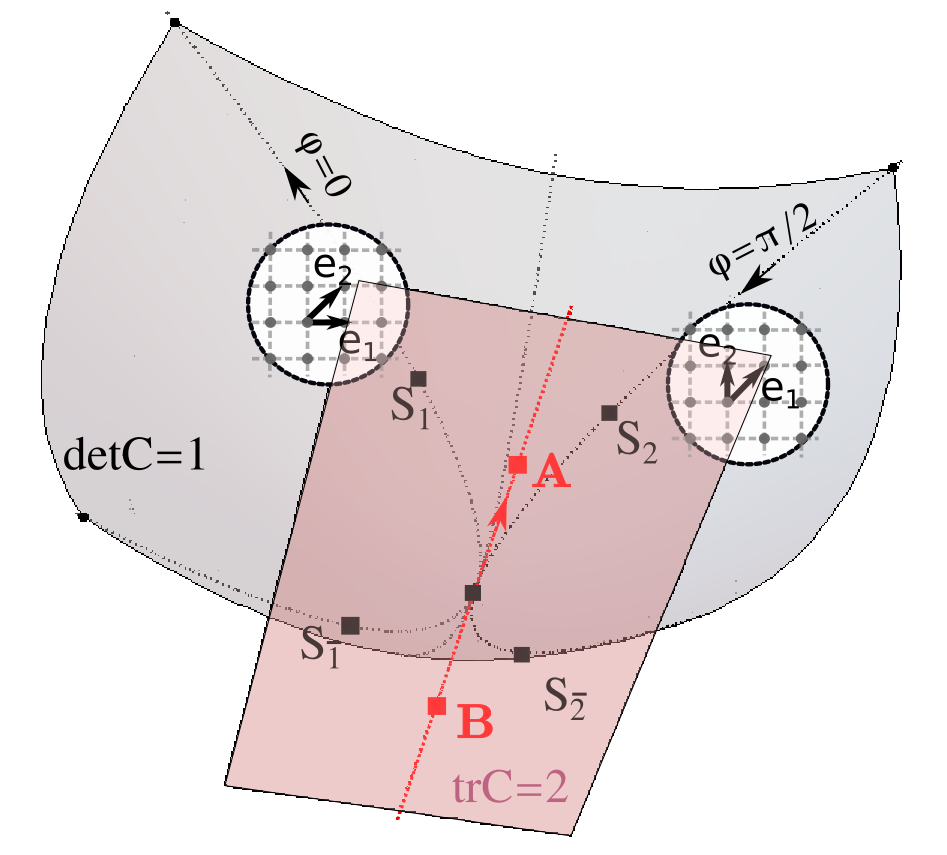}
%{/home/rbaggio/Documents/Bozze/answers_CR/images/corr.png}\\
\caption{\scriptsize {The effect of the geometric linearization of MTM around  the square energy well $S_0$. In such  approximation the (gray)  hyperbolic surface  $\det{\bold C}=1$ is replaced by the (pink) tangent plane $\tr {\bf C}=2$. The two simple shear paths with $\phi=0$ and $\phi=\pi/2$ as well as the pure shear path (shown by a separating gray dashed line) collapse into a single  path shown by the red straight line $AB$. Here a single point $A$ describes the two  square lattices $S_1$ and $S_2$, while a single  point $B$ stands for  the two  square lattices $S_{\bar1}$ and $S_{\bar 2}$}.
\label{fig:corr}}
\end{figure}

The fact that this first unstable mode is not the one associated with $\theta_c\approx \pi/2$, which would mean a   single slip in the  `horizontal'  direction,  may be thought as counter-intuitive based on the naive application of the  plastic 'mechanisms'  approach of CP.  And   indeed,     the mode  with $\theta_c\approx1.5746\text{rad}$, which  corresponds   to the anticipated   `horizontal' direction of slip,   is  destabilized  right after  the mode  with $\theta_c\approx 3.035\text{rad}$. 
%The implied near degeneracy of the bifurcation is an indicator that both  slip planes  may be activated almost simultaneously.  
The  \emph{splitting}  of these two modes  is the  effect of geometric nonlinearity. Thus,  if kinematics is linearized, which essentially means that the terms of the order of $\alpha^2$ are neglected, simple shears along the tensorial directions $\phi=0$ and $\phi=\pi/2$  become indistinguishable 
%when written in terms of the metric tensor  
%$ {\bf C(\alpha,\phi)}$
%$=(1/2)( {\bf F}^T(\alpha,\phi) {\bf F}(\alpha,\phi)-{\bf I})$ 
since  
\begin{equation}
\label{eqn:Ezernov}
\mathbf{C}(\alpha,0)=\left[\begin{array}{cc}
1 & \alpha\\
\alpha & 1+\alpha^{2}
\end{array}\right],\qquad\mathbf{C}(-\alpha,\pi/2)=\left[\begin{array}{cc}
1+\alpha^{2} & \alpha\\
\alpha & 1
\end{array}\right].
%\mathbf{C}(\alpha,0)=
% %\frac{1}{2}
% \left[\begin{array}{cc}
% 1 & \alpha\\                     
% \alpha & 1+\alpha^{2}\end{array}\right]
% ,\,\,\,\,\,\,
% \mathbf{C}(\alpha,\pi/2)=\left[\begin{array}{cc}
% 1+\alpha^{2} & -\alpha\\
% -\alpha & 1
% \end{array}\right]\,.
 \end{equation}
 %\mathbf{C}(-\alpha,\pi/2)=\left[\begin{array}{cc}
%1+\alpha^{2} & \alpha\\
%\alpha & 1
% \end{array}\right]
%%\bf{C}(\alpha,0)=
%%%\frac{1}{2}
%%\left[\begin{array}{cc}
%%1 & \alpha\\                     
%%\alpha & 1+\alpha^{2}
%%\end{array}\right]\qquad \bf{C}(\alpha,\pi/2)=
%%%\frac{1}{2}
%%\left[\begin{array}{cc}
%%1+\alpha^{2} & \alpha\\                     
%%\alpha & 1
%%\end{array}\right]

The geometrically  linearized theories, neglecting the terms \emph{quadratic} in $\alpha$,   are   confined to the plane $\tr {\bf C} =2$,  which is tangential to the hyperboloid $\det{\bold C}=1$ at $S_0$, see Fig. \ref{fig:corr}. 
In such theories, see, for instance,   \cite{Onuki2003-ln,Carpio2005-fl,Minami2007-ew,Geslin2014-ad}, the strain paths corresponding to the simple shearing deformations $\mathbf{\bar{F}}(\alpha,\phi=0)$ and $\mathbf{\bar{F}}(-\alpha, \phi=\frac{\pi}{2})$ would merge into a single line $C_{11}=C_{22}=1$. As a result, for instance, the two minima $S_1$ and $S_{2}$ would collapse on each other, see Fig. \ref{fig:corr}.   In our  geometrically nonlinear approach the  applied simple shear  biases one of these modes  and, as we have seen above,  the two   plastic `mechanisms' end up  activated \emph{consecutively} rather than \emph{simultaneously} as the linearized theory (as well as classical CP) would suggest.  Similar degeneracy  occurs  also  in the case of triangular lattices if the  configurational space is reduced to the plane tangent to the point $T_0$ as in \cite{Onuki2005-xp}.

\begin{figure}[h!]
\centering
\includegraphics[scale=.23]{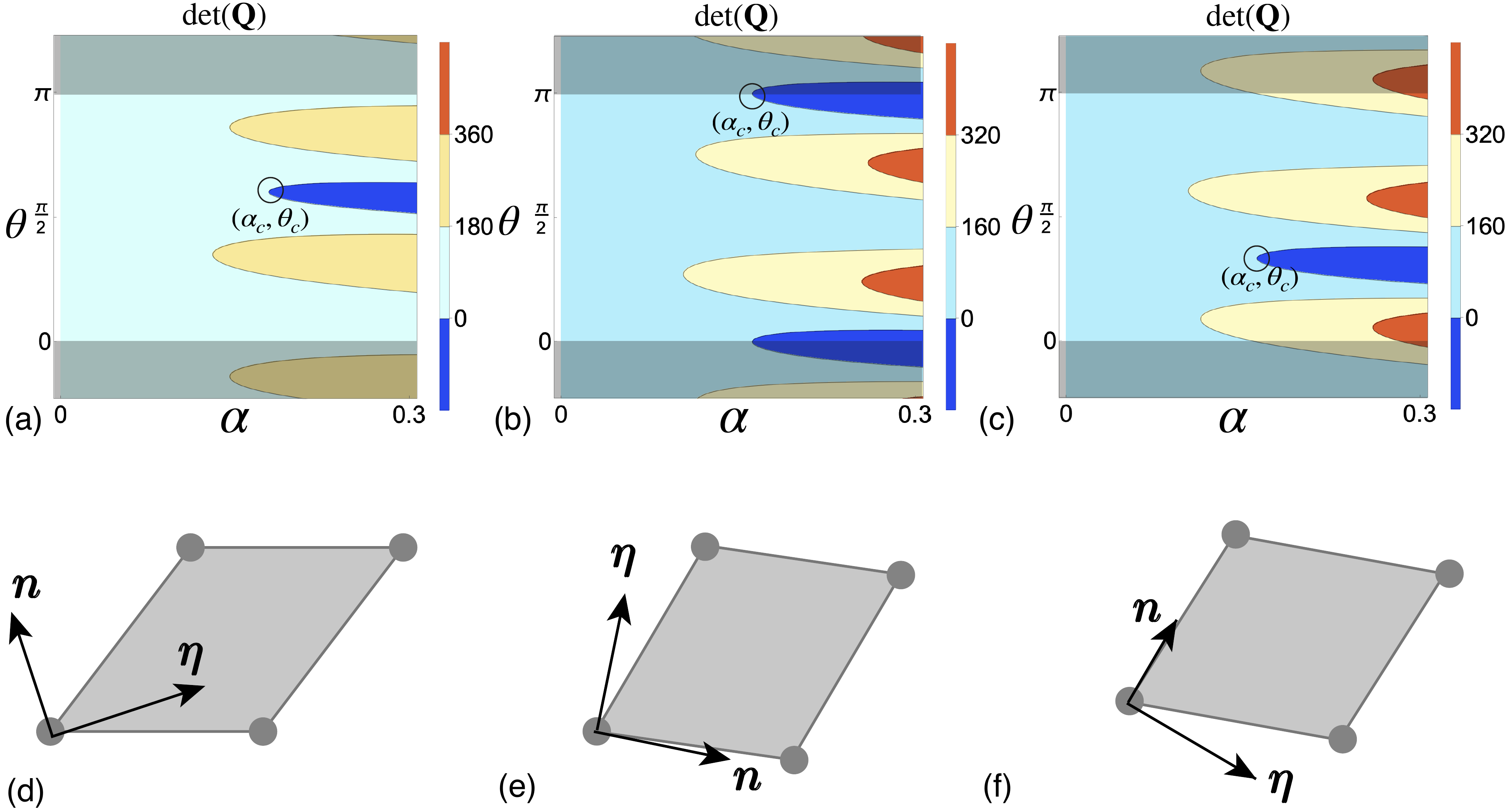}
\caption{\scriptsize {Instability limits for triangular  lattices: (a-c) level sets for the determinant of the Eulerian acoustic tensor, (d-f) parameters of the unstable modes vis a vis the deformed lattice vectors.
%function versus  the angle $\phi$ for  crystal orientations:
% (a) $\phi=0$, (b) $\phi=\arctan{1/2}$ and (c) $\phi=\pi/4$. The  parameters   ($\alpha_c$,$\theta_c$) are calculated from the condition $\det(\bold Q)=0$.
%Triangular crystal: determinant of Eulerian acoustic tensor function as a function of $\alpha$ and $\phi$ for  shearing directions: 
(a) $\phi=0$, (b) $\phi=\pi/3$ and (c) $\phi=\pi/2$. The  parameters  ($\alpha_c$,$\theta_c$) are calculated from the condition $\det(\pmb{\mathcal Q})=0$.}
 \label{fig:at_hex}}
\end{figure}

For our two other loading  paths with  $\phi=\arctan(1/2)$ and  $\phi=\pi/4$, see 
 Fig.~\ref{fig:at_sq}(b,c),  a single unstable mode  can be clearly isolated.  The  vectors  $\pmb{n}_c $ and $ \pmb{\eta}_c\approx\pmb{n}_c^\perp$  are again almost parallel  to the deformed lattice vectors in the case of $\phi=\arctan(1/2)$ path, see Fig.~\ref{fig:at_sq}(e). The  ensuing instability is close to the one along the path  $\phi=0$.
 In particular,   a  single slip system is    activated (`vertical') and initially nucleated dislocation dipoles  is  of one type only. The post-bifurcational localization can be then again expected to  take the form of  the nucleation of dislocation pairs  along the  plane selected by  the condition of continuum instability as in Peierls-Nabarro model \cite{Nabarro2002-js}. 
 
  In the case of $\phi=\pi/4$ path, see Fig.~\ref{fig:at_sq}(f), the situation is different because the pre-instability lattice is strongly distorted and is almost rectangular. The  instability mode is no longer aligned with the  low energy valleys of the energy landscape which should lead to the simultaneous activation of both  slip systems and the emergence of both 'vertical' and 'horizontal' dislocations.   Overall, as we move away from  the `soft' path $\phi=0$, the  loading directions become progressively  `harder'  and  along such paths the instability takes place at higher values of energy density.  Given that  the corresponding  the unstable modes are  not perfectly aligned with deformed lattice vectors,  the ensuing geometrical frustration should lead to higher complexity of the  dislocation patterns.

%For the paths which evolve in the low energy basins   instability occurs at relatively low values of the load $\alpha$. Moreover two unstable modes may appear almost simultaneously suggesting the simultaneous activation of the two plastic mechanisms corresponding to the two low energy valleys associated with $\phi=0$  and $\phi=\pi/2$ simple shears. Instead, paths evolving through higher energy barriers are characterized by higher values of $\alpha_c$, instability takes place at higher values of energy density   and the associated unstable mode is not aligned with lattice directions. 

In the case of triangular lattice, we again consider non-generic  simple shear loading paths along the dense crystallographic planes with  $\phi=0$ and $\pi/3$  and  a generic path  with $\phi=\pi/2$, see~\eqref{pathshex}. The first  two   represent  `soft' shearing directions, while the last one is the  `hard'  one  facing potentially higher energy barrier.  Figure~\ref{fig:at_hex}(a,b,c)  show the level sets  of the determinant of the Eulerian acoustic tensor  in the plane of parameters  $\alpha$ and  $\theta$  for  each of the three paths.  

Note first that,  in contrast to the case of   square symmetry,   here  the instability thresholds  along different paths do not differ considerably.  Moreover, the apparently `hard' path with $\phi=\pi/2$, which does not correspond to any  straightforward  plastic `mechanism', gets destabilized before the apparently  `easy' ones with $\phi=0$ and $\phi=\pi/3$. We reiterate  that such quasi-uniformity  of the instability conditions  is a natural property  of higher symmetry lattices with all   thresholds  collapsing onto one in the case of ideal isotropic solids (with superimposed fluctuations in the case of amorphous glasses or polycrystals).

We also observe that  in all three cases the instability modes are  characterized by the vectors $\pmb{n}_c $ and $ \pmb{\eta}_c\approx\pmb{n_c}^\perp$  that are no longer co-linear  with the orientation of the deformed lattice basis. This suggests that the initial patterning,  controlled by the  continuum instability,   is incommensurate with lattice slips and therefore the  mechanism of dislocation  nucleation   may be rather different from the one predicted  by the Peierls-Nabarro model \cite{Nabarro2002-js}. The fact that still $\pmb{\eta}_c\approx\pmb{n}_c^\perp$  suggests that the macroscopic instability   will appear in the form of simple shear, however,  the misalignment of the instability mode with  crystallographic slip planes makes the prediction of the actual activated slip planes  difficult. 

\begin{figure}[h!]
\centering
\includegraphics[scale=.35]{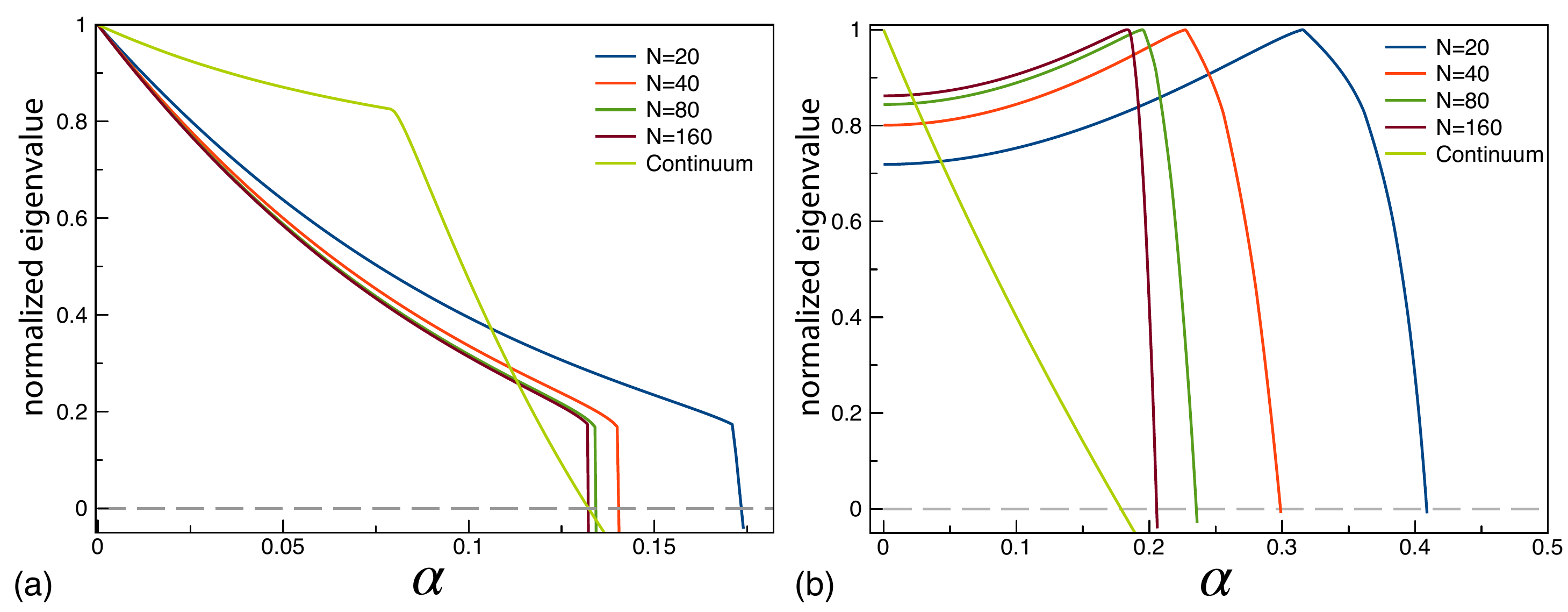}
\caption{\scriptsize {Smallest eigenvalue of the discrete stiffness  matrix ${\bf K}$ as a function of the loading parameter $\alpha$ for different values of element size $h=1/N$: (a) square, (b) triangular lattice.   The  crystal orientation is always $\phi=0$.  In the continuum case,  we  show  the smallest eigenvalue of the  acoustic tensor  $\pmb{\mathcal Q}({\bf F},\pmb{n}_c)$.}
 \label{fig:smalleigen}}
\end{figure}

To show the effect of finite $N$, we also studied the stability limits in the original discrete setup, see also  \cite{Sanderson2016-ht,Sanderson2018-rt}. To this end we   computed the   smallest eigenvalue of the $2N^2\times 2N^2$ Hessian  matrix $K^{ab}_{ij}$.   Our   Fig.~\ref{fig:smalleigen} shows   that although the critical value of the loading parameter  $\alpha_c^*(N)$ in the discretized problem, chosen by the condition that such eigenvalue is equal to zero,  is slightly larger than the ideal shear strength $\alpha_c$ in the continuum problem,  the two approach quickly  as the element size $h=1/N$ decreases. In the case of square lattices  the gap  closes up already at $N\sim 100$  as the instability directions are almost perfectly  aligned with the deformed crystal, see  Fig.~\ref{fig:smalleigen}(a).  In the case of triangular lattices, the misalignement is much stronger and the finite size effect in the form of a gap between the predictions of discrete and continuum theories remains apparent for much smaller element sizes, see Fig.~\ref{fig:smalleigen}(b). 
\begin{figure}[h!]
\centering
\includegraphics[scale=.4]{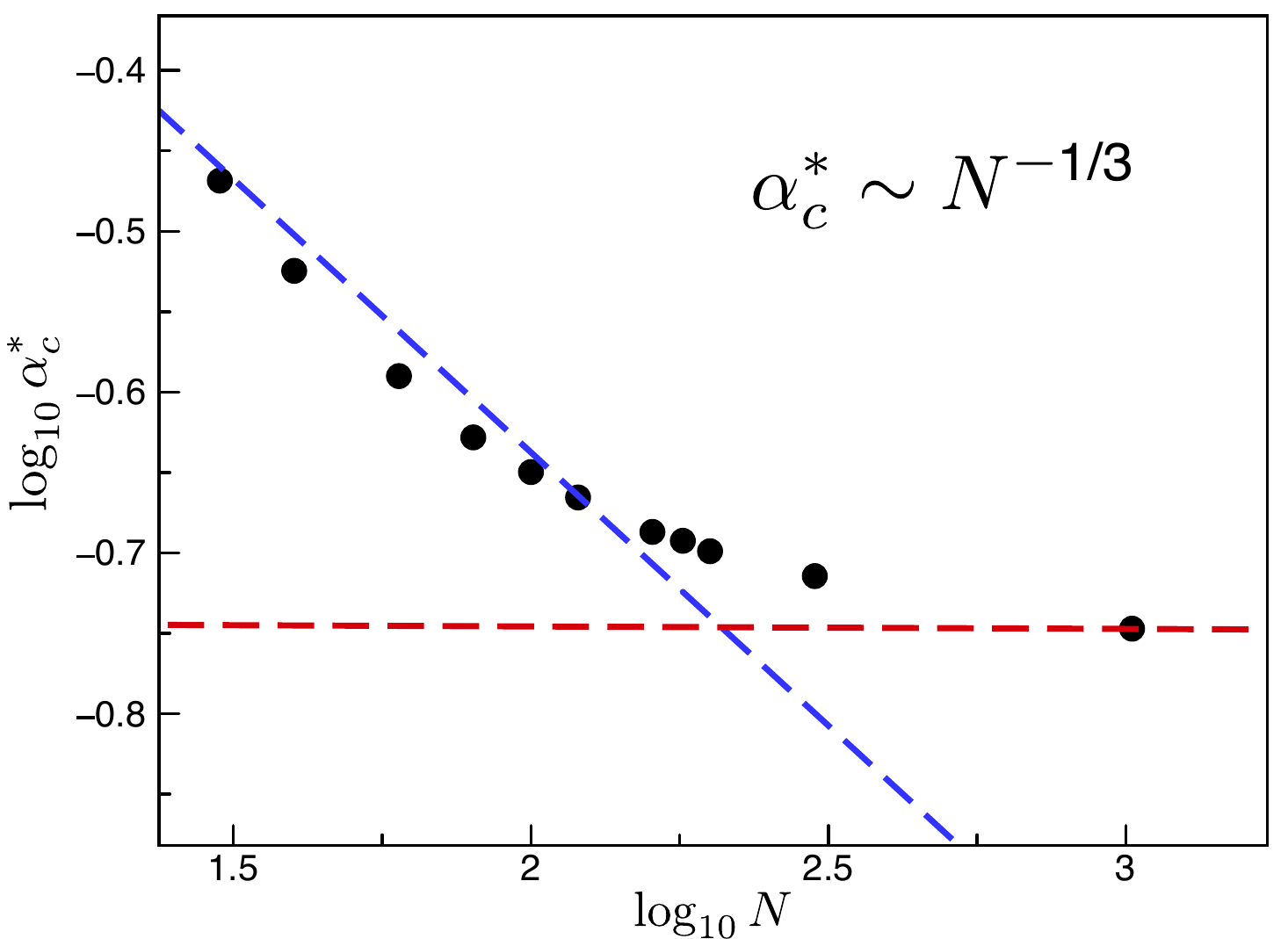}
\caption{\scriptsize {System size dependence  of the critical loading parameter $\alpha_c^*(N)$ in numerical experiments with  triangular crystals oriented at $\phi=0$.}
 \label{fig:smalleigen1}}
\end{figure}
Note that the asymptotic behavior of the smallest eigenvalues near zero, shown in Fig. \ref{fig:smalleigen},  is different  from the  prediction of the theory of amorphous plasticity where the corresponding eigenvectors are  quasi-localized~\cite{Dasgupta2012-uj,Bonfanti2019-xi,PhysRevMaterials.4.113609}. The linear response of pristine crystalline solids with zero  disorder is  global and that is why its elastic instabilities  can be largely captured by the classical continuum theory.

The dependence of the instability threshold $\alpha_c^*$  on the system size $N$ is illustrated in Fig. \ref{fig:smalleigen1}. The theoretical limit, effectively reached  by the samples with $N=1024$, is shown by the red dashed line. At small $N$ we observe the emergence of  a nontrivial  asymptotics $\alpha_c^* \sim N^{-1/3}$. It suggest that the  spatial scale of the dislocation microstructure  is of the order $ \sim  h^{1/3}L^{2/3}$, which implies  hierarchical 'domain splitting'  near the  external boundaries.

\begin{figure}[hbt!]
\centering
\includegraphics[scale=.15 ]{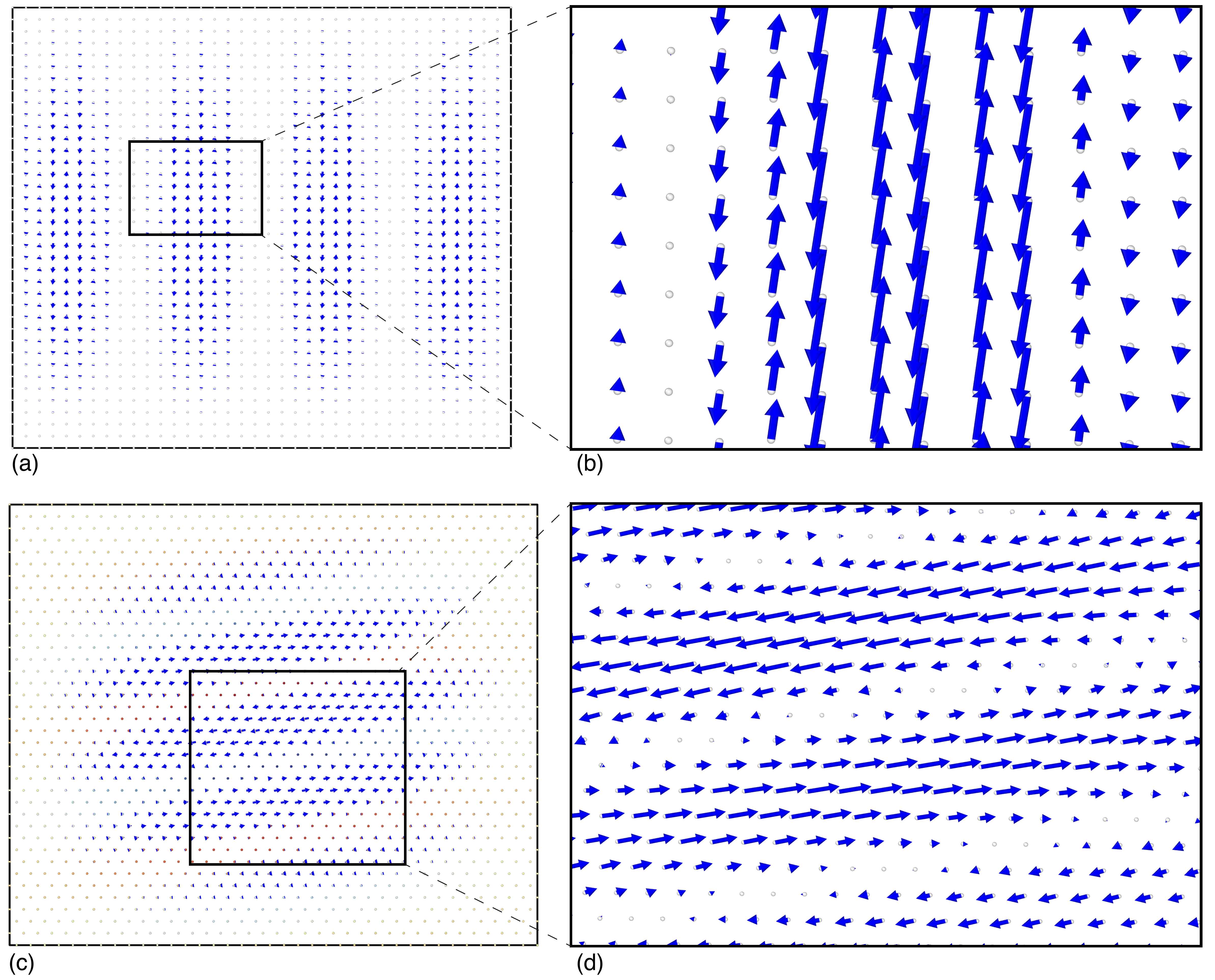}
\caption{\scriptsize {Unstable modes at the instability point   for  square  (a) and triangular (b) lattices. Blue arrows show the   modes associated with the lowest eigenvalue of the stiffness matrix $\bf{K}$. Here $N=40$ and $\phi=0$. 
%Note that in the square lattice case the instability develops on  two  spatial  scales;  of the discrete elements  and  of the  whole domain. 
}
 \label{fig:modes_sq}}
\end{figure}
%

%\begin{figure}[h!]
%\centering
%\includegraphics[scale=.17]{./fig18.pdf}
%\caption{\scriptsize {The first three unstable modess in the triangular crystal at ${\bf F}^c$. Red arrows indicate the amplitude and direction of the displacement field.}
% \label{fig:modes_tr}}
%\end{figure}
For square and triangular lattices with  $N=40$ we illustrate in Fig. \ref{fig:modes_sq}(a,b),   the $2N^2$ dimensional eigenvectors $u_i^a$ of the corresponding Hessian matrices $K^{ab}_{ij}$  when their lowest eigenvalues  crosses zero at $\alpha_c^*$. These eigenvectors are presented as vector fields with the homogeneous component subtracted. In both cases, the modulations have orientations $\pmb n_c^*$ that are close to the ones predicted by the continuum theory. However, if the unstable wavenumber remains arbitrary in the scale-free continuum theory, the discrete theory selects a particular length scale. The strong misalignment, in the case of triangular lattices between the orientation of the macro modulations and the lattice vectors, see Fig. \ref{fig:modes_sq}(b), can delay the transition to the ultimate lattice scale instability culminating in the nucleation of dislocation dipoles. In the case of square lattices, shown in Fig. \ref{fig:modes_sq}(a), we observe lattice scale modulations present already in the original unstable mode.  They correspond to the wave vectors at the boundary of the Brillouin zone, see some elementary examples of such multiscale instabilities in \cite{Truskinovsky2005-gx,Bertoldi2008-au}.

To summarize, our numerical experiments show that pristine sub-micron crystals exposed to affine deformation yield discontinuously near the thresholds of theoretical strength. In the case of simple shear loading,  the obtained thresholds show dependence on the orientation of the sample in the loading device, which is in qualitative agreement with experimental results, e.g.  \cite{Ziegenhain2010-tu,Bagheripoor2020-qm}.  Our study also reveals a strong link between  the structure of the apparent yield surface and  the lattice symmetry.

\section{Discontinuous yielding}
 
In this Section, we  turn to the study of the fine structure of the microscopic response  following the loss of stability of the homogeneous state. The goal is to reveal the detailed unfolding of the catastrophic dislocation nucleation avalanche and trace how the final configuration, containing a large number of defects, emerges in the process of energy minimization. In particular, we show that complementary pictures of the evolving defect configurations emerge in the real and the configurational spaces. 

All numerical experiments start with a dislocation-free crystal $\bold u =0$, and we drive the system using an athermal quasistatic protocol. We use hard device boundary condition controlling the positions of surface nodes and impose in this way an affine deformation $\bar{\bf F}$ on the boundary of a square domain and choose the loading step (strain increment) sufficiently small to ensure continuity of the minima outside the avalanches. The macroscopic stress-strain and energy-strain curves obtained in such numerical experiments are summarized in Fig.~\ref{fig:energydrops}(a) and Fig.~\ref{fig:energydrops}(b) for the square and triangular crystals, respectively.

For $\alpha<\alpha^*_c$ the picture is rather simple. In real space, the deformation is affine ${\bf F(\bf x)}\equiv\bar{\bf F}$ and in the configurational space of the metric tensors ${\bf C}$  we observe the perfect overlap for all elements.    As we have already mentioned, there is an excellent agreement between the numerical value of $\alpha^*_c$ and the predicted value of ideal shear strength, see Fig.~\ref{fig:energydrops}. As soon as stability is lost, the homogeneous configuration breaks down. The pristine crystal transforms into a highly defected one, with most of the dislocations eventually escaping from the bulk and forming various pile-ups near the fixed boundaries. The configurational points, corresponding to different elastic elements, spread in the space of metric tensors ${\bf C}$, reaching the neighborhoods of distant energy wells. The emerging dislocation patterns, which we discuss in detail below,  were obtained from the simulations involving  more than a million elements.

%
%We observe that, in each case, at the critical value of the loading parameter $\alpha$, the affine elastic branch becomes unstable and a new equilibrium branch is reached  after an abrupt drop in the energy  whose amplitude depends on the direction of loading $\phi$. In what follows,  we study the mechanical response of the crystals following the loss of stability of  homogeneous states and show how in each case a perfect crystal transforms into a `defected' one.
%
%
%In order to ensure independence of the system size, we performed the simulations considering different numbers of nodes, ranging from $10^4$ to $10^6$.
% We also used different boundary conditions, experimenting with both fixed and periodic conditions\footnote{In order to apply periodic boundary conditions, we assign as neighbor to each boundary node the corresponding node on the opposite side. Positions of these periodic neighbors are then adjusted, keeping in mind that they have to be translated by the domain dimension deformed accordingly with the imposed deformation gradient $\bf F$. In correspondence of every loading step, boundary nodes are moved in agreement with the imposed deformation gradient, but are then let free to relax. }.
%In the performed simulations, no significant dependence on system size $N$ has been observed for what concerns the main features of the post-instability pattern.

\subsection{Square lattices}

\begin{figure}[h!]
\centering
\includegraphics[scale=.35]{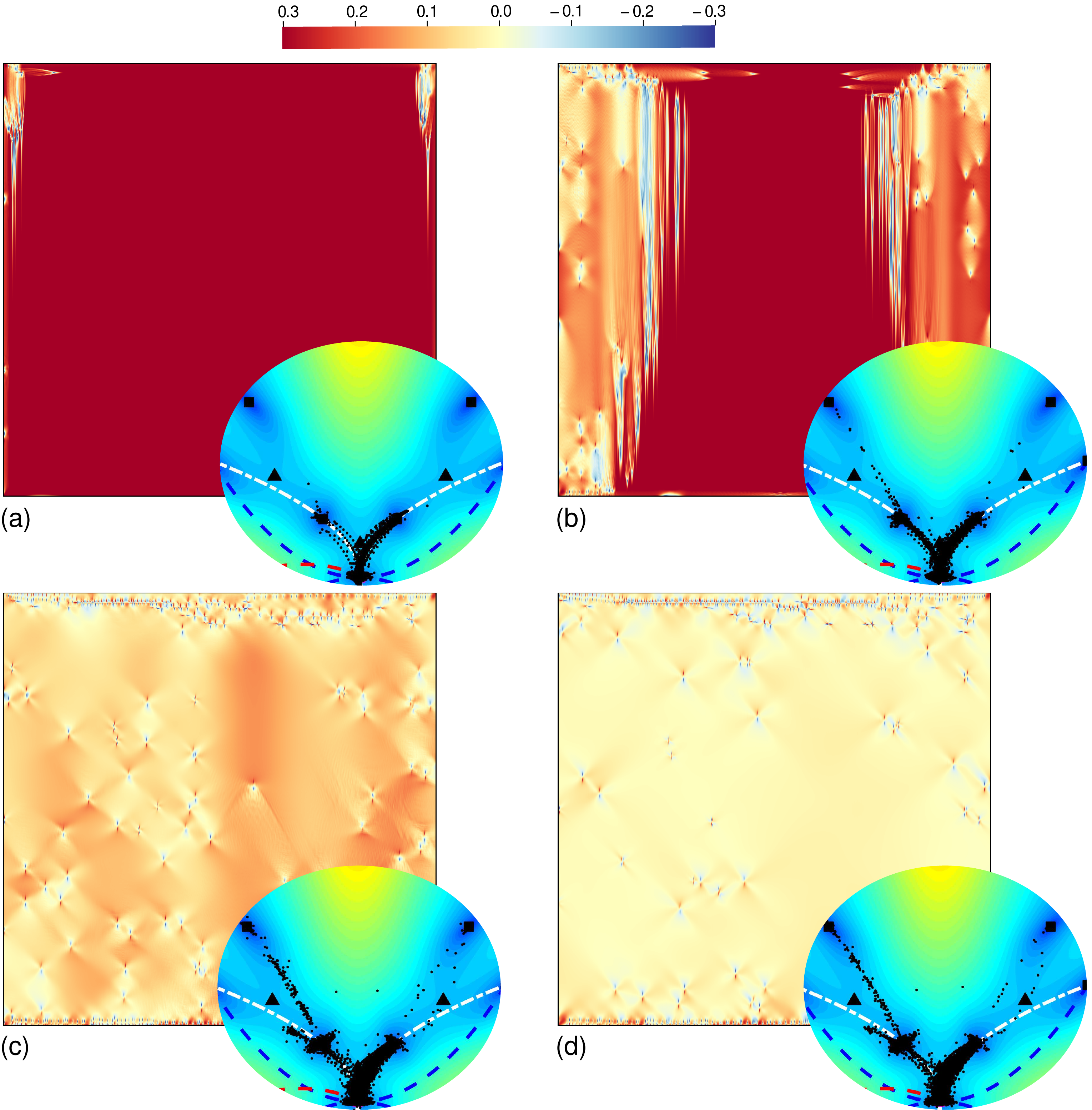}
\caption{\scriptsize {Snapshots of the stress field during discontinuous yield in pristine square crystals. Colors indicate the level of the shear component of the Cauchy stress. Insets show the same images in the configurational space. Black dots indicate the value of the metric tensor in individual finite elements.
%Square crystal: dislocation nucleation   following the loss of stability as it evolves during the different stages (a-h) of the energy minimisation. 
 %; (i) the final mechanical equilibrium configuration. 
 Here $N=1024$ and $\phi=0$. }
 \label{fig:sq0}}
\end{figure}

Consider first a square crystal loaded in a  crystallographically exact  simple shear with $\phi=0$.  In Fig. \ref{fig:sq0}, we show the snapshots of the distribution of the shear component of the Cauchy stress  ${\bf \sigma} = (\det{\bf F})^{-1}{\bf F}{\bf P}^T$ in the physical space as the crystal evolves (in fast computational time) through the   avalanche after the loading parameter  has reached  the   value $\alpha_c^* $. In the insets,  we illustrate  the concurrent evolution  of the cloud of configurational points in the space of metric tensors which tracks different stages of  the implied energy minimization process.

At  positive values of the loading parameter $\alpha$   the crystal  is driven along the energy valley  away  from the energy minimum $S_0$   towards   the minima  $S_1$, $S_{11}, $ etc.  This  could suggest the development of plastic slip along the horizontal  plane, however, as we have seen,  the first instability is `vertical'   which means activation of the slip system  represented by the minima  $S_2$, $S_{22},$ etc., see Fig. \ref{fig:at_sq}. While the initial states of the avalanche  are indeed dominated by the activation of the vertical dislocation dipoles, see Fig.~\ref{fig:sq0} (a), horizontal dipoles  appear later as well  because of local stress concentration and consequent involvement of the second slip system  predicted by the instability analysis to be activated at slightly larger loading levels, see Fig.~\ref{fig:sq0}(b). 

Note that   the   two low energy valleys merge around the point $T_0$, which describes a  triangular lattice. The structure  of the energy landscape around this point  is close to   a   monkey saddle  \footnote{As discussed in detail in ~\cite{Baggio2019-rs},   for our choice of the elastic potential it is not exactly a  degenerate 'monkey saddle' but rather  a shallow energy maximum (triangular lattice)   surrounded by   three closely located non-degenerate (classical) saddles (rhombic lattices). } and passing of configurational points through such region  creates non-trivial coupling between horizontal   and vertical  slip systems.   As a result, even though most of the units move towards  the energy well $S_2$ some of the configurational points eventually  also populate   the energy well $S_1$.  As the crystal is driven  further,  it is pulled towards another (almost) monkey saddle  describing the  triangular lattices  $T_1$.  Around this intersection  of the low energy valleys,  the configurational points take a turn  towards the energy well $S_{12}$ instead of continuing towards apparently more natural well $S_{11}$. Similarly, around the (almost) monkey saddle $T_2$ the configurational flow is directed towards the energy well $S_{21}$, see Fig.~\ref{fig:sq0}(b-d). All these  choices depend  sensitively on the boundary conditions and will be rationalized in a separate publication.

%where we see that the second nearest energy wells $S_{12}$ and $S_{21}$, are also filled because of the secondary interaction of two perpendicular slip planes which again involves a monkey saddle type of flat barrier associated with triangular lattices  $T_1$ and $T_2$. 
%the initial partial slips transform into identifiable individual dislocations (Fig.~\ref{fig:sq0}(b-c)). From the perspective of our model, the image of dislocation cores in the configurational space is composed of the metric coordinates of the finite elements units located  between the bottoms of the   equivalent configurations. 
%The final mechanical equilibrium configuration is shown with its configuration space image in Fig.~\ref{fig:sq0}(d)

\begin{figure}[h!]
\centering
\includegraphics[scale=.15]{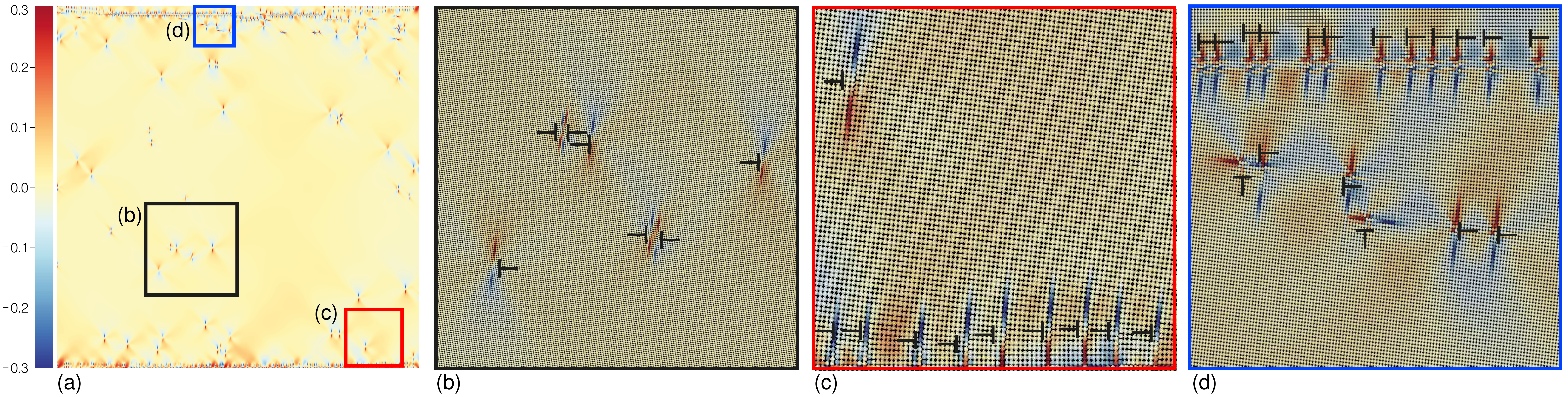}
\caption{\scriptsize {Final stress configuration after the  discontinuous yield in pristine square crystals: (b-d) show the enlarged versions of the square indicated by the same letters  in (a).
%: final equilibrium complex dislocation pattern shown in Fig. \ref{fig:sq0}(d). We observe examples of single dislocations and dipoles (b), and dislocation pile-ups close to fixed boundaries (c);  junctions form as well  due to the interaction  of two perpendicular different gliding directions (d).
 Here $N=1024$ and $\phi=0$.}
 \label{fig:sq0_2}}
\end{figure}

\begin{figure}[h!]
\centering
\includegraphics[scale=.35]{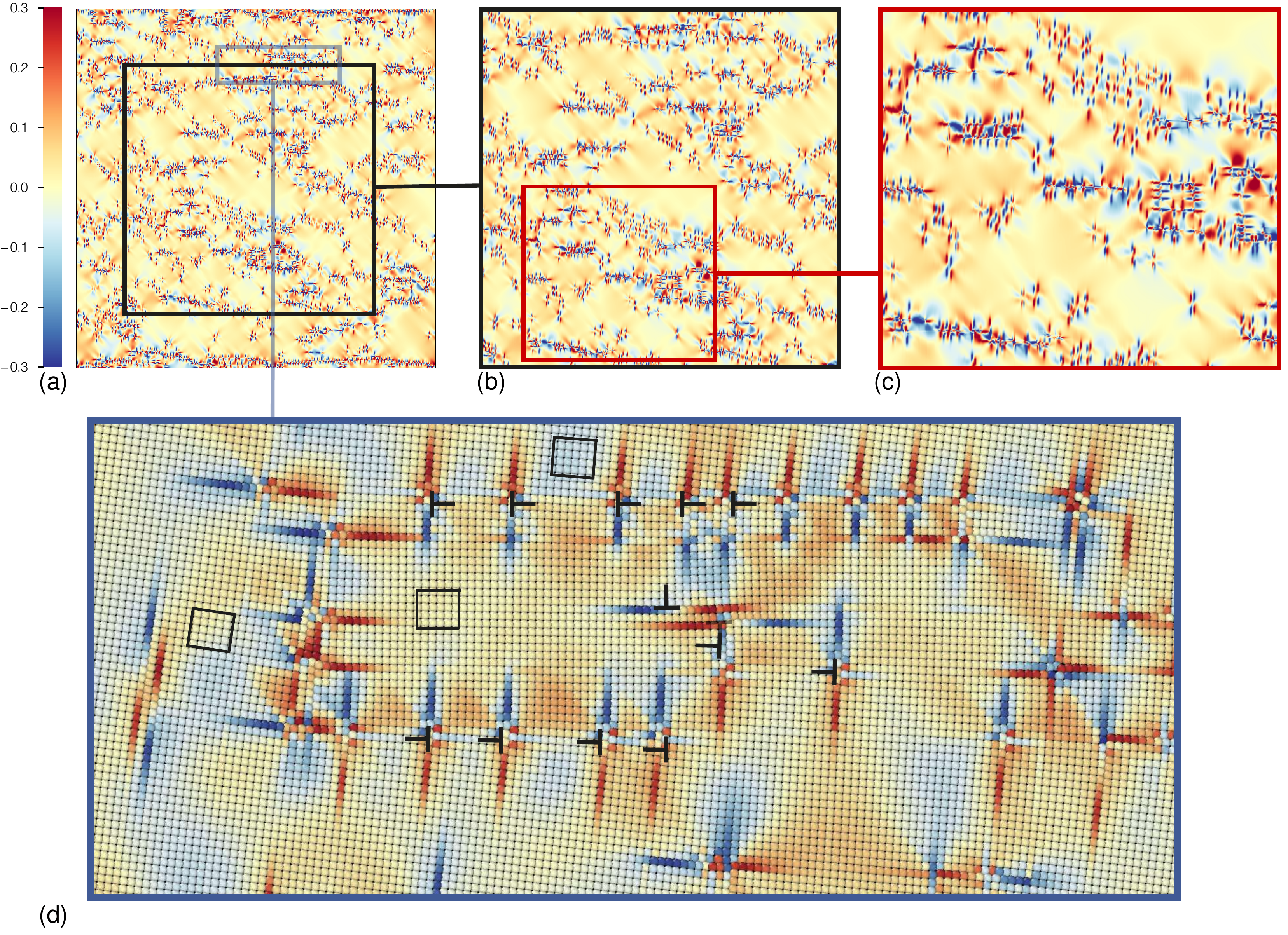}
\caption{\scriptsize {Final stress configuration after the  discontinuous yield in a square crystal  which was  randomly perturbed at the point of instability. Colors indicate the level of the  shear component of the Cauchy stress. 
%The presence of disorder induces simultaneous activation of 2 gliding planes leading to the formation distinct dislocation-rich and -poor regions. 
Squares show   the rotated versions of the unstressed  lattice. Here $N=1024$ and $\phi=0$.}
 \label{fig:disorder}}
\end{figure}

To see how the spreading of the configuration points in the space of metric tensors transforms when we move into   the physical space, we focused  in   Fig. \ref{fig:sq0_2} on three  randomly selected  fragments of  the post-avalanche  equilibrium configuration shown in Fig.~\ref{fig:sq0}(d). The emerging equilibrium pattern contains dislocational pile ups near the boundaries and few locked up dislocational dipoles   in the bulk of the specimen. Note, in particular, the   pairs of dislocations on two parallel slip planes blocking each other, see  Fig. \ref{fig:sq0_2}(b), and on two perpendicular slip planes forming   characteristic locks, see  Fig. \ref{fig:sq0_2}(d). The fact   that after the system spanning avalanche most of the nucleated defects manage to either annihilate  or  escape  to the boundaries of the crystal, explains why the crystal  becomes almost pristine  after such a devastating  discontinuous yield event.

To check the generality of these conclusions, we  also performed   numerical simulations of the discontinuous yield  at $\alpha \sim \alpha^*_c$ with  initially \emph{perturbed} displacement field which was `dirtied'  by  random Gaussian  perturbations  with  zero mean and small variance ($ \sim 0.0001$). Such disorder is too small  to suppress brittleness \cite{Zhang2020-ax} and, we show in Fig. \ref{fig:disorder} the  equilibrium stress field after the catastrophic avalanche. Since the pre-avalanche state was not `pristine', the energy minimizing self organization of dislocations is compromised by the geometrical frustration  imposed by the disorder. The  resulting dislocation  pattern appears to be structurally similar at several scales, see the successive insets shown in  Fig. \ref{fig:disorder}(b,c). The two main perpendicular slip planes are now activated at almost the same level.  Note  the formation of self screening dislocation-rich walls, reminiscent of low-angle grain boundaries,  which separate   dislocation-poor, low-stress  domains  where  the lattice  undergoes simple rotations, see small rotated squares in Fig. \ref{fig:disorder}(d);  all these structural features have been previously recorded in  physical experiments. Note also, that since  samples containing imperfections   show diminished brittleness~\cite{Zhang2020-ax}, the  discontinuous yield can disappear   in cyclic loading already after few cycles.

\begin{figure}[h!]
\centering
\includegraphics[scale=.35]{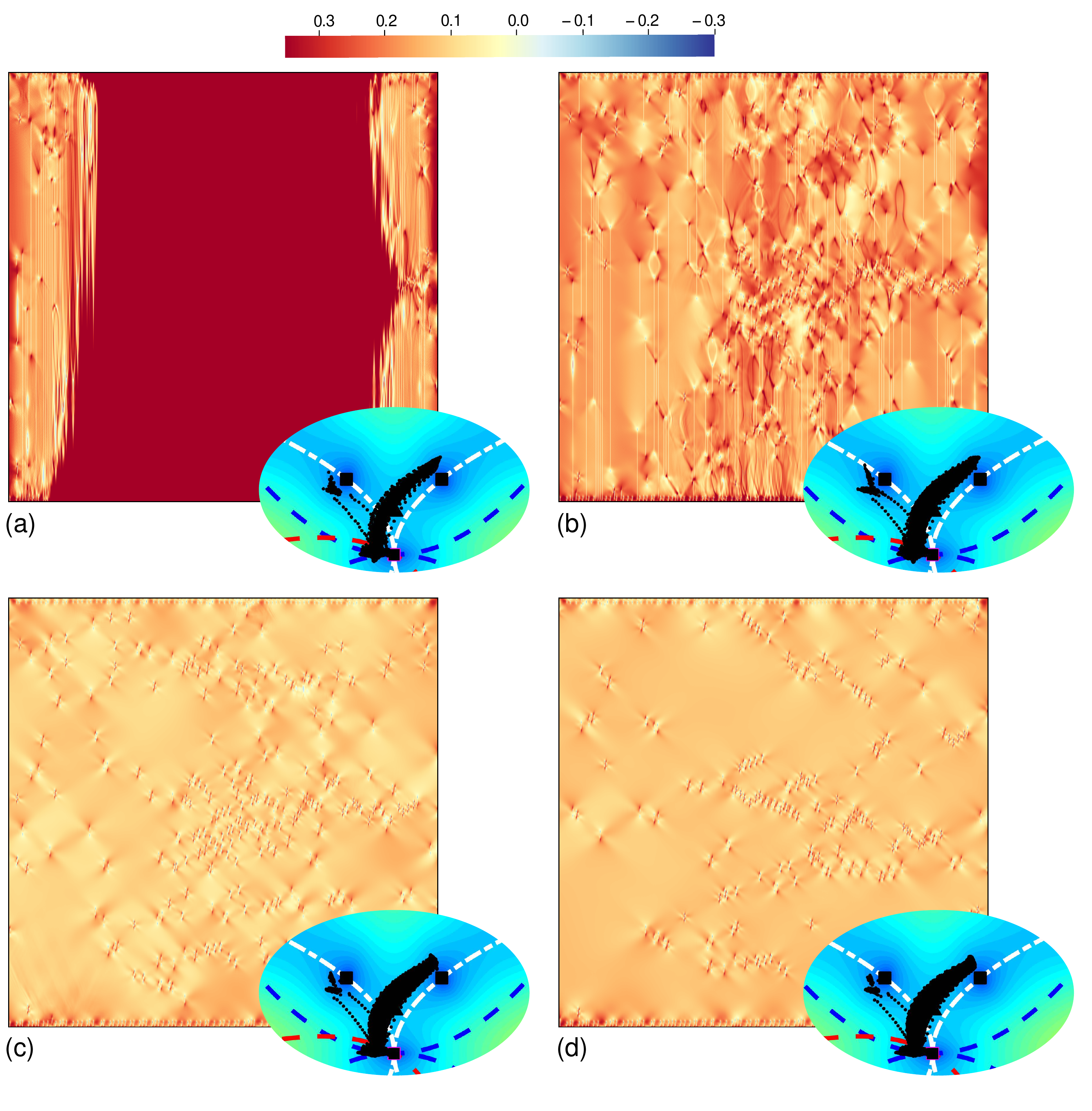}
\caption{\scriptsize {Snapshots of the stress field during discontinuous yield in pristine square crystal. Colors indicate the level of the shear component of the Cauchy stress. Insets show the same images in the configurational space. Black dots indicate the value of the metric tensor in individual finite elements.
%Square crystal : final equilibrium complex dislocation nucleation pattern. Colors indicate the level of the Cauchy stresses $\sigma_{xy}$.  
Here $N=1024$ and $\phi=\arctan( 1/2)$.  }
 \label{fig:th26}}
\end{figure}

Consider next  a generic  `hard'   shear loading path with  $\phi=\arctan( 1/2)$ which  does not follow the low energy  valley.    As we have already mentioned, in this case a homogeneously deformed crystal will  store  a bit more  elastic energy before  instability than in the case of the symmetric   `soft' paths. The  breakdown scenario is shown in  Fig. \ref{fig:th26}: after reaching the (regularized) ideal strength limit  $ \bar{\bf F}(\alpha_c^*,\arctan{\frac{1}{2}})$, dislocations nucleate along the `vertical' slip plane  which  agrees with the prediction of the stability analysis.  In the configurational space of metric tensors the stream of points describing such dislocations is directed  from  $S_0$ to  $S_{2}$, see Fig. \ref{fig:th26}(a). Eventually few `horizontal' dislocations form as well, see Fig. \ref{fig:th26}(b,c), and because of geometrical frustration  they are not all expelled to the boundaries of the crystal, see Fig. \ref{fig:th26}(d). Therefore the system is left with considerable residual energy and the stress drop is relatively small.
  
\begin{figure}[h!]
\centering
\includegraphics[scale=.35]{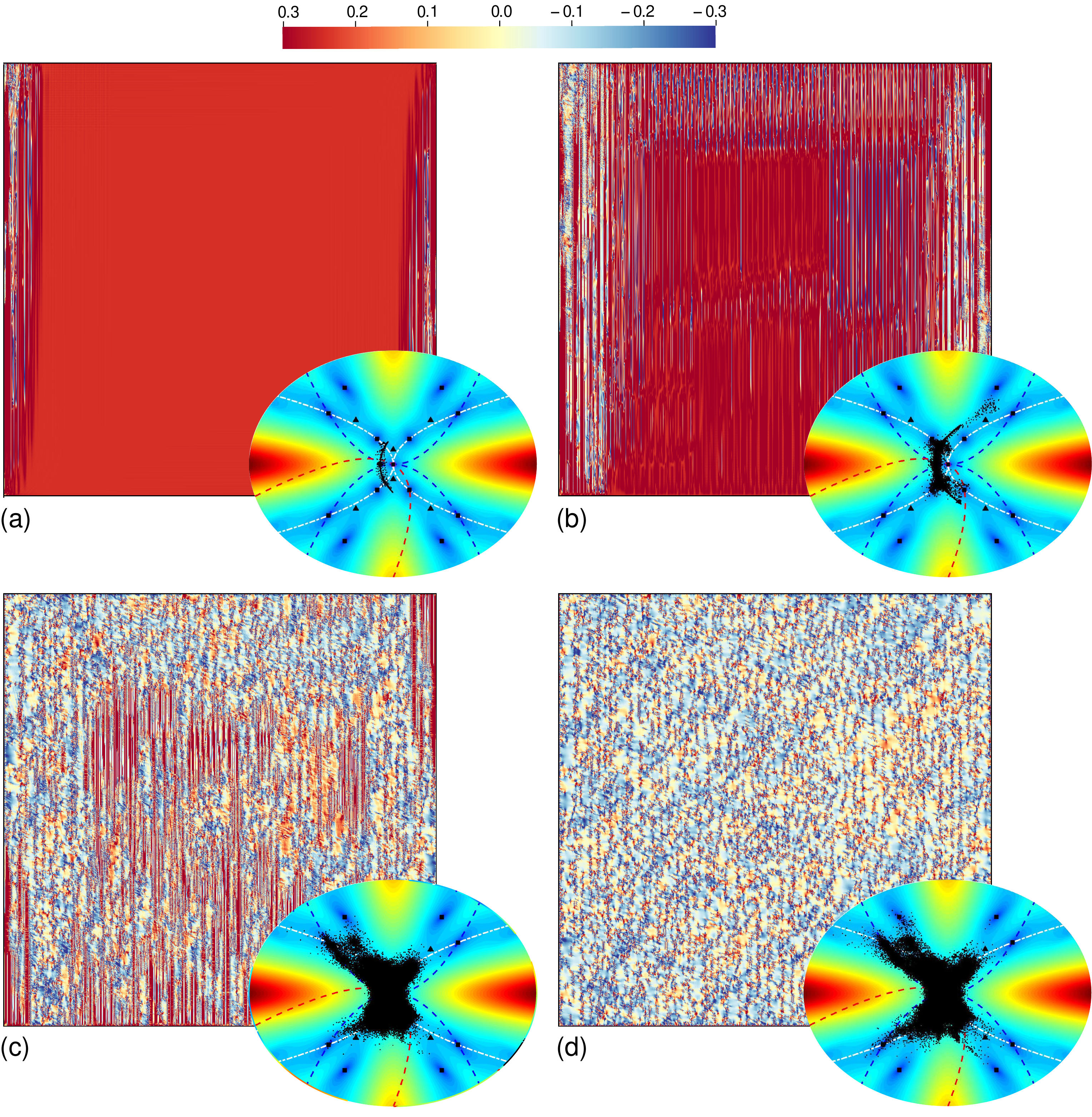}
\caption{\scriptsize {Snapshots of the stress field during discontinuous yield in pristine square crystal. Colors indicate the level of the shear component of the Cauchy stress. Insets show the same images in the configurational space.  Black dots indicate the value of the metric tensor in individual finite elements. Here $N=1024$ and $\phi=\pi/4$.  }
 \label{fig:th45}}
\end{figure}

 The most interesting yield scenario is obtained  for the non-generic $\phi=\pi/4$ loading  path   which leads   from the reference state  $S_0$  to the second closest  energy minimum $S_{11}$ after passing a much higher  energy barrier than in the case $\phi=0$. Here even more elastic energy is stored  before the  instability than in the case $\phi=\arctan( 1/2)$. The breakdown takes place at    $ \bar{\bf F}(\alpha_c^*,\pi/4)$  with a single
instability mode activated, see \ref{fig:th45} (a). However, the ensuing process of  collective dislocation nucleation quickly becomes very complex.  Due to the presence of the considerable amount of stored elastic energy and attending geometrical frustration, almost instantaneously, in addition to the original energy well $S_0$,  all   four neighboring square wells  ($S_1$, $S_{\bar1}$) and ($S_{2}$, $S_{\bar2}$)  become   engaged  and few elements even  reach more distant square  wells, see Fig.~\ref{fig:th45}(b,c).  More specifically,  
%During such instabilities a large amount of  elastic   energy is released  and the connection between the initial instability mode and the developing deformation pattern becomes quickly lost. 
the  flow  of configurational points in the space of metric tensors is so 'intense' that it  passes through   a series of (almost)  monkey saddles  corresponding to triangular lattice configurations  to  accesses  not only  the    nearest  but also  the  next nearest square  energy wells  $S_{11}$ and $S_{\bar1\bar1}$ which represent  \emph{compositions} of   shears  from both main slip systems. 
  
\begin{figure}[h!]
\centering
\includegraphics[scale=.3]{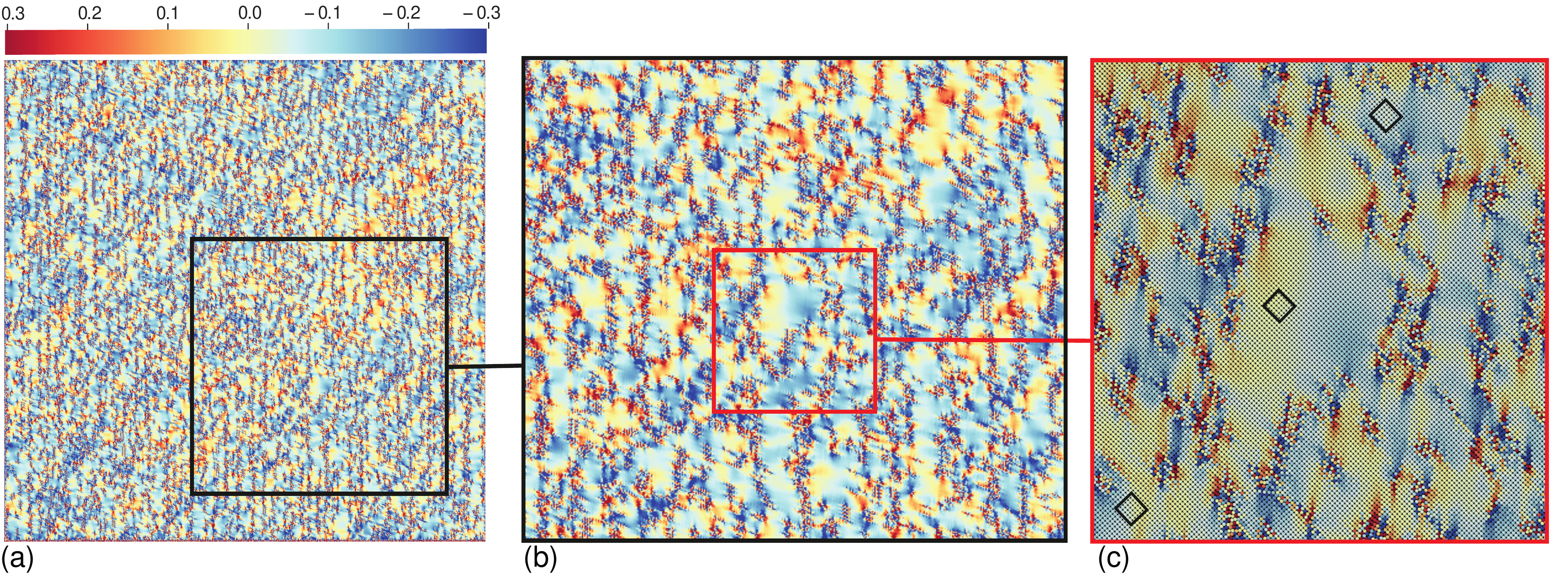}
\caption{\scriptsize {Stress field after the avalanche for a pristine square crystal with orientation  $\phi=\pi/4$  (a);  zoomed view of the marked area (b);  further zoom to the scale of the discrete finite element nodes (c). 
%We observe regions along which dislocation spacing is close to the grid size between differently oriented dislocation-poor regions.
}
 \label{fig:path45zoomed}}
\end{figure}

 The post-avalanche  dislocation pattern is then multi-slip  with minimal pile up on the boundaries and most of dislocations self-organizing in the bulk of the crystal, see Fig. \ref{fig:th45} (d).   A close-up of the highlighted  region shown in Fig.~\ref{fig:path45zoomed}(a-b) down to the scale of finite element nodes puts in evidence the formation of  dislocation walls separating almost unstressed \emph{cells} with  correlated   lattice mis-orientations. In other words, we observe the formation of dislocation-rich bands which  self screen long-range elastic fields allowing the strains  inside the dislocation-poor regions to relax by reducing the deformation gradient to pure rotation.  In Figure \ref{fig:path45zoomed}(c), we show such differently re-oriented crystal (almost) stress free regions in more detail. Interestingly, molecular dynamics simulations also suggest that at almost theoretical strength, dislocations alone can no longer release the elastic energy, and the reorientation of   crystal lattices emerges  as an important  mechanism of plastic  response~\cite{Zepeda-Ruiz2017-gu}. In other words, a perfect sub-micron crystal appears to be yielding by catastrophic strain-induced reorientation of the crystal lattice.
 
To summarize,  our  numerical experiments conducted on square lattices suggests that while in \emph{specially oriented} crystals, pristine-to-pristine brittle yield at ideal shear strength threshold is a possibility, the \emph{generically oriented},  dislocation free sub-micron crystals,  can be expected to  yield discontinuously with  massive dislocation nucleation culminating in the formation of complex dislocational patterns.  During such instabilities a large amount of stored elastic energy is released and the connection between the initial  instability  mode   and the developing deformation pattern is quickly  lost.  To minimize elastic energy,   dislocations do not only nucleate cooperatively but also self organize  hierarchically   forming multi-scale  cell structures with presumably complex statistical properties, to be studied separately using, for instance,  the approach  presented in \cite{Zhang2020-ax}.

%\begin{figure}[tbp]
%\centering
%\includegraphics[scale=.4]{./figures_new/square0nobar_10000.pdf}
%\caption{\scriptsize {Square crystal:  Here $N=1024$ and $\phi=0$. }
% \label{fig:sq0}}
%\end{figure}

%
%The orientation of these inhomogeneous bands is fully consistent with the direction of instability $ \pmb{n}^\perp$ obtained from the strong ellipticity condition. 
%
%

\subsection{Triangular lattices}

Consider next the case of higher symmetry triangular lattices. As we have already seen, in this case the macroscopic mechanical response is in overall agreement with the analytical yield surface, modulo  the fact that, differently from what we have seen in the case of square lattices, discontinuous yield  is observed  for values $\alpha_c^*$ which were  about a $10\%$ higher than the corresponding analytical predictions $\alpha_c$, see Fig. \ref{fig:smalleigen}(b). This gap, which is a  function of the finite element size $h$,   results from  the disregistry between the orientation of the unstable modes and the  `soft' directions of  the energy landscape. 

\begin{figure}[h!]
\centering
\includegraphics[scale=.35]{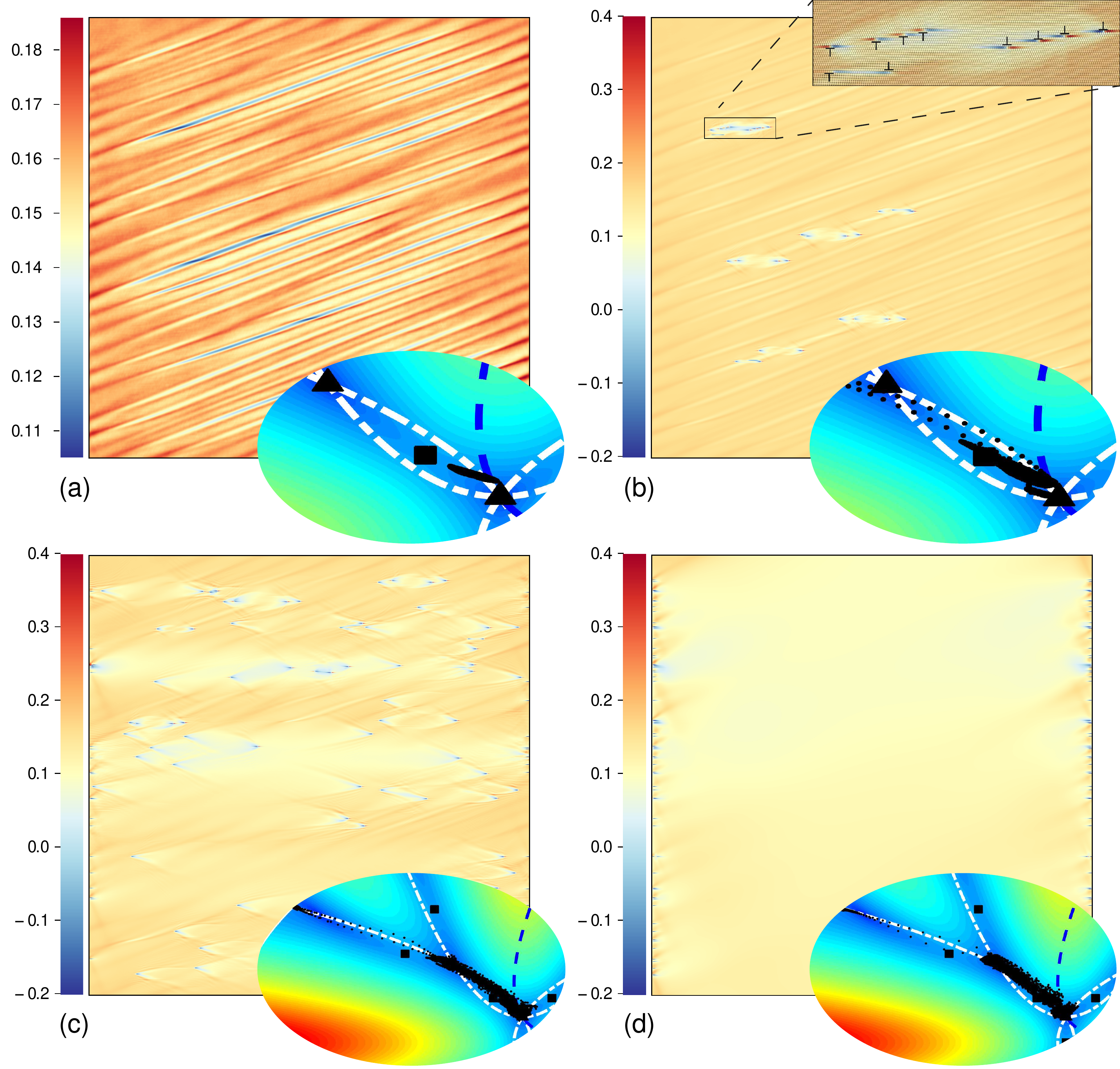}
\caption{\scriptsize {
Snapshots of the stress field during discontinuous yield in pristine triangular  crystal. Colors indicate the level of the Cauchy shear stress $\sigma_{xy}$. Insets show the same images in the configurational space.  Black dots indicate the value of the metric tensor in individual finite elements.
%Triangular crystal:  dislocation nucleation   following the loss of stability as it evolves during the different stages (a-c) of the energy minimisation. Colors indicate the level of the Cauchy stresses $\sigma_{xy}$; (d) the final mechanical equilibrium configuration.  
Here $N=1024$ and $\phi=0$.  }
 \label{fig:th0hex}}
\end{figure}

Consider first    the  loading path corresponding to a simple shear   \eqref{pathshex}  with  $\phi=0$.  An application of the idea of plastic `mechanisms'  suggests that crystal should evolve  from the initial energy well  $T_0$ to the next closest triangular well $T_{1}$; the corresponding 
 low energy valley   is shown in Fig. \ref{fig:detc1enhex}  as the path $1-1$. However, as we have already seen,  the actual  instability of the  homogeneously deformed  lattice state $T_0$ occurs strictly after  the determinant of the continuum acoustic tensor has become  negative for some orientation  $\pmb n_c$.  To deal with encountered soft directions,  the  implied dislocation  nucleation is postponed. The anticipated low energy valleys in the energy landscape cannot be  used  because the apparently low  barriers  leading towards the state $T_1$ along the path $1-1$ are still too  high. Instead the softness of the energy landscape  drives the system  in the direction of the saddle point  $S_1$ which corresponds to the square lattice while still remaining within  the fundamental domain.  Therefore in our numerical experiments   we observed that during the gradient descent type energy minimization at  $\alpha_c$  that periodically-spaced modulations develop prior to the dislocation nucleation, see Fig.~\ref{fig:th0hex}(a). The orientation of the  modulation bands   agrees with the directions $\pmb n_c$   obtained using  the continuum  instability  condition.

As the instability develops, the non-equilibrium configurations found by the minimization algorithm  show the sharpening of the   band boundaries and then the  secondary symmetry breaking  through the homogeneous  nucleation of dislocation dipoles, see Fig.~\ref{fig:th0hex}(b). Note that the 
 orientation of these dipoles agrees with the lattice ($T_0$ to  $T_{1}$ path) but not with the orientation of the bands ($T_0$ to  $S_{1}$ path).   Note also that the implied pairs of dislocations with opposite signs  nucleate at the centers  of the modulation bands   where the displacement gradients are  large  which apparently helps the system to overcome the energy barriers  leading  to  $T_{1}$ lattice state. The emerging dislocation nucleation scenario is slightly different from the one suggested by the Peierls-Nabarro model where, effectively,  the bands are assumed to be  atomically sharp  and lattice oriented.

In   Fig. \ref{fig:th0hex}(b-d)) we see that the modulation eventually completely breaks down  with only  one slip system  ending up being  activated.  Due to the misalignment of the bands with  the slip planes and the associated geometrical frustration, the nucleated dislocations  interact with each other  strongly causing  the  activation of  the double   slip  represented by the triangular lattice state   $T_{11}$. With dislocations either annihilating or escaping to the boundaries,  the resulting picture is in basic agreement with the prediction of a single slip CP theory.

\begin{figure}[h!]
\centering
\includegraphics[scale=.35]{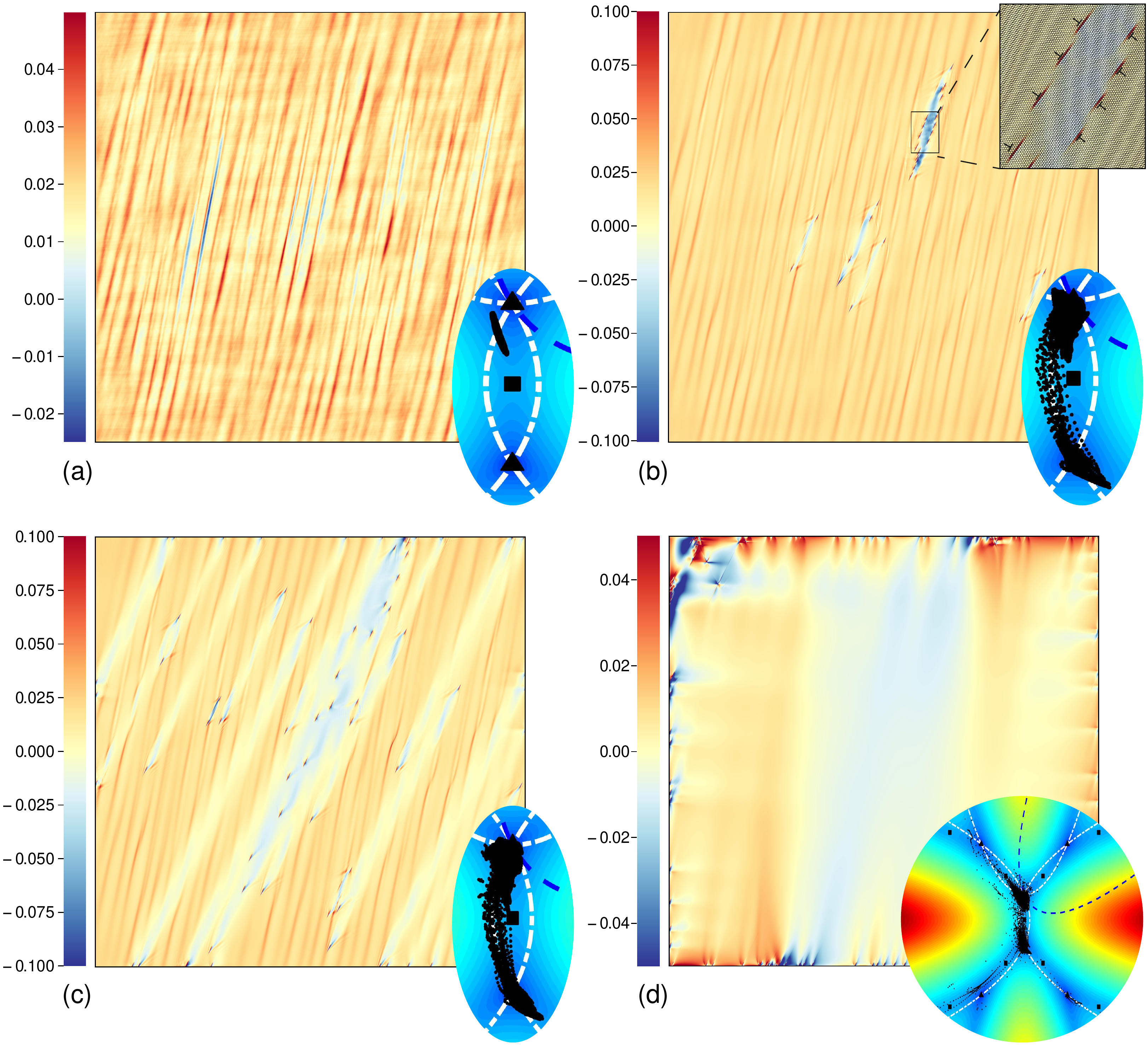}
\caption{\scriptsize {Snapshots of the stress field during discontinuous yield in pristine triangular  crystal. Colors indicate the level of the  shear component of the Cauchy stress. Insets show the same images in the configurational space.  Black dots are placed according to the value of the  metric tensor's components in individual finite elements.
%Triangular crystal: collective dislocation nucleation   following the loss of stability as it evolves during the different stages (a-c) of the energy minimisation. Colors indicate the level of the Cauchy stresses $\sigma_{xy}$.  (d) the final mechanical equilibrium configuration.  
Here $N=1024$ and $\phi=\pi/3$.  }
 \label{fig:th60hex}}
\end{figure}
 
Along  the other symmetric loading paths with $\phi =\pi/3$ and $\phi =2\pi/3$, which also describe  shears on  dense planes of the triangular lattice, the general scenario of discontinuous yield remains  the same. While the degree of misalignment of the  initial modulation bands with the lattice does not change,
the orientation of these bands relative to the square computational box is slightly different in each of these cases which (mildly) affects the outcome of the instability. The succession of events, constituting the system spanning avalanche  in the case of  loading orientation  $\phi=\pi/3$, is illustrated in  Figs.~\ref{fig:th60hex}(a-d). Here again  the macroscopic  modulations create the nucleation sites responsible for   the secondary instability which leads to the formation of almost ideal dislocational dipolar mats~\cite{Hansen1986-rs}, see  Figs. \ref{fig:th60hex}(b).  Dislocations eventually self organize into bands, see  Figs.~\ref{fig:th60hex}(c), but eventually all end up on the surface of the crystal, see  Figs.~\ref{fig:th60hex}(d).

Here again the primary, purely elastic  instability develops in the direction from triangular well $T_0$ to the square well $S_0$. The secondary instability associated with barrier crossing  leads to the flow of configurational points from  $T_0$ to  $T_{\bar 1}$ which  therefore  creates plastic slip. As dislocations pile up near the boundaries, they   activate the secondary  slip  associated with    the   well    $T_{\bar 1\bar 1}$ and  even the second slip system represented by the well $T_{\bar 1\bar 2}$.  Ultimately, all four triangular lattices $T_1$, $T_2$,  $T_{1\bar1}$ and $T_{\bar1\bar2}$ get involved, however, the ensuing complexity remains localized near the sharp corners of the sample. At the end of the avalanche the bulk of the sample appear to be free of dislocations which explains the dramatic stress drop experienced by the crystal.
\begin{figure}[h!]
\centering
\includegraphics[scale=.35]{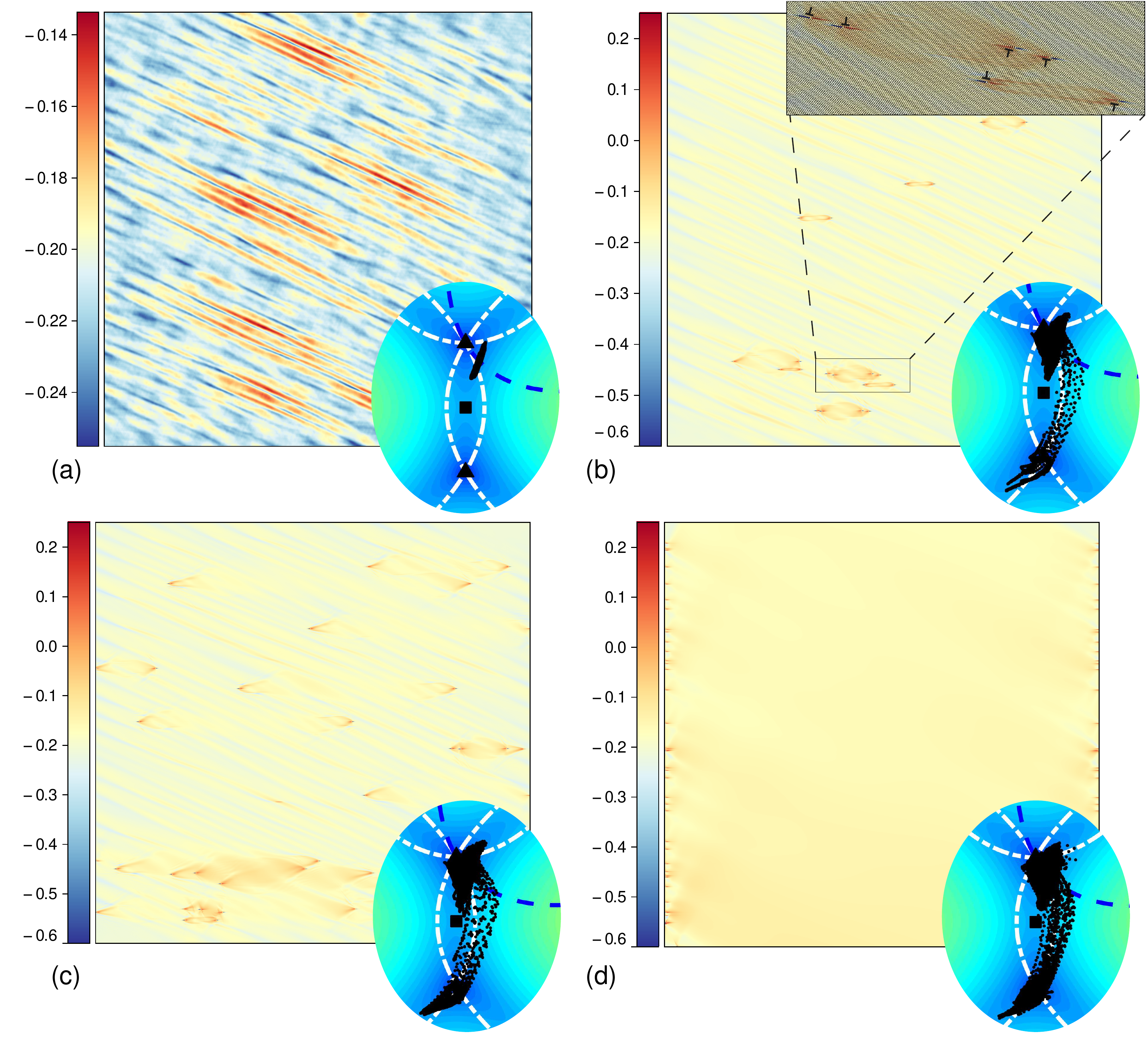}
\caption{\scriptsize {Snapshots of the stress field during discontinuous yield in pristine triangular  crystal. Colors indicate the level of the shear component of the Cauchy   stress. Insets show the same images in the configurational space.  Black dots indicate the value of the metric tensor in individual finite elements. Here $N=1024$ and $\phi=\pi/2$.  }
 \label{fig:th90hex}}
\end{figure}
In view of the high symmetry of the triangular lattice, the structure of the yielding avalanche  for the   generic path with $ \phi =\pi/2$,   which also starts  in the   original  energy well wells $T_0$,  is not very different from what we have already seen in the case of the non-generic path with $ \phi=0$. 
The influence of the neighboring square well $S_0$ is again felt in the   structure of the initial elastic modulations, see Fig.  \ref{fig:th90hex}(a).  The deformation ultimately localizes inside the low energy valley connecting the triangular wells  $T_0$ and $T_{\bar 1}$.  Transient dislocation entanglements eventually get dissociated with dislocations either annihilating or escaping to the boundaries, however,  the stored elastic energy is not fully relaxed by such limited plasticity, see Fig.  \ref{fig:th90hex}(b-d). In general, this example suggests  that in sub-micron crystals with high  symmetry and accordingly `round'   surfaces of theoretical strength,  the outcome of the discontinuous yield is decided not so much  by   the orientation of the sample in the loading machine, but rather by the orientation of the sample boundaries vis-a-vis  lattice directions.

To summarize, using MTM, we were able to show that, similar to what is observed in experiments, pristine sub-micron crystals  undergo discontinuous yielding close to theoretical strength. During the catastrophic avalanche,  massive nucleation of dislocations first transforms such crystals from defect-free to defect-saturated, but then most of the defects either annihilate or escape to the fixed boundaries producing again almost pristine crystals. The access through MTM to \emph{transient nonequilibrium states} allowed us to study for the first time the sensitive dependence of the phenomenon of discontinuous yield on crystal symmetry, crystal orientation in the loading machine, and even crystal shape.  

The general  feature  of the \emph{simulated} discontinuous yield is that  the instability   starts as  a long-wave purely elastic modulation which breaks the affinity of the original homogeneous state.  The actual dislocation nucleation emerges  as a \emph{secondary} instability,  triggered by   non-affinity and   proceeding  collectively. A large number of defects appear almost simultaneously and the  `auto-catalytic' nature of such avalanche  is due to the fact that   already nucleated dislocations trigger new  nucleation events. The process terminates  only when the elastic repulsion from the walls  finally blocks the nucleation in the bulk.

The pristine-to-pristine yield takes place if a non-generic loading allows the system to completely avoid frustration and channel almost all dislocations away from the bulk. Otherwise the avalanche \emph{jams} with dislocations forming \emph{self-locking} patterns. Such patterns can be interpreted as cell structures because relatively narrow  dislocation  walls   bound extended  domains where lattice undergoes simple \emph{rotation}.  In the case of open boundaries, the dislocation pile-up, playing an important role in this scenario, will be  replaced by the global  strain \emph{ localization}. %
\begin{figure}[h!]
\centering
\includegraphics[scale=.35]{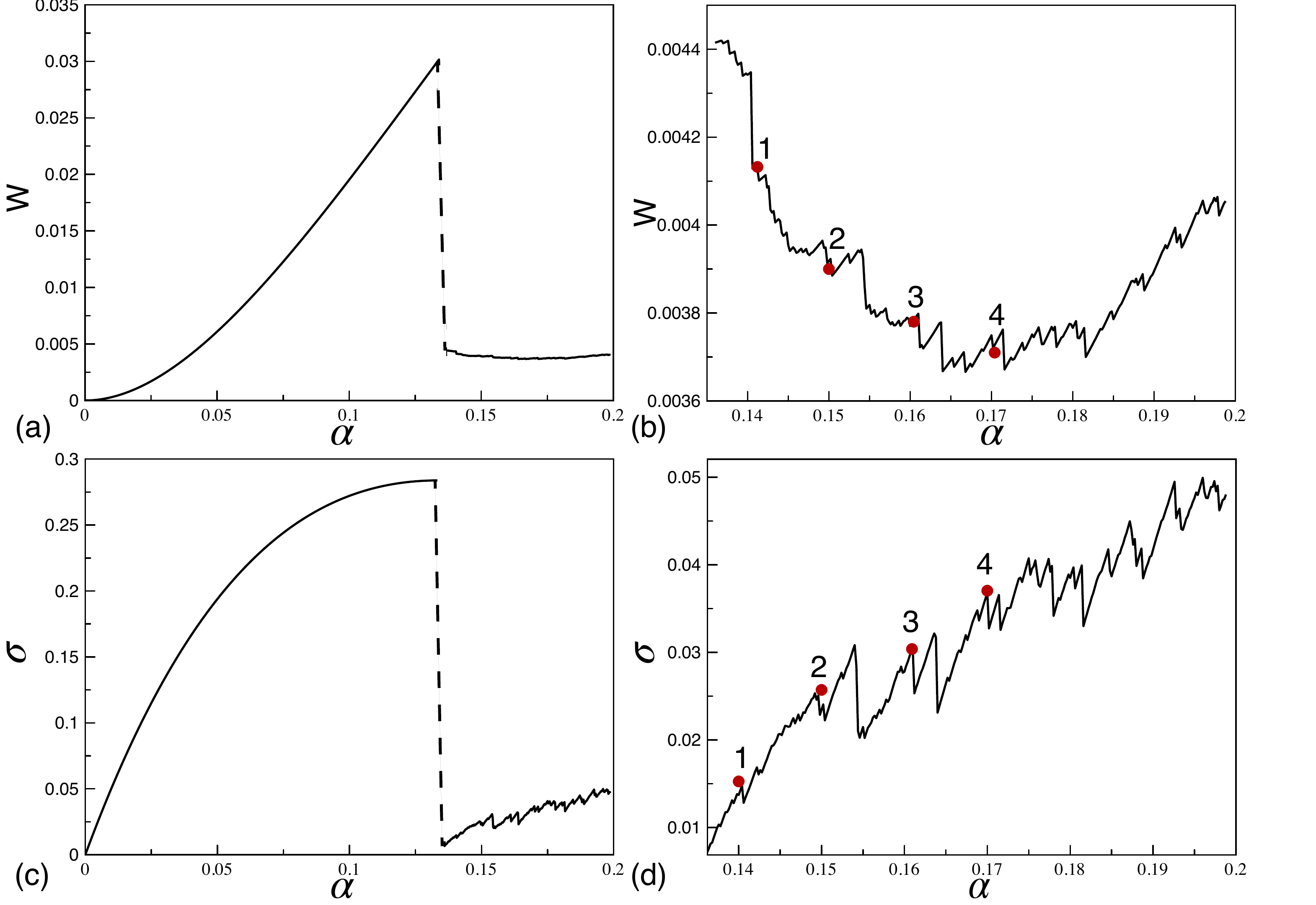}
\caption{\scriptsize{Macroscopic mechanical  response of square crystal upon loading beyond  the catastrophic system-size avalanche.  Here $N=500$ and $\phi=0$.}
\label{fig:beyond}}
\end{figure}
\section{Beyond the principal  avalanche}

The analytic and numerical results presented in the previous Sections are relevant for conceptualizing the puzzling discontinuous response of nominally ductile crystals. Such brittle-like response is generic for ultra-small and, therefore, dislocation-starved crystals in the sense that it is observed routinely in compression and nano-indentation experiments ~\cite{Corcoran1997-vt,Bei2007-vk,Chisholm2012-ki,wang2012pristine}.

As we have shown,  to ensure such pseudo-brittle behavior, the crystals should be structurally perfect (dislocation and defect-free). In this case, the main effect is an explosive system spanning avalanche at the end of the elastic regime with massive collective nucleation of dislocations, which results in a catastrophic load drop known as discontinuous yield. In this Section, we briefly discuss what happens if the loading resumes and the crystal, whose purity is now compromised, continues to yield. We limit the analysis to just a few observations, while the detailed study will be presented elsewhere.

\begin{figure}[h!]
\centering
\includegraphics[scale=.35]{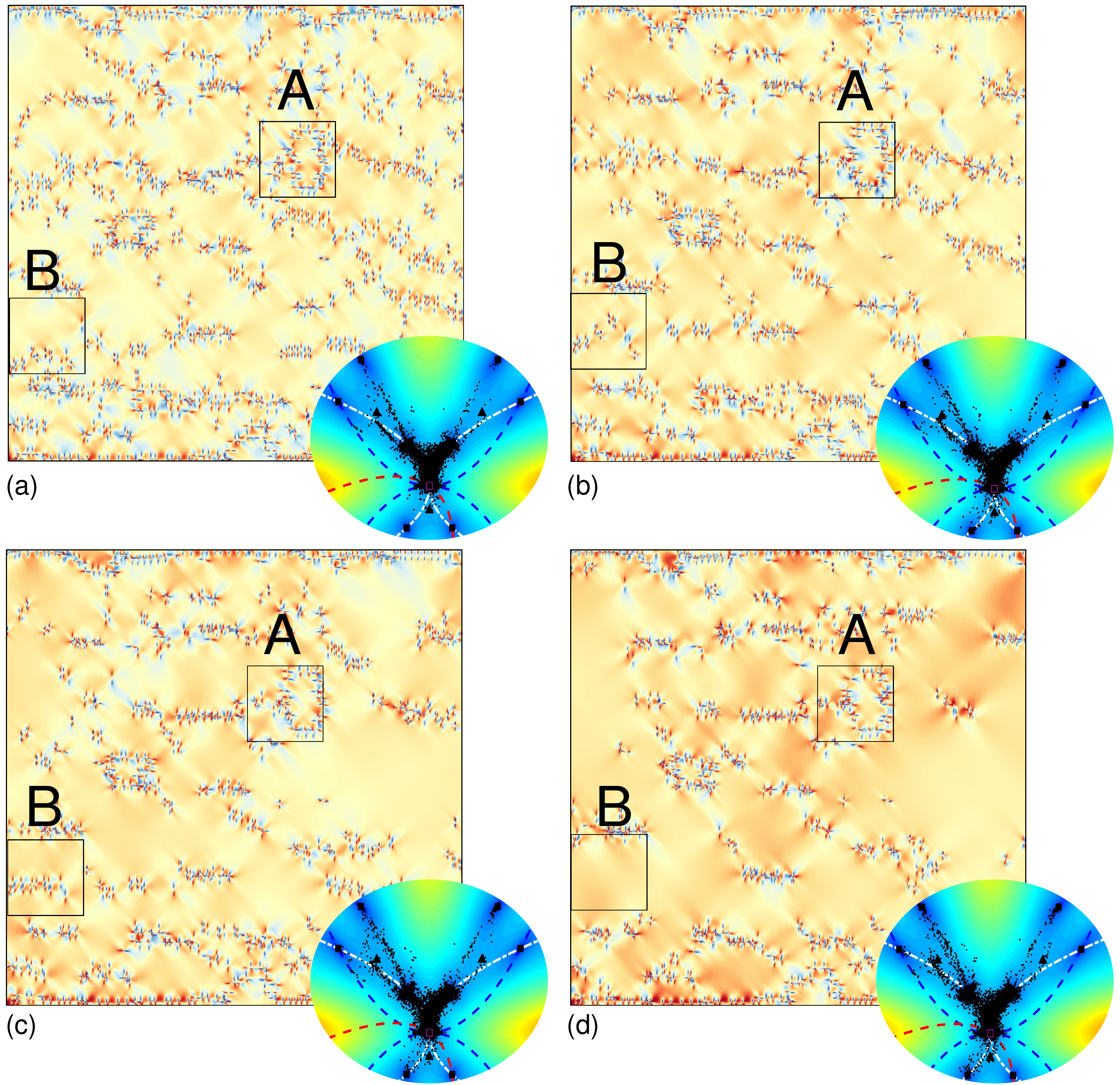}
\caption{\scriptsize{Evolution of the shear component of Cauchy stress in physical space after the catastrophic system-size avalanche. The snapshots correspond to the selected points   in Fig.~\ref{fig:beyond}  marked by numbers  $1, 2, 3, 4$. The corresponding distributions in the configurational space  are shown in the insets. Here $N=500$ and $\phi=0$.}
 \label{fig:beyond2}}
\end{figure}

As we have seen, after the termination of the system size avalanche and  the  re-stabilization of the inhomogeneous state, the  generic crystalline sample   can be expected to contain a highly correlated  configuration of lattice defects.  If the quasi-static loading continues,  such an imperfect crystal  will first undergo  another quasi-elastic deformation which will be succeeded  by another avalanche, necessarily less dramatic in view of the ubiquity  of   nucleation sources and the availability of   the  locking sites. Subsequent monotone  loading   will  be punctuated by an  intermittent succession of such avalanches,  with most of them  small,   but some   reaching again  the size of the whole system \cite{Friedman2012-ie,Zaiser2013-ff,Ispanovity2014-ra,Derlet2016-mp,Sethna2017-tm,Papanikolaou2017-ld,Cui2018-lq,Sparks2019-ie,Zhang2020-ax}.
To illustrate these general observations and show that MTM approach is well suited to study the  effects of intermittent discontinuous yield,  we now  present few illustrations of  the   mechanical response of square crystals subjected to simple shear with $\phi=0$  following directly the initial  plastic avalanche. Since, naturally,  the complexity of  the dislocational patterns  increases with subsequent loading,    we   reduce the precision of the description by  adopting  a smaller value of  $N=512$ but preserve the small \emph{annealed disorder} as in the experiment shown in Fig.  \ref{fig:disorder}. 

% 
%  As before we consider a single crystal, loaded in simple shear.  For simplicity we'll be dealing with the  simplest square symmetry and consider the simplest easy loading path with orientation  $\phi=0$ and a smaller size $N=512$.

%

The simulated macroscopic response is  summarized in  Fig.~\ref{fig:beyond}, where we show the computed  energy-strain, see  Fig.~\ref{fig:beyond}(a),  and the stress-strain, see Fig.~\ref{fig:beyond}(c),  relations. The corresponding zoom ins are shown in Fig.~\ref{fig:beyond}(b,d), respectively. 

We first observe that after the major stress/energy drop, the smooth elastic response turns into jerky plastic yield. More specifically, the post-catastrophic response can be decomposed into conservative elastic steps that are interdigitated by dissipative stress/ energy drops representing dislocation avalanches broadly distributed in size, see Fig.~\ref{fig:beyond}(b,d). While the averaged stress response in this range shows monotone hardening, the averaged energy first decreases and then increases. This is due to the fact that the energy losses cannot even be seen on such stress-strain curves, see \cite{Puglisi2005-lg} for more details. The jerkiness of the response originates in transitory elastic deformation of self-locked microstructures which always ends with partial or complete unlocking and dynamical restructuring, accompanied by energy dissipation.

\begin{figure}[h!]
\centering
\includegraphics[scale=.055]{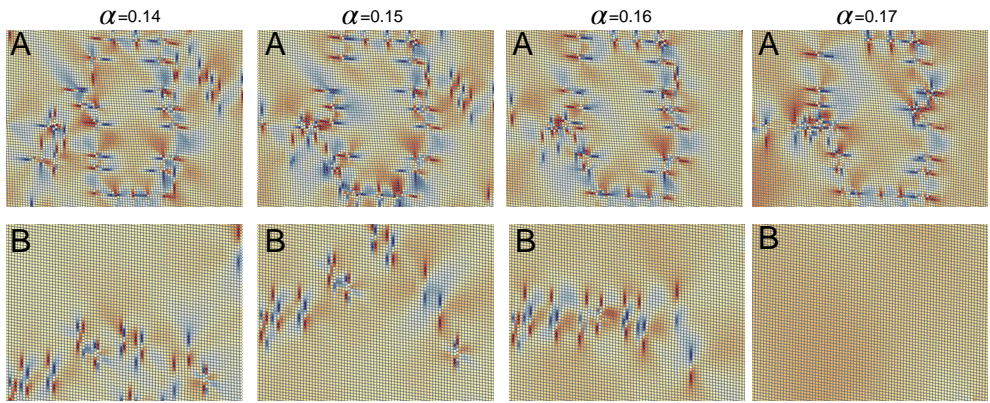}
\caption{\scriptsize {Enlarged view of the dislocation configurations inside the square regions  (A) and (B) shown in Fig.~\ref{fig:beyond2} as they evolve with  loading. Colors indicate the shear component of the Cauchy stress tensor.
}
\label{fig:zooms}}
\end{figure}

To trace such restructuring,   we show in  Fig. \ref{fig:beyond2}(a-d) a succession of four snapshots of the stress field in the arbitrary chosen moments of `time'  as   indicated in by dots  marked by numbers $1, 2, 3, 4$ in Fig. \ref{fig:beyond}(b,d). Afterwards, in Fig. \ref{fig:zooms} we zoom into two spatial regions $A$ and $B$  and follow  the `time'  evolution of  the particular  groups of dislocations over several avalanches.  The numerical experiment, illustrated by these figures,  is similar to the one   shown in Fig.  \ref{fig:sq0_2}  but with even smaller annealed initial disorder. We observe that continuing  loading is accommodated   by  additional nucleation of dislocations which remain in the bulk due to apparent repulsion from the already saturated boundaries.  The configurational points in the space of metric tensors spread more uniformly among the three equivalent square energy wells $S_0$, $S_1$ and $S_2$, being distributed almost evenly by the (almost) monkey saddle $T_0$. The ensuing  dislocation pattern includes  individual dislocation locks as well as dislocation walls separating domains where the  lattice is almost unstressed. This pattern  results from dynamic self-organization of dislocations and can be  viewed as generic.
  
%  It is obvious, that the crystal is no longer pristine, however, the stress has dropped almost to zero which means that dislocation nucleation and dislocation patterning  was extremely effective in relaxing the  considerable stress accumulated during the preceding elastic deformation.

Upon further loading, the crystal continues to  harden as seen in Fig.~\ref{fig:beyond}(d). The  plastic yield takes the form  of  an irregular sequence of stress-drops associated with partial restructuring of the dislocation pattern, see Fig. \ref{fig:beyond2}(b-d).  The configurational points continue to spread over the metric space reaching  distant wells corresponding to various compositions of simple shears. This means that despite the simple shear strain applied on the boundary, the deformation in the bulk of the sample takes the manifestly multi-slip form.   Interestingly, the   patterns identified  in Fig. \ref{fig:beyond2}(b-d) are reminiscent of the ones obtained by  x-ray diffraction method  in \cite{Jakobsen2006-ai} where the authors observed the formation of   dislocation boundaries separating (nearly) dislocation-free regions with almost perfect lattices. Such  subgrains were shown to exhibit  intermittent dynamics,   appearing and disappearing  as well as displaying transient splitting behavior.

%  The drops separate  (close to  linear) elastic stages where almost no dislocation nucleation or rearrangement  occurs. Interestingly, while the stress in average increases, the energy can both increase and decrease. 
 
%To follow the dislocation rearrangements, including both nucleation and annihilation of dislocation pairs during the intermittent stress drops,   we present  in Figs.~\ref{fig:beyond}(e-h) the evolution, in $1\%$ increments of the applied shear amplitude,  of the  Cauchy stress distribution. 

Note that most small-sized avalanches are associated with the transitional motion of depinning dislocations between the dislocation-rich wall-like patterns. During such transitions, dislocations either get locked again or annihilate, however, at least some of the existing large-scale dislocation structures persist during the loading. To illustrate this conclusion, we show in Fig.~\ref{fig:zooms} the enlarged view of the time evolution for the two rather arbitrarily selected square sub-domains, which we identify as $A$ and $B$. In $A$, we see an almost rectangular dislocation structure based on dipolar walls, which ensure an ideal screening of the long-range elastic fields~\cite{Zaiser2013-ff}. With loading, this structure deforms elastically but otherwise remains relatively stable, experiencing only minor rearrangements (addition and subtraction of dislocations from local entanglements). Instead, the incomplete dipolar wall shown in $B$ does not survive the loading, and finally, the participating dislocations disappear entirely due to an abrupt collective annihilation. Usually, such restructuring events give rise to bigger avalanches, and we checked that the largest avalanche between the deformation states (c) and (d) shown in Fig.~\ref{fig:beyond}(b,d), can be indeed linked to the disappearance of the multi-dislocation structure $B$, see Fig.~\ref{fig:zooms}.

 %see, for instance, the square  region $A$  in Figs.~\ref{fig:beyond}(e-h) and first row of Fig.~\ref{fig:zooms} for an enlarged view.

%At the same time some apparently self locked dislocation structures occasionally rearrange and may even completely disappear during plastic avalanches, see for instance the square region $B$ in Figs.~\ref{fig:beyond}(e-h) and  second row of Fig.~\ref{fig:zooms} for an enlarged view. 

We observe, however, that the  avalanches  associated with partial restructuring of the dislocation pattern,  are  still much smaller in   size than  the system spanning avalanche responsible for the major restructuring behind the original discontinuous yield of pristine crystal. This is associated with a `special preparation' of the pristine (homogeneous) crystal whose  high  degree of  degeneracy causes instability to happen simultaneously all over the sample. The dynamic randomization produced by the discontinuous yield generates annealed disorder and  reduces the degeneracy,    producing more conventional yield represented by mostly small avalanches. However, due to long-range elastic interactions, strong  correlations in such apparently random system  remain, and they lead to occasional recurrence of the system size avalanches, known as `dragon-king event' \cite{Sornette2012-kh}.  This leads to the supercritical avalanche distribution in sub-micron crystals which was observed in experiment and has been recently  successfully simulated in the framework of  a scalar version of  MTM \cite{Zhang2020-ax}.

% associated with the 'brittle-like` dislocation nucleation event occurring in pristine crystals. Therefore such discontinuous  yield event is fundamentally different from the regular yield built of many mostly small and only occasionally large avalanches. If statistically, the restructuring events are responsible for the critical,  power-law distributed avalanches, the discontinuous yield can be associated with the supercritical spinodal nucleation also known as a  All these issue will be addressed   separately.

%Indeed, this latter creates a considerable amount of 'disorder` in the form of self-locked dislocation configurations suppressing any possibility for another full-system size cooperation in this 'broken and divided' crystal characterized in our theory by the spreading of metrics among the energy wells in hyperbolic space. 
 
\section{Conclusions}

In this paper, we presented the first systematic demonstration of the \emph{effectiveness} of the mesoscopic tensorial model (MTM)  by showing that it can effectively simulate complex plastic phenomena in crystals involving a large number of interacting dislocations. A characteristic feature of the MTM is that it resolves lattice scale dislocation cores while operating with the engineering concept of stress and strain. It can be viewed as a tensorial version of the scalar phase-field model, accounting for large rotations and finite lattice invariant shears in \emph{geometrically precise} way. Rather remarkably, the MTM can deal adequately with short-range interactions of dislocations while accounting for full crystallographic symmetry, without any direct use of interatomic potentials or other ab-initio methods. 

The key feature of the MTM is that it is formulated as a nonlinear \emph{field theory} of Landau-type respecting the discrete $GL(3,\mathbb Z)$ symmetry. It can also be viewed as a geometrically and physically nonlinear anisotropic elasticity theory.  It may first sound paradoxical that  \emph{elasticity} framework was chosen to describe \emph{plasticity}. However, this viewpoint is fully consistent with the fact that dislocations are elastic defects whose interaction is largely elastic. Under quasi-static loading, inelastic \emph{dissipation} takes place during fast switching of elastic branches.  Such branch switching events are ubiquitous due to the \emph{rugged} nature of the generic MTM energy landscape and the presence of elastic driving.

% In case of quasistatic loading, such exchanges  take place instantaneously,  giving rise to rate independent dissipation~\cite{Puglisi2005-lg}. They occur unceasingly due to  the presence of lattice invariant transformations ensuring that  the  elastic energy landscape has an infinite  number of wells, arranged periodically in a way that is  compatible with tensorial $GL(3,\mathbb  Z)$ symmetry.   Furthermore, this assumption is needed to place the energy density in a continuum hyperelastic framework. 

To generate such  energy landscape, the  MTM relies on  local validity of the  \emph{Cauchy-Born} rule  even though some features of the nonlocal atomistic  description are then  inevitably lost. Behind the emergence of the continuum stresses and strains is then the  \emph{spatial averaging} over a microstructural length scale: the latter   characterizes the size of a   cluster of atoms  (an element) assumed to   deform   homogeneously.   The elastic branches are switched  when, as a result  of elastic instabilities,  such  \emph{elements}  switch between the neighboring wells of the globally periodic Landau  energy.

An important feature of the MTM approach to crystal plasticity is that  plastic `mechanisms' are not assigned a priori, but take the form of low energy valleys  connecting the  wells of Landau energy. Plastically deformed state emerges in this representation
as a \emph{mixture} of equivalent Landau phases, with dislocations appearing as \emph{incompatible }components of phase boundaries.  In such  a model slip directions are not rigidly pre-defined and loading along one slip system  can, by itself,  initiate another slip system. Topological transitions, like the ones involved in the activity of Frank-Read and  SA  (single arm) sources, are also incorporated  automatically.

In this  paper we used the simplest  2D version of the MTM   to simulate numerically  a  \emph{benchmark} phenomenon:  the catastrophic discontinuous yielding   in pristine sub-micron crystals due to  cooperative nucleation of dislocations. We used Dirichlet type  boundary conditions (hard device) to be consistent with our definition of the theoretical strength, but we also checked that switching to periodic boundary conditions   changes the results only quantitatively.
%The advantage of the  proposed method is that the description of dislocation flow is obtained in a fully coherent framework with minimal assumptions. 
We  showed that  in  simulations, the development of the collective  `brittle-like' dislocation nucleation event is similar to  the one  observed in  experiments  with imperfection-free sub-micron crystals. The use of the MTM allowed us   to  quantify, for the first time, the dependence of  the ensuing dislocation patterning  on the crystal \emph{symmetry} and on the \emph{orientation} of the crystal in the loading device.  The critical load was shown to be in agreement  with  analytical predictions.    Moreover,  the  yield criterion  based on the  Legendre-Hadamard   (strong ellipticity) condition was shown to be successful  in predicting  the  overall orientation  of the non-affine patterns.  

Vastly different dislocation  nucleation scenarios  were observed during  the discontinuous yield    in square and triangular lattices. In the case of \emph{square} symmetry,  shearing  in the `soft' directions  causes  pristine-to-pristine transitions with stress dropping almost to zero as a result of the  catastrophic avalanche. Instead, loading of the same crystals in the  `hard' directions leads to the formation of a complex, highly correlated  dislocation pattern and results in residual stress. In   higher symmetry   \emph{hexagonal} crystals,  the theoretical strength  threshold  depends only weakly on the orientation of the crystal. Due to ensuing  geometrical frustration,  the nucleation of dislocations is preceded  by  the development of long-wave elastic modulations. While most of the dislocations manage to escape  to the boundaries of the crystal, the post avalanche  stress remains  considerable independently of the sample orientation. In both square and hexagonal cases, lattice configurations  with alternative  symmetries (hexagonal and square, accordingly)  appear  as saddle points of the Landau energy and   play an important role in fomenting  \emph{multi-slip} by re-directing the flow of configurational points between converging  energy valleys.

Our study shows the \emph{potential} of the MTM to deal with plastic flows, involving \emph{strongly interacting }dislocations, while relying only on minimal phenomenological assumptions. It opens new paths in the study of spatial and temporal \emph{complexity} associated with developed plastic flows. Already the first concrete results,  obtained in this paper, represent promising steps towards harnessing brittle events in sub-micron crystals compromising forming processes, endangering the load-carrying capacity of micro-machine parts, and jeopardizing reliability in various other micromechanical applications.

\section{ Acknowledgments}
The first two authors contributed to this paper equally.  All authors acknowledge  helpful discussions with S. Conti, M. Jędrychowski, N. Greneche, J. Weiss, P. Zhang, S. Queyreau   and I. R. Ioanescu. The work was partially supported by the grants ANR-17-CE08-0047 (O. U. S. and L. T.) and ANR-19-CE08-0010-01 (O. U. S). The possibility to access  computational resources (MAGI)   provided by  Université Sorbonne Paris Nord is highly appreciated.
\bibliographystyle{crplain}

%  This inserts the bib file
%\bibliography{plasticity}

\def\bysame{\leavevmode ---------\thinspace}
\makeatletter\if@francais\providecommand{\og}{<<~}\providecommand{\fg}{~>>}
\else\gdef\og{``}\gdef\fg{''}\fi\makeatother
\def\cdrandname{\&}
\providecommand\cdrnumero{no.~}
\providecommand{\cdredsname}{eds.}
\providecommand{\cdredname}{ed.}
\providecommand{\cdrchapname}{chap.}
\providecommand{\cdrmastersthesisname}{Memoir}
\providecommand{\cdrphdthesisname}{PhD Thesis}
\begin{thebibliography}{100}

\bibitem{Stein2018-gr}
{\og Crystal Plasticity and Evolution of Polycrystalline Microstructure\fg}, in
  \emph{Encyclopedia of Computational Mechanics Second Edition} (E.~Stein,
  R.~de~Borst, T.~J.~R. Hughes, \cdredsname), CISM Courses and Lectures No.
  292, vol.~44, John Wiley \& Sons, Ltd, Chichester, UK, 2018, p.~1-23.

\bibitem{Acharya2006-nm}
A.~Acharya, A.~Roy, {\og Size effects and idealized dislocation microstructure
  at small scales: Predictions of a Phenomenological model of Mesoscopic Field
  Dislocation Mechanics: Part {I}\fg}, \emph{J. Mech. Phys. Solids} \textbf{54}
  (2006), p.~1687-1710.

\bibitem{Akinwande2017-mf}
D.~Akinwande, C.~J. Brennan, J.~S. Bunch, P.~Egberts, J.~R. Felts, H.~Gao,
  R.~Huang, J.-S. Kim, T.~Li, Y.~Li, K.~M. Liechti, N.~Lu, H.~S. Park, E.~J.
  Reed, P.~Wang, B.~I. Yakobson, T.~Zhang, Y.-W. Zhang, Y.~Zhou, Y.~Zhu, {\og A
  review on mechanics and mechanical properties of {2D} materials---Graphene
  and beyond\fg}, \emph{Extreme Mechanics Letters} \textbf{13} (2017),
  p.~42-77.

\bibitem{Anderson2017-qz}
P.~M. Anderson, J.~P. Hirth, J.~Lothe, \emph{Theory of Dislocations}, Cambridge
  University Press, 2017 (en).

\bibitem{Andrews2019-qp}
L.~C. Andrews, H.~J. Bernstein, N.~K. Sauter, {\og A space for lattice
  representation and clustering\fg}, \emph{Acta Crystallogr A Found Adv}
  \textbf{75} (2019), \cdrnumero Pt 3 (en), p.~593-599.

\bibitem{Antolovich2014-ki}
S.~D. Antolovich, R.~W. Armstrong, {\og Plastic strain localization in metals:
  origins and consequences\fg}, \emph{Prog. Mater Sci.} \textbf{59} (2014),
  p.~1-160.

\bibitem{Argon2013-fv}
A.~S. Argon, {\og Strain avalanches in plasticity\fg}, \emph{Philos. Mag.}
  \textbf{93} (2013), \cdrnumero 28-30, p.~3795-3808.

\bibitem{R_Baggio}
R.~Baggio, {\og Th{\'e}orie de la Plasticit{\'e} Cristalline Tenant Compte de
  la Sym{\'e}trie GL(2, Z)\fg}, \cdrphdthesisname, Universit{\'e} Sorbonne
  Paris Nord, 2019.

\bibitem{Baggio2019-rs}
R.~Baggio, E.~Arbib, P.~Biscari, S.~Conti, L.~Truskinovsky, G.~Zanzotto, O.~U.
  Salman, {\og {Landau-Type} Theory of Planar Crystal Plasticity\fg},
  \emph{Phys. Rev. Lett.} \textbf{123} (2019), \cdrnumero 20 (en), p.~205501.

\bibitem{Bagheripoor2018-nt}
M.~Bagheripoor, R.~Klassen, {\og Length Scale Plasticity: A Review from the
  Perspective of Dislocation Nucleation\fg}, \emph{Rev. Adv. Mater. Sci.}
  \textbf{56} (2018), \cdrnumero 1, p.~21-61.

\bibitem{Bagheripoor2020-gs}
\bysame , {\og The effect of crystal anisotropy and pre-existing defects on the
  incipient plasticity of {FCC} single crystals during nanoindentation\fg},
  \emph{Mech. Mater.} \textbf{143} (2020), p.~103311.

\bibitem{Bagheripoor2020-qm}
\bysame , {\og Effect of crystal orientation on the size effects of nano-scale
  fcc metals\fg}, \emph{Mater. Sci. Technol.} \textbf{36} (2020), \cdrnumero
  17, p.~1829-1850.

\bibitem{Bei2007-vk}
H.~Bei, S.~Shim, M.~K. Miller, G.~M. Pharr, E.~P. George, {\og Effects of
  focused ion beam milling on the nanomechanical behavior of a molybdenum-alloy
  single crystal\fg}, \emph{Appl. Phys. Lett.} \textbf{91} (2007), \cdrnumero
  11, p.~111915.

\bibitem{bei2008effects}
H.~Bei, S.~Shim, G.~M. Pharr, E.~P. George, {\og Effects of pre-strain on the
  compressive stress--strain response of Mo-alloy single-crystal
  micropillars\fg}, \emph{Acta Mater.} \textbf{56} (2008), \cdrnumero 17,
  p.~4762-4770.

\bibitem{Benzerga2009-ny}
A.~A. Benzerga, {\og Micro-pillar plasticity: 2.5D mesoscopic simulations\fg},
  \emph{J. Mech. Phys. Solids} \textbf{57} (2009), \cdrnumero 9, p.~1459-1469.

\bibitem{Bertin2019-bb}
N.~Bertin, S.~Aubry, A.~Arsenlis, W.~Cai, {\og {GPU-accelerated} dislocation
  dynamics using subcycling time-integration\fg}, \emph{Modell. Simul. Mater.
  Sci. Eng.} \textbf{27} (2019), \cdrnumero 7 (en), p.~075014.

\bibitem{Bertin2020-hv}
N.~Bertin, R.~B. Sills, W.~Cai, {\og Frontiers in the Simulation of
  Dislocations\fg}, \emph{Annu. Rev. Mater. Res.} \textbf{50} (2020),
  \cdrnumero 1, p.~437-464.

\bibitem{Bertoldi2008-au}
K.~Bertoldi, M.~C. Boyce, {\og Wave propagation and instabilities in monolithic
  and periodically structured elastomeric materials undergoing large
  deformations\fg}, \emph{Phys. Rev. B} \textbf{78} (2008), \cdrnumero 18,
  p.~184107.

\bibitem{Beyerlein2016-av}
I.~J. Beyerlein, A.~Hunter, {\og Understanding dislocation mechanics at the
  mesoscale using phase field dislocation dynamics\fg}, \emph{Philos. Trans. A
  Math. Phys. Eng. Sci.} \textbf{374} (2016), \cdrnumero 2066 (en).

\bibitem{Bhimanapati2015-bw}
G.~R. Bhimanapati, Z.~Lin, V.~Meunier, Y.~Jung, J.~Cha, S.~Das, D.~Xiao,
  Y.~Son, M.~S. Strano, V.~R. Cooper, L.~Liang, S.~G. Louie, E.~Ringe, W.~Zhou,
  S.~S. Kim, R.~R. Naik, B.~G. Sumpter, H.~Terrones, F.~Xia, Y.~Wang, J.~Zhu,
  D.~Akinwande, N.~Alem, J.~A. Schuller, R.~E. Schaak, M.~Terrones, J.~A.
  Robinson, {\og Recent Advances in {Two-Dimensional} Materials beyond
  Graphene\fg}, \emph{ACS Nano} \textbf{9} (2015), \cdrnumero 12 (en),
  p.~11509-11539.

\bibitem{Bigoni2012-by}
D.~Bigoni, \emph{Nonlinear Solid Mechanics: Bifurcation Theory and Material
  Instability}, Cambridge University Press, 2012 (en).

\bibitem{Biscari2016-qy}
P.~Biscari, M.~F. Urbano, A.~Zanzottera, G.~Zanzotto, {\og Intermittency in
  crystal plasticity informed by lattice symmetry\fg}, \emph{J. Elast.} (2016).

\bibitem{Bittencourt2019-tg}
E.~Bittencourt, {\og Interpretation of the size effects in micropillar
  compression by a strain gradient crystal plasticity theory\fg}, \emph{Int. J.
  Plast.} \textbf{116} (2019), p.~280-296.

\bibitem{Bochkanov2013-lk}
S.~Bochkanov, V.~Bystritsky, {\og Alglib\fg}, \emph{Available from: www alglib
  net} (2013).

\bibitem{Bonfanti2019-xi}
S.~Bonfanti, R.~Guerra, C.~Mondal, I.~Procaccia, S.~Zapperi, {\og Elementary
  plastic events in amorphous silica\fg}, \emph{Phys Rev E} \textbf{100}
  (2019), \cdrnumero 6-1 (en), p.~060602.

\bibitem{Bonilla2007-jk}
L.~L. Bonilla, A.~Carpio, I.~Plans, {\og Dislocations in cubic crystals
  described by discrete models\fg}, \emph{Physica A: Statistical Mechanics and
  its Applications} \textbf{376} (2007), p.~361-377.

\bibitem{Brenner1956-rx}
S.~S. Brenner, {\og Tensile Strength of Whiskers\fg}, \emph{J. Appl. Phys.}
  \textbf{27} (1956), \cdrnumero 12, p.~1484-1491.

\bibitem{Brenner1958-dr}
\bysame , {\og Growth and properties of ``whiskers''\fg}, \emph{Science}
  \textbf{128} (1958), \cdrnumero 3324, p.~569-575.

\bibitem{bulatov1994stochastic}
V.~Bulatov, A.~Argon, {\og A stochastic model for continuum elasto-plastic
  behavior. I. Numerical approach and strain localization\fg}, \emph{Modelling
  and Simulation in Materials Science and Engineering} \textbf{2} (1994),
  \cdrnumero 2, p.~167.

\bibitem{Cai2006-fe}
W.~Cai, A.~Arsenlis, C.~R. Weinberger, V.~V. Bulatov, {\og A non-singular
  continuum theory of dislocations\fg}, \emph{J. Mech. Phys. Solids}
  \textbf{54} (2006), \cdrnumero 3, p.~561-587.

\bibitem{Carpio2003-yu}
A.~Carpio, L.~L. Bonilla, {\og Edge dislocations in crystal structures
  considered as traveling waves in discrete models\fg}, \emph{Phys. Rev. Lett.}
  \textbf{90} (2003), \cdrnumero 13 (en), p.~135502.

\bibitem{Carpio2005-fl}
\bysame , {\og Discrete models of dislocations and their motion in cubic
  crystals\fg}, \emph{Phys. Rev. B Condens. Matter} \textbf{71} (2005),
  \cdrnumero 13, p.~134105.

\bibitem{Cayron2019-vs}
C.~Cayron, {\og The transformation matrices (distortion, orientation,
  correspondence), their continuous forms and their variants\fg}, \emph{Acta
  Crystallogr A Found Adv} \textbf{75} (2019), \cdrnumero Pt 3 (en),
  p.~411-437.

\bibitem{Cazacu2013-no}
O.~Cazacu, \emph{Multiscale Modeling of Heterogenous Materials: From
  Microstructure to {Macro-Scale} Properties}, John Wiley \& Sons, 2013 (en).

\bibitem{Chan2010-ha}
P.~Y. Chan, G.~Tsekenis, J.~Dantzig, K.~A. Dahmen, N.~Goldenfeld, {\og
  Plasticity and dislocation dynamics in a phase field crystal model\fg},
  \emph{Phys. Rev. Lett.} \textbf{105} (2010), \cdrnumero 1, p.~015502.

\bibitem{Chen2016-bd}
J.~Chen, G.~Schusteritsch, C.~J. Pickard, C.~G. Salzmann, A.~Michaelides, {\og
  Two Dimensional Ice from First Principles: Structures and Phase
  Transitions\fg}, \emph{Phys. Rev. Lett.} \textbf{116} (2016), \cdrnumero 2
  (en), p.~025501.

\bibitem{Chen2002-cn}
L.-Q. Chen, {\og Phase-field models for microstructure evolution\fg},
  \emph{Annu. Rev. Mater. Res.} \textbf{32} (2002), \cdrnumero 1, p.~113-140.

\bibitem{Chen2018-gt}
Y.~Chen, Z.~Fan, Z.~Zhang, W.~Niu, C.~Li, N.~Yang, B.~Chen, H.~Zhang, {\og
  {Two-Dimensional} Metal Nanomaterials: Synthesis, Properties, and
  Applications\fg}, \emph{Chem. Rev.} \textbf{118} (2018), \cdrnumero 13 (en),
  p.~6409-6455.

\bibitem{Chen2010-sq}
Y.~S. Chen, W.~Choi, S.~Papanikolaou, J.~P. Sethna, {\og Bending crystals:
  emergence of fractal dislocation structures\fg}, \emph{Phys. Rev. Lett.}
  \textbf{105} (2010), \cdrnumero 10 (en), p.~105501.

\bibitem{Chisholm2012-ki}
C.~Chisholm, H.~Bei, M.~B. Lowry, J.~Oh, S.~A. Syed~Asif, O.~L. Warren, Z.~W.
  Shan, E.~P. George, A.~M. Minor, {\og Dislocation starvation and exhaustion
  hardening in Mo alloy nanofibers\fg}, \emph{Acta Mater.} \textbf{60} (2012),
  \cdrnumero 5, p.~2258-2264.

\bibitem{chrobak2011deconfinement}
D.~Chrobak, N.~Tymiak, A.~Beaber, O.~Ugurlu, W.~W. Gerberich, R.~Nowak, {\og
  Deconfinement leads to changes in the nanoscale plasticity of silicon\fg},
  \emph{Nat. Nanotechnol.} \textbf{6} (2011), \cdrnumero 8 (en), p.~480-484.

\bibitem{Cia2012-pj}
W.~Cia, J.~Li, S.~Yip, {\og Comprehensive Nuclear Materials, Vol. by {RJM}
  Konings\fg}, 2012.

\bibitem{Clouet2015-zk}
E.~Clouet, D.~Caillard, N.~Chaari, F.~Onimus, D.~Rodney, {\og Dislocation
  locking versus easy glide in titanium and zirconium\fg}, \emph{Nat. Mater.}
  \textbf{14} (2015), \cdrnumero 9 (en), p.~931-936.

\bibitem{Conti2004-sv}
S.~Conti, G.~Zanzotto, {\og A Variational Model for Reconstructive Phase
  Transformations in Crystals, and their Relation to Dislocations and
  Plasticity\fg}, \emph{Arch. Ration. Mech. Anal.} \textbf{173} (2004),
  \cdrnumero 1, p.~69-88.

\bibitem{Corcoran1997-vt}
S.~G. Corcoran, R.~J. Colton, E.~T. Lilleodden, W.~W. Gerberich, {\og Anomalous
  plastic deformation at surfaces: Nanoindentation of gold single crystals\fg},
  \emph{Phys. Rev. B Condens. Matter} \textbf{55} (1997), \cdrnumero 24,
  p.~R16057-R16060.

\bibitem{Cottrell2002-dh}
A.~H. Cottrell, {\og Commentary. A brief view of work hardening\fg}, in
  \emph{Dislocations in Solids} (F.~R.~N. Nabarro, M.~S. Duesbery,
  \cdredsname), vol.~11, Elsevier, 2002, p.~vii-xvii.

\bibitem{Csikor2007-ua}
F.~F. Csikor, C.~Motz, D.~Weygand, M.~Zaiser, S.~Zapperi, {\og Dislocation
  Avalanches, Strain Bursts, and the Problem of Plastic Forming at the
  Micrometer Scale\fg}, \emph{Science} \textbf{318} (2007), \cdrnumero 5848,
  p.~251-254.

\bibitem{Cui2018-lq}
Y.~Cui, N.~Ghoniem, {\og Spatio-temporal plastic instabilities at the
  nano/micro scale\fg}, \emph{J. Micromech. Mol. Phys.} \textbf{03} (2018),
  \cdrnumero 03n04, p.~1840006.

\bibitem{Cui2017-xn}
Y.~Cui, G.~Po, N.~Ghoniem, {\og Influence of loading control on strain bursts
  and dislocation avalanches at the nanometer and micrometer scale\fg},
  \emph{Phys. Rev. B} \textbf{95} (2017), \cdrnumero 6, p.~064103.

\bibitem{Dasgupta2012-uj}
R.~Dasgupta, S.~Karmakar, I.~Procaccia, {\og Universality of the plastic
  instability in strained amorphous solids\fg}, \emph{Phys. Rev. Lett.}
  \textbf{108} (2012), \cdrnumero 7 (en), p.~075701.

\bibitem{Derlet2016-mp}
P.~M. Derlet, R.~Maa{\ss}, {\og The stress statistics of the first pop-in or
  discrete plastic event in crystal plasticity\fg}, \emph{J. Appl. Phys.}
  \textbf{120} (2016), \cdrnumero 22, p.~225101.

\bibitem{Devincre1992-mu}
B.~Devincre, V.~Pontikis, Y.~Brechet, G.~Canova, M.~Condat, L.~Kubin, {\og
  {Three-Dimensional} Simulations of Plastic Flow in Crystals\fg}, in
  \emph{Microscopic Simulations of Complex Hydrodynamic Phenomena}
  (M.~Mareschal, B.~L. Holian, \cdredsname), Springer US, Boston, MA, 1992,
  p.~413-423.

\bibitem{Differt1986-qi}
K.~Differt, U.~Esmann, H.~Mughrabi, {\og A model of extrusions and intrusions
  in fatigued metals {II}. Surface roughening by random irreversible slip\fg},
  \emph{Philos. Mag. A} \textbf{54} (1986), \cdrnumero 2, p.~237-258.

\bibitem{Dimiduk2006-fz}
D.~M. Dimiduk, C.~Woodward, R.~Lesar, M.~D. Uchic, {\og Scale-free intermittent
  flow in crystal plasticity\fg}, \emph{Science} \textbf{312} (2006),
  \cdrnumero 5777 (en), p.~1188-1190.

\bibitem{Dmitrieva2010-qn}
O.~Dmitrieva, J.~V. Svirina, E.~Demir, D.~Raabe, {\og Investigation of the
  internal substructure of microbands in a deformed copper single crystal:
  experiments and dislocation dynamics simulation\fg}, \emph{Modell. Simul.
  Mater. Sci. Eng.} \textbf{18} (2010), \cdrnumero 8 (en), p.~085011.

\bibitem{Dobson2007-ax}
M.~Dobson, R.~S. Elliott, M.~Luskin, E.~B. Tadmor, {\og A multilattice
  quasicontinuum for phase transforming materials: Cascading Cauchy Born
  kinematics\fg}, \emph{Journal of Computer-Aided Materials Design} \textbf{14}
  (2007), p.~219.

\bibitem{El-Awady2016-ea}
J.~A. El-Awady, H.~Fan, A.~M. Hussein, {\og Advances in Discrete Dislocation
  Dynamics Modeling of {Size-Affected} Plasticity\fg}, in \emph{Multiscale
  Materials Modeling for Nanomechanics}, vol. 245, 2016, p.~337.

\bibitem{El-Azab2020-fu}
A.~El-Azab, G.~Po, {\og Continuum Dislocation Dynamics: Classical Theory and
  Contemporary Models\fg}, in \emph{Handbook of Materials Modeling: Methods:
  Theory and Modeling} (W.~Andreoni, S.~Yip, \cdredsname), Springer
  International Publishing, Cham, 2020, p.~1583-1607.

\bibitem{Elder2002-qt}
K.~R. Elder, M.~Katakowski, M.~Haataja, M.~Grant, {\og Modeling elasticity in
  crystal growth\fg}, \emph{Phys. Rev. Lett.} \textbf{88} (2002), \cdrnumero 24
  (en), p.~245701.

\bibitem{Engel1986-tm}
P.~Engel, \emph{Geometric Crystallography: An Axiomatic Introduction to
  Crystallography}, Springer Netherlands, 1986.

\bibitem{Ericksen1970-rx}
J.~L. Ericksen, {\og Nonlinear elasticity of diatomic crystals\fg}, \emph{Int.
  J. Solids Struct.} \textbf{6} (1970), \cdrnumero 7, p.~951-957.

\bibitem{Ericksen1973-yt}
\bysame , {\og Loading Devices and Stability of Equilibrium\fg}, in
  \emph{Nonlinear Elasticity} (R.~W. Dickey, \cdredname), Academic Press, 1973,
  p.~161-173.

\bibitem{Ericksen1977-pj}
\bysame , {\og Special Topics in {Elastostatics††The} research work herein
  reported was supported by a grant from the National Science Foundation\fg},
  in \emph{Advances in Applied Mechanics} (C.-S. Yih, \cdredname), vol.~17,
  Elsevier, 1977, p.~189-244.

\bibitem{Ericksen1980-km}
\bysame , {\og Some phase transitions in crystals\fg}, \emph{Arch. Ration.
  Mech. Anal.} \textbf{73} (1980), \cdrnumero 2, p.~99-124.

\bibitem{Ericksen2008-kx}
\bysame , {\og On the {Cauchy---Born} Rule\fg}, \emph{Math. Mech. Solids}
  \textbf{13} (2008), \cdrnumero 3-4, p.~199-220.

\bibitem{Finel2010-rd}
A.~Finel, Y.~Le~Bouar, A.~Gaubert, U.~Salman, {\og Phase field methods:
  Microstructures, mechanical properties and complexity\fg}, \emph{C. R. Phys.}
  (2010).

\bibitem{Finel2000-um}
A.~Finel, D.~Rodney, {\og Phase field methods and dislocations\fg}, \emph{MRS
  Fall Meeting, Boston, MA} \textbf{652} (2000), \cdrnumero Y4.9.

\bibitem{itensor}
M.~Fishman, S.~R. White, E.~M. Stoudenmire, {\og The \mbox{ITensor} Software
  Library for Tensor Network Calculations\fg}, 2020,
  \url{https://arxiv.org/abs/2007.14822}.

\bibitem{Folkins1991-em}
I.~Folkins, {\og Functions of two‐dimensional Bravais lattices\fg}, \emph{J.
  Math. Phys.} \textbf{32} (1991), \cdrnumero 7, p.~1965-1969.

\bibitem{Fonseca1987-pd}
I.~Fonseca, {\og Variational methods for elastic crystals\fg}, \emph{Arch.
  Ration. Mech. Anal.} \textbf{97} (1987), \cdrnumero 3, p.~189-220.

\bibitem{Forest2019-au}
S.~Forest, K.~Ammar, B.~Appolaire, V.~d. Rancourt, S.~Wulfinghoff, {\og
  Generalized Continua and {Phase-Field} Models: Application to Crystal
  Plasticity\fg}, in \emph{Mesoscale Models: From {Micro-Physics} to
  {Macro-Interpretation}} (S.~Mesarovic, S.~Forest, H.~Zbib, \cdredsname),
  Springer International Publishing, Cham, 2019, p.~299-344.

\bibitem{Forest2019}
S.~Forest, J.~R. Mayeur, D.~L. McDowell, {\og Micromorphic Crystal
  Plasticity\fg}, p.~643-686, Springer International Publishing, Cham, 2019,
  \url{https://doi.org/10.1007/978-3-319-58729-5_49}.

\bibitem{Franciosi1985-cu}
P.~Franciosi, {\og The concepts of latent hardening and strain hardening in
  metallic single crystals\fg}, \emph{Acta Metall.} \textbf{33} (1985),
  \cdrnumero 9, p.~1601-1612.

\bibitem{Frenkel1939-tk}
J.~Frenkel, T.~Kontorova, {\og On the theory of plastic deformation and
  twinning\fg}, \emph{Izv. Akad. Nauk, Ser. Fiz.} \textbf{1} (1939),
  p.~137-149.

\bibitem{Friedman2012-ie}
N.~Friedman, A.~T. Jennings, G.~Tsekenis, J.-Y. Kim, M.~Tao, J.~T. Uhl, J.~R.
  Greer, K.~A. Dahmen, {\og Statistics of dislocation slip avalanches in
  nanosized single crystals show tuned critical behavior predicted by a simple
  mean field model\fg}, \emph{Phys. Rev. Lett.} \textbf{109} (2012), \cdrnumero
  9 (en), p.~095507.

\bibitem{Friesecke2002-hb}
{Friesecke}, {Theil}, {\og Validity and Failure of the {Cauchy-Born} Hypothesis
  in a {Two-Dimensional} {Mass-Spring} Lattice\fg}, \emph{J. Nonlinear Sci.}
  \textbf{12} (2002), \cdrnumero 5, p.~445-478.

\bibitem{Gao2020-qy}
Y.~Gao, {\og A Cayley graph description of the symmetry breaking associated
  with deformation and structural phase transitions in metallic materials\fg},
  \emph{Materialia} \textbf{9} (2020), p.~100588.

\bibitem{Gao2020-zy}
Y.~Gao, T.~Yu, Y.~Wang, {\og Phase Transformation Graph and Transformation
  Pathway Engineering for Shape Memory Alloys\fg}, \emph{Shape Memory and
  Superelasticity} \textbf{6} (2020), \cdrnumero 1, p.~115-130.

\bibitem{Gao2020-bx}
Y.~Gao, Y.~Zhang, Y.~Wang, {\og Determination of twinning path from broken
  symmetry: A revisit to deformation twinning in bcc metals\fg}, \emph{Acta
  Mater.} \textbf{196} (2020), p.~280-294.

\bibitem{Garg2016-kz}
A.~Garg, C.~E. Maloney, {\og Universal scaling laws for homogeneous dislocation
  nucleation during nano-indentation\fg}, \emph{J. Mech. Phys. Solids}
  \textbf{95} (2016), p.~742-754.

\bibitem{Geslin2014-ad}
P.-A. Geslin, B.~Appolaire, A.~Finel, {\og Investigation of coherency loss by
  prismatic punching with a nonlinear elastic model\fg}, \emph{Acta Mater.}
  \textbf{71} (2014), p.~80-88.

\bibitem{Geslin2017-fb}
P.-A. Geslin, R.~Gatti, B.~Devincre, D.~Rodney, {\og Implementation of the
  nudged elastic band method in a dislocation dynamics formalism: Application
  to dislocation nucleation\fg}, \emph{J. Mech. Phys. Solids} \textbf{108}
  (2017), p.~49-67.

\bibitem{Van_der_Giessen2020-gl}
E.~van~der Giessen, P.~A. Schultz, N.~Bertin, V.~V. Bulatov, W.~Cai,
  G.~Cs{\'a}nyi, S.~M. Foiles, M.~G.~D. Geers, C.~Gonz{\'a}lez, M.~H{\"u}tter,
  {Others}, {\og Roadmap on multiscale materials modeling\fg}, \emph{Modell.
  Simul. Mater. Sci. Eng.} \textbf{28} (2020), \cdrnumero 4, p.~043001.

\bibitem{Gomez-Garcia2006-ur}
D.~G{\'o}mez-Garc{\'\i}a, B.~Devincre, L.~P. Kubin, {\og Dislocation Patterns
  and the Similitude Principle: 2.5D Mesoscale Simulations\fg}, \emph{Phys.
  Rev. Lett.} \textbf{96} (2006), \cdrnumero 12.

\bibitem{Grabovsky2014-fb}
Y.~Grabovsky, L.~Truskinovsky, {\og Normality Condition in Elasticity\fg},
  \emph{J. Nonlinear Sci.} \textbf{24} (2014), \cdrnumero 6, p.~1125-1146.

\bibitem{Greer2011-av}
J.~R. Greer, J.~T.~M. De~Hosson, {\og Plasticity in small-sized metallic
  systems: Intrinsic versus extrinsic size effect\fg}, \emph{Prog. Mater Sci.}
  \textbf{56} (2011), \cdrnumero 6, p.~654-724.

\bibitem{Greer2005-ak}
J.~R. Greer, W.~C. Oliver, W.~D. Nix, {\og Size dependence of mechanical
  properties of gold at the micron scale in the absence of strain
  gradients\fg}, \emph{Acta Mater.} \textbf{53} (2005), \cdrnumero 6,
  p.~1821-1830.

\bibitem{Grosse-Kunstleve2004-oi}
R.~W. Grosse-Kunstleve, N.~K. Sauter, P.~D. Adams, {\og Numerically stable
  algorithms for the computation of reduced unit cells\fg}, \emph{Acta
  Crystallogr. A} \textbf{60} (2004), \cdrnumero Pt 1 (en), p.~1-6.

\bibitem{Gurtin2009-bl}
M.~E. Gurtin, L.~Anand, {\og Thermodynamics applied to gradient theories
  involving the accumulated plastic strain: The theories of Aifantis and Fleck
  and Hutchinson and their generalization\fg}, \emph{J. Mech. Phys. Solids}
  \textbf{57} (2009), \cdrnumero 3, p.~405-421.

\bibitem{Han2005-uh}
C.~S. Han, H.~Gao, Y.~Huang, W.~D. Nix, {\og Mechanism-based strain gradient
  crystal plasticity - I. Theory\fg}, \emph{J. Mech. Phys. Solids} \textbf{53}
  (2005), \cdrnumero 5, p.~1188-1203.

\bibitem{Han2015-db}
W.-Z. Han, L.~Huang, S.~Ogata, H.~Kimizuka, Z.-C. Yang, C.~Weinberger, Q.-J.
  Li, B.-Y. Liu, X.-X. Zhang, J.~Li, E.~Ma, Z.-W. Shan, {\og From ``Smaller is
  Stronger'' to ``{Size-Independent} Strength Plateau'': Towards Measuring the
  Ideal Strength of Iron\fg}, \emph{Adv. Mater.} \textbf{27} (2015), \cdrnumero
  22 (en), p.~3385-3390.

\bibitem{Hansen1986-rs}
N.~Hansen, D.~Kuhlmann-Wilsdorf, {\og Low energy dislocation structures due to
  unidirectional deformation at low temperatures\fg}, \emph{Int. J. Green
  Nanotech. Materials Sci. Eng.} \textbf{81} (1986), p.~141-161.

\bibitem{He2016-bi}
Y.~He, L.~Zhong, F.~Fan, C.~Wang, T.~Zhu, S.~X. Mao, {\og In situ observation
  of shear-driven amorphization in silicon crystals\fg}, \emph{Nat.
  Nanotechnol.} \textbf{11} (2016), \cdrnumero 10 (en), p.~866-871.

\bibitem{Hill1962-fb}
R.~Hill, {\og Acceleration waves in solids\fg}, \emph{J. Mech. Phys. Solids}
  \textbf{10} (1962), \cdrnumero 1, p.~1-16.

\bibitem{Hoang2016-hm}
V.~V. Hoang, N.~T. Hieu, {\og Formation of {Two-Dimensional} Crystals with
  Square Lattice Structure from the Liquid State\fg}, \emph{J. Phys. Chem. C}
  \textbf{120} (2016), \cdrnumero 32, p.~18340-18347.

\bibitem{Hochrainer2014-sk}
T.~Hochrainer, S.~Sandfeld, M.~Zaiser, P.~Gumbsch, {\og Continuum dislocation
  dynamics: Towards a physical theory of crystal plasticity\fg}, \emph{J. Mech.
  Phys. Solids} \textbf{63} (2014), p.~167-178.

\bibitem{Hu2001-cp}
S.~Y. Hu, L.~Q. Chen, {\og Solute segregation and coherent nucleation and
  growth near a dislocation---a phase-field model integrating defect and phase
  microstructures\fg}, \emph{Acta Mater.} \textbf{49} (2001), \cdrnumero 3,
  p.~463-472.

\bibitem{Hu2018-yn}
Y.~Hu, L.~Shu, Q.~Yang, W.~Guo, P.~K. Liaw, K.~A. Dahmen, J.-M. Zuo, {\og
  Dislocation avalanche mechanism in slowly compressed high entropy alloy
  nanopillars\fg}, \emph{Communications Physics} \textbf{1} (2018), \cdrnumero
  1 (en), p.~1-8.

\bibitem{Hunter2011-vp}
A.~Hunter, I.~J. Beyerlein, T.~C. Germann, M.~Koslowski, {\og Influence of the
  stacking fault energy surface on partial dislocations in fcc metals with a
  three-dimensional phase field dislocations dynamics model\fg}, \emph{Phys.
  Rev. B Condens. Matter} \textbf{84} (2011), \cdrnumero 14, p.~144108.

\bibitem{Ispanovity2014-ra}
P.~D. Isp{\'a}novity, L.~Laurson, M.~Zaiser, I.~Groma, S.~Zapperi, M.~J. Alava,
  {\og Avalanches in {2D} dislocation systems: plastic yielding is not
  depinning\fg}, \emph{Phys. Rev. Lett.} \textbf{112} (2014), \cdrnumero 23
  (en), p.~235501.

\bibitem{Issa2015-mq}
I.~Issa, J.~Amodeo, J.~R{\'e}thor{\'e}, L.~Joly-Pottuz, C.~Esnouf,
  J.~Morthomas, M.~Perez, J.~Chevalier, K.~Masenelli-Varlot, {\og In situ
  investigation of {MgO} nanocube deformation at room temperature\fg},
  \emph{Acta Mater.} \textbf{86} (2015), p.~295-304.

\bibitem{Jakobsen2006-ai}
B.~Jakobsen, H.~F. Poulsen, U.~Lienert, J.~Almer, S.~D. Shastri, H.~O.
  S{\o}rensen, C.~Gundlach, W.~Pantleon, {\og Formation and subdivision of
  deformation structures during plastic deformation\fg}, \emph{Science}
  \textbf{312} (2006), \cdrnumero 5775 (en), p.~889-892.

\bibitem{Javanbakht2016-dr}
M.~Javanbakht, V.~I. Levitas, {\og Phase field approach to dislocation
  evolution at large strains: Computational aspects\fg}, \emph{Int. J. Solids
  Struct.} \textbf{82} (2016), p.~95-110.

\bibitem{Jin2018-ty}
J.~Jin, P.~Yang, J.~Cao, S.~Li, Q.~Peng, {\og Quasicontinuum Simulation of the
  Effect of {Lotus-Type} Nanocavity on the Onset Plasticity of Single Crystal
  Al during Nanoindentation\fg}, \emph{Nanomaterials (Basel)} \textbf{8}
  (2018), \cdrnumero 10 (en).

\bibitem{Jin2001-fw}
Y.~M. Jin, A.~G. Khachaturyan, {\og Phase field microelasticity theory of
  dislocation dynamics in a polycrystal: Model and three-dimensional
  simulations\fg}, \emph{Philos. Mag. Lett.} \textbf{81} (2001), \cdrnumero 9,
  p.~607-616.

\bibitem{kaxiras1994energetics}
E.~Kaxiras, L.~Boyer, {\og Energetics of large lattice strains: Application to
  silicon\fg}, \emph{Physical Review B} \textbf{50} (1994), \cdrnumero 3,
  p.~1535.

\bibitem{Kiener2011-hc}
D.~Kiener, A.~M. Minor, {\og Source-controlled yield and hardening of Cu(100)
  studied by in situ transmission electron microscopy\fg}, \emph{Acta Mater.}
  \textbf{59} (2011), \cdrnumero 4, p.~1328-1337.

\bibitem{Knap2003-pp}
J.~Knap, M.~Ortiz, {\og Effect of indenter-radius size on Au(001)
  nanoindentation\fg}, \emph{Phys. Rev. Lett.} \textbf{90} (2003), \cdrnumero
  22 (en), p.~226102.

\bibitem{Kochmann2016-bv}
D.~M. Kochmann, J.~S. Amelang, {\og The Quasicontinuum Method: Theory and
  Applications\fg}, in \emph{Multiscale Materials Modeling for Nanomechanics}
  (C.~R. Weinberger, G.~J. Tucker, \cdredsname), Springer International
  Publishing, Cham, 2016, p.~159-193.

\bibitem{Kohnert2021-ts}
A.~A. Kohnert, L.~Capolungo, {\og Spectral discrete dislocation dynamics with
  anisotropic short range interactions\fg}, \emph{Comput. Mater. Sci.}
  \textbf{189} (2021), p.~110243.

\bibitem{Koslowski2002-dn}
M.~Koslowski, A.~M. Cuiti{\~n}o, M.~Ortiz, {\og A phase-field theory of
  dislocation dynamics, strain hardening and hysteresis in ductile single
  crystals\fg}, \emph{J. Mech. Phys. Solids} \textbf{50} (2002), \cdrnumero 12,
  p.~2597-2635.

\bibitem{Koslowski2004-sa}
M.~Koslowski, R.~LeSar, R.~Thomson, {\og Avalanches and scaling in plastic
  deformation\fg}, \emph{Phys. Rev. Lett.} \textbf{93} (2004), \cdrnumero 12
  (en), p.~125502.

\bibitem{Kovalev1993-se}
A.~S. Kovalev, A.~D. Kondratyuk, A.~M. Kosevich, A.~I. Landau, {\og Theoretical
  Description of the Crowdion in an Anisotropic Crystal Based on the
  {Frenkel-Kontorova} Model Including and Elastic {Three-Dimensional}
  Medium\fg}, \emph{phys. stat. sol. (b)} \textbf{177} (1993), \cdrnumero 1,
  p.~117-127.

\bibitem{Kryuchkov2018-ud}
N.~P. Kryuchkov, S.~O. Yurchenko, Y.~D. Fomin, E.~N. Tsiok, V.~N. Ryzhov, {\og
  Complex crystalline structures in a two-dimensional core-softened system\fg},
  \emph{Soft Matter} \textbf{14} (2018), \cdrnumero 11 (en), p.~2152-2162.

\bibitem{Kubin1992-bs}
L.~P. Kubin, G.~Canova, {\og The modelling of dislocation patterns\fg},
  \emph{Scripta Metallurgica et Materialia} \textbf{27} (1992), \cdrnumero 8,
  p.~957-962.

\bibitem{kubin2013dislocations}
L.~Kubin, \emph{Dislocations, mesoscale simulations and plastic flow}, vol.~5,
  Oxford University Press, 2013.

\bibitem{Kumar2015-xo}
S.~Kumar, D.~M. Parks, {\og On the hyperelastic softening and elastic
  instabilities in graphene\fg}, \emph{Proc. R. Soc. A} \textbf{471} (2015),
  \cdrnumero 2173 (en), p.~20140567.

\bibitem{Landu1994-wt}
A.~I. Landau, {\og Application of a model of interacting atomic chains for the
  description of edge dislocations\fg}, \emph{phys. stat. sol. (b)}
  \textbf{183} (1994), \cdrnumero 2, p.~407-417.

\bibitem{Lee2011-oi}
S.-W. Lee, S.~Aubry, W.~D. Nix, W.~Cai, {\og Dislocation junctions and jogs in
  a free-standing {FCC} thin film\fg}, \emph{Modell. Simul. Mater. Sci. Eng.}
  \textbf{19} (2011), \cdrnumero 2 (en), p.~025002.

\bibitem{Lee2009-as}
S.-W. Lee, S.~M. Han, W.~D. Nix, {\og Uniaxial compression of fcc Au
  nanopillars on an {MgO} substrate: The effects of prestraining and
  annealing\fg}, \emph{Acta Mater.} \textbf{57} (2009), \cdrnumero 15,
  p.~4404-4415.

\bibitem{Lee2020-nx}
S.~Lee, A.~Vaid, J.~Im, B.~Kim, A.~Prakash, J.~Gu{\'e}nol{\'e}, D.~Kiener,
  E.~Bitzek, S.~H. Oh, {\og In-situ observation of the initiation of plasticity
  by nucleation of prismatic dislocation loops\fg}, \emph{Nat. Commun.}
  \textbf{11} (2020), \cdrnumero 1 (en), p.~2367.

\bibitem{Li2002-tw}
L.~X. Li, Y.~Lou, L.~B. Yang, D.~S. Peng, K.~P. Rao, {\og Flow stress behavior
  and deformation characteristics of {Ti-3Al-5V-5Mo} compressed at elevated
  temperatures\fg}, \emph{Mater. Des.} \textbf{23} (2002), \cdrnumero 5,
  p.~451-457.

\bibitem{Li2011-nf}
P.~Li, S.~X. Li, Z.~G. Wang, Z.~F. Zhang, {\og Fundamental factors on formation
  mechanism of dislocation arrangements in cyclically deformed fcc single
  crystals\fg}, \emph{Prog. Mater Sci.} \textbf{56} (2011), \cdrnumero 3,
  p.~328-377.

\bibitem{Li2016-cz}
R.~Li, H.~Kang, Z.~Chen, G.~Fan, C.~Zou, W.~Wang, S.~Zhang, Y.~Lu, J.~Jie,
  Z.~Cao, T.~Li, T.~Wang, {\og A promising structure for fabricating high
  strength and high electrical conductivity copper alloys\fg}, \emph{Sci. Rep.}
  \textbf{6} (2016) (en), p.~20799.

\bibitem{Li2017-dz}
Y.~Li, S.~Hu, X.~Sun, M.~Stan, {\og A review: applications of the phase field
  method in predicting microstructure and property evolution of irradiated
  nuclear materials\fg}, \emph{npj Computational Materials} \textbf{3} (2017),
  \cdrnumero 1 (en), p.~16.

\bibitem{Lilleodden2006-jq}
E.~T. Lilleodden, W.~D. Nix, {\og Microstructural length-scale effects in the
  nanoindentation behavior of thin gold films\fg}, \emph{Acta Mater.}
  \textbf{54} (2006), \cdrnumero 6, p.~1583-1593.

\bibitem{Lim2018-bj}
H.~Lim, C.~C. Battaile, C.~R. Weinberger, {\og Simulating dislocation
  plasticity in bcc metals by integrating fundamental concepts with macroscale
  models\fg}, \emph{Integrated Computational Materials Engineering (ICME) for
  Metals: Concepts and Case Studies} (2018), p.~71-106.

\bibitem{lomdahl1986dislocation}
P.~Lomdahl, D.~Srolovitz, {\og Dislocation generation in the two-dimensional
  Frenkel-Kontorova model at high stresses\fg}, \emph{Phys. Rev. Lett.}
  \textbf{57} (1986), \cdrnumero 21, p.~2702.

\bibitem{Louchez2017-ui}
M.-A. Louchez, L.~Thuinet, R.~Besson, A.~Legris, {\og Microscopic {Phase-Field}
  modeling of hcp|fcc interfaces\fg}, \emph{Comput. Mater. Sci.} \textbf{132}
  (2017), p.~62-73.

\bibitem{Lu2016-ag}
K.~Lu, {\og Stabilizing nanostructures in metals using grain and twin boundary
  architectures\fg}, \emph{Nature Reviews Materials} \textbf{1} (2016),
  \cdrnumero 5, p.~16019.

\bibitem{Lu2011-cu}
Y.~Lu, J.~Song, J.~Y. Huang, J.~Lou, {\og Surface dislocation nucleation
  mediated deformation and ultrahigh strength in sub-10-nm gold nanowires\fg},
  \emph{Nano Res.} \textbf{4} (2011), \cdrnumero 12, p.~1261-1267.

\bibitem{Ma2020-mi}
R.~Ma, D.~Cao, C.~Zhu, Y.~Tian, J.~Peng, J.~Guo, J.~Chen, X.-Z. Li, J.~S.
  Francisco, X.~C. Zeng, L.-M. Xu, E.-G. Wang, Y.~Jiang, {\og Atomic imaging of
  the edge structure and growth of a two-dimensional hexagonal ice\fg},
  \emph{Nature} \textbf{577} (2020), \cdrnumero 7788 (en), p.~60-63.

\bibitem{Maas2018-qu}
R.~Maa{\ss}, P.~M. Derlet, {\og Micro-plasticity and recent insights from
  intermittent and small-scale plasticity\fg}, \emph{Acta Mater.} \textbf{143}
  (2018), p.~338-363.

\bibitem{Maas2013-tq}
R.~Maa{\ss}, P.~M. Derlet, J.~R. Greer, {\og Small-scale plasticity: Insights
  into dislocation avalanche velocities\fg}, \emph{Scr. Mater.} \textbf{69}
  (2013), \cdrnumero 8, p.~586-589.

\bibitem{Maas2012-ib}
R.~Maa{\ss}, L.~Meza, B.~Gan, S.~Tin, J.~R. Greer, {\og Ultrahigh strength of
  dislocation-free {Ni3Al} nanocubes\fg}, \emph{Small} \textbf{8} (2012),
  \cdrnumero 12 (en), p.~1869-1875.

\bibitem{Madec2002-hk}
R.~Madec, B.~Devincre, L.~P. Kubin, {\og From dislocation junctions to forest
  hardening\fg}, \emph{Phys. Rev. Lett.} \textbf{89} (2002), \cdrnumero 25
  (en), p.~255508.

\bibitem{Marano2019-yo}
A.~Marano, L.~G{\'e}l{\'e}bart, S.~Forest, {\og Intragranular localization
  induced by softening crystal plasticity: Analysis of slip and kink bands
  localization modes from high resolution {FFT-simulations} results\fg},
  \emph{Acta Mater.} \textbf{175} (2019), p.~262-275.

\bibitem{Marconi2005-cp}
V.~I. Marconi, E.~A. Jagla, {\og Diffuse interface approach to brittle
  fracture\fg}, \emph{Phys. Rev. E Stat. Nonlin. Soft Matter Phys.} \textbf{71}
  (2005), \cdrnumero 3 Pt 2A (en), p.~036110.

\bibitem{Merabet2018-cv}
A.~Merabet, M.~Texier, C.~Tromas, S.~Brochard, L.~Pizzagalli, L.~Thilly,
  J.~Rabier, A.~Talneau, Y.-M. Le~Vaillant, O.~Thomas, J.~Godet, {\og
  Low-temperature intrinsic plasticity in silicon at small scales\fg},
  \emph{Acta Mater.} \textbf{161} (2018), p.~54-60.

\bibitem{Merodio}
J.~Merodio, R.~Ogden, {\og Material instabilities in fiber-reinforced
  nonlinearly elasti solids under plane deformation\fg}, \emph{Archives of
  Mechanics} \textbf{54} (2002), \cdrnumero 5.

\bibitem{Mielke2011-ck}
A.~Mielke, L.~Truskinovsky, {\og From Discrete {Visco-Elasticity} to Continuum
  {Rate-Independent} Plasticity: Rigorous Results\fg}, \emph{Arch. Ration.
  Mech. Anal.} \textbf{203} (2011), \cdrnumero 2, p.~577-619.

\bibitem{Miguel2001-fi}
M.~C. Miguel, A.~Vespignani, S.~Zapperi, J.~Weiss, J.~R. Grasso, {\og
  Intermittent dislocation flow in viscoplastic deformation\fg}, \emph{Nature}
  \textbf{410} (2001), \cdrnumero 6829 (en), p.~667-671.

\bibitem{Miller2008-rr}
R.~E. Miller, D.~Rodney, {\og On the nonlocal nature of dislocation nucleation
  during nanoindentation\fg}, \emph{J. Mech. Phys. Solids} \textbf{56} (2008),
  \cdrnumero 4, p.~1203-1223.

\bibitem{Miller2002-pa}
R.~E. Miller, E.~B. Tadmor, {\og The Quasicontinuum Method: Overview,
  applications and current directions\fg}, \emph{Journal of Computer-Aided
  Materials Design} \textbf{9} (2002), \cdrnumero 3, p.~203-239.

\bibitem{Minami2007-ew}
A.~Minami, A.~Onuki, {\og Nonlinear elasticity theory of dislocation formation
  and composition change in binary alloys in three dimensions\fg}, \emph{Acta
  Mater.} \textbf{55} (2007), \cdrnumero 7, p.~2375-2384.

\bibitem{Mordehai2018-qm}
D.~Mordehai, O.~David, R.~Kositski, {\og {Nucleation‐Controlled} Plasticity
  of Metallic Nanowires and Nanoparticles\fg}, \emph{Adv. Mater.} (2018).

\bibitem{Mordehai2011-ys}
D.~Mordehai, M.~Kazakevich, D.~J. Srolovitz, E.~Rabkin, {\og Nanoindentation
  size effect in single-crystal nanoparticles and thin films: A comparative
  experimental and simulation study\fg}, \emph{Acta Mater.} \textbf{59} (2011),
  \cdrnumero 6, p.~2309-2321.

\bibitem{Mordehai2011-to}
D.~Mordehai, S.-W. Lee, B.~Backes, D.~J. Srolovitz, W.~D. Nix, E.~Rabkin, {\og
  Size effect in compression of single-crystal gold microparticles\fg},
  \emph{Acta Mater.} \textbf{13} (2011), \cdrnumero 59 (en), p.~5202-5215.

\bibitem{Nabarro2002-js}
F.~R.~N. Nabarro, {\og Dislocations in a simple cubic lattice\fg}, \emph{Proc.
  Phys. Soc. London} \textbf{59} (1947), \cdrnumero 2, p.~256.

\bibitem{Ng2009-cc}
K.~S. Ng, A.~H.~W. Ngan, {\og Effects of trapping dislocations within small
  crystals on their deformation behavior\fg}, \emph{Acta Mater.} \textbf{57}
  (2009), \cdrnumero 16, p.~4902-4910.

\bibitem{Niiyama2015-gv}
T.~Niiyama, T.~Shimokawa, {\og Atomistic mechanisms of intermittent plasticity
  in metals: dislocation avalanches and defect cluster pinning\fg}, \emph{Phys.
  Rev. E Stat. Nonlin. Soft Matter Phys.} \textbf{91} (2015), \cdrnumero 2
  (en), p.~022401.

\bibitem{Nix1998-xr}
W.~D. Nix, H.~Gao, {\og Indentation size effects in crystalline materials: A
  law for strain gradient plasticity\fg}, \emph{J. Mech. Phys. Solids}
  \textbf{46} (1998), \cdrnumero 3, p.~411-425.

\bibitem{ogden_nl}
R.~Ogden, \emph{Non-Linear Elastic Deformations}, John Wiley and Sons, 1984.

\bibitem{ogden1997non}
R.~Ogden, {\og Non-linear elastic deformations. Ellis Horwood, Chichester
  1984.\fg}, 1997.

\bibitem{Oh2009-uf}
S.~H. Oh, M.~Legros, D.~Kiener, G.~Dehm, {\og In situ observation of
  dislocation nucleation and escape in a submicrometre aluminium single
  crystal\fg}, \emph{Nat. Mater.} \textbf{8} (2009), \cdrnumero 2 (en),
  p.~95-100.

\bibitem{Onuki2003-ln}
A.~Onuki, {\og Plastic flow in two-dimensional solids\fg}, \emph{Phys. Rev. E
  Stat. Nonlin. Soft Matter Phys.} \textbf{68} (2003), \cdrnumero 6 Pt 1 (en),
  p.~061502.

\bibitem{Onuki2005-xp}
A.~Onuki, A.~Furukawa, A.~Minami, {\og Sheared solid materials\fg},
  \emph{Pramana} \textbf{64} (2005), \cdrnumero 5, p.~661-677.

\bibitem{Ortiz1999-pp}
M.~Ortiz, E.~a. Repetto, {\og Nonconvex energy minimization and dislocation
  structures in ductile single crystals\fg}, \emph{J. Mech. Phys. Solids}
  \textbf{47} (1999), \cdrnumero 2, p.~397-462.

\bibitem{Papanikolaou2017-ld}
S.~Papanikolaou, Y.~Cui, N.~Ghoniem, {\og Avalanches and plastic flow in
  crystal plasticity: an overview\fg}, \emph{Modell. Simul. Mater. Sci. Eng.}
  \textbf{26} (2017), \cdrnumero 1 (en), p.~013001.

\bibitem{Parakh2020-mk}
A.~Parakh, S.~Lee, K.~A. Harkins, M.~T. Kiani, D.~Doan, M.~Kunz, A.~Doran,
  L.~A. Hanson, S.~Ryu, X.~W. Gu, {\og Nucleation of Dislocations in 3.9 nm
  Nanocrystals at High Pressure\fg}, \emph{Phys. Rev. Lett.} \textbf{124}
  (2020), \cdrnumero 10, p.~106104.

\bibitem{Parry1998-sv}
G.~P. Parry, {\og {Low‐Dimensional} Lattice Groups for the Continuum
  Mechanics of Phase Transitions in Crystals\fg}, \emph{Arch. Ration. Mech.
  Anal.} \textbf{145} (1998), \cdrnumero 1, p.~1-22.

\bibitem{Parry1976-zt}
G.~P. Parry, {\og On the elasticity of monatomic crystals\fg}, \emph{Math.
  Proc. Cambridge Philos. Soc.} \textbf{80} (1976), \cdrnumero 1, p.~189-211.

\bibitem{Peierls2002-xo}
R.~Peierls, {\og The size of a dislocation\fg}, \emph{Proc. Phys. Soc. London}
  \textbf{52} (1940), \cdrnumero 1, p.~34.

\bibitem{pitteri2002continuum}
M.~Pitteri, G.~Zanzotto, \emph{Continuum models for phase transitions and
  twinning in crystals}, Chapman and Hall/CRC, 2002.

\bibitem{Plans2007-cx}
I.~Plans, A.~Carpio, L.~L. Bonilla, {\og Homogeneous nucleation of dislocations
  as bifurcations in a periodized discrete elasticity model\fg}, \emph{EPL}
  \textbf{81} (2007), \cdrnumero 3 (en), p.~36001.

\bibitem{Po2014-qu}
G.~Po, M.~Lazar, D.~Seif, N.~Ghoniem, {\og Singularity-free dislocation
  dynamics with strain gradient elasticity\fg}, \emph{J. Mech. Phys. Solids}
  (2014).

\bibitem{Podio-Guidugli2010-hx}
P.~Podio-Guidugli, {\og On ({Andersen--)Parrinello--Rahman} molecular dynamics,
  the related metadynamics, and the use of the Cauchy--born rule\fg}, \emph{J.
  Elast.} \textbf{100} (2010), \cdrnumero 1-2, p.~145-153.

\bibitem{Puglisi2005-lg}
G.~Puglisi, L.~Truskinovsky, {\og Thermodynamics of rate-independent
  plasticity\fg}, \emph{J. Mech. Phys. Solids} \textbf{53} (2005), \cdrnumero
  3, p.~655-679.

\bibitem{Qiu2019-ue}
D.~Qiu, P.~Zhao, C.~Shen, W.~Lu, D.~Zhang, M.~Mrovec, Y.~Wang, {\og Predicting
  grain boundary structure and energy in {BCC} metals by integrated atomistic
  and phase-field modeling\fg}, \emph{Acta Mater.} \textbf{164} (2019),
  p.~799-809.

\bibitem{Rao2008-df}
S.~I. Rao, D.~M. Dimiduk, T.~A. Parthasarathy, {Uchic}, M.~Tang, C.~Woodward,
  {\og Athermal mechanisms of size-dependent crystal flow gleaned from
  three-dimensional discrete dislocation simulations\fg}, \emph{Acta Mater.}
  \textbf{56} (2008), \cdrnumero 13, p.~3245-3259.

\bibitem{Read1955-ps}
W.~T. Read, Jr., H.~Brooks, {\og Dislocations in Crystals\fg}, \emph{Phys.
  Today} \textbf{8} (1955), \cdrnumero 2, p.~17-18.

\bibitem{Rice1976-ta}
J.~R. Rice, {\og Localization of plastic deformation\fg}, Tech. report, Brown
  Univ., Providence, RI (USA). Div. of Engineering, 1976.

\bibitem{PhysRevMaterials.4.113609}
D.~Richard, M.~Ozawa, S.~Patinet, E.~Stanifer, B.~Shang, S.~A. Ridout, B.~Xu,
  G.~Zhang, P.~K. Morse, J.-L. Barrat, L.~Berthier, M.~L. Falk, P.~Guan, A.~J.
  Liu, K.~Martens, S.~Sastry, D.~Vandembroucq, E.~Lerner, M.~L. Manning, {\og
  Predicting plasticity in disordered solids from structural indicators\fg},
  \emph{Phys. Rev. Materials} \textbf{4} (2020), p.~113609.

\bibitem{Rodney2003-wy}
D.~Rodney, Y.~Le~Bouar, A.~Finel, {\og Phase field methods and
  dislocations\fg}, \emph{Acta Mater.} \textbf{51} (2003), \cdrnumero 1,
  p.~17-30.

\bibitem{Rodney1999-em}
D.~Rodney, R.~Phillips, {\og Structure and Strength of Dislocation Junctions:
  An Atomic Level Analysis\fg}, \emph{Phys. Rev. Lett.} \textbf{82} (1999),
  \cdrnumero 8, p.~1704-1707.

\bibitem{Roters2010-cw}
F.~Roters, P.~Eisenlohr, L.~Hantcherli, D.~D. Tjahjanto, T.~R. Bieler,
  D.~Raabe, {\og Overview of constitutive laws, kinematics, homogenization and
  multiscale methods in crystal plasticity finite-element modeling: Theory,
  experiments, applications\fg}, \emph{Acta Mater.} \textbf{58} (2010),
  \cdrnumero 4, p.~1152-1211.

\bibitem{Ruan2019-bx}
Q.~Ruan, M.~Yang, W.~Liu, A.~Godfrey, {\og Plastic yielding and tensile
  strength of near-micrometer grain size pure iron\fg}, \emph{Materials Science
  and Engineering: A} \textbf{744} (2019), p.~764-772.

\bibitem{Ruffini2017-fs}
A.~Ruffini, Y.~Le~Bouar, A.~Finel, {\og Three-dimensional phase-field model of
  dislocations for a heterogeneous face-centered cubic crystal\fg}, \emph{J.
  Mech. Phys. Solids} \textbf{105} (2017), p.~95-115.

\bibitem{salman_thesis}
O.~U. Salman, {\og Modeling of spatio-temporal dynamics and patterning
  mechanisms of martensites by phase-field and Lagrangian methods. (2009)\fg},
  \cdrphdthesisname, Universit{\'e} Pierre et Marie Curie, 2009.

\bibitem{Salman2012-oa}
O.~U. Salman, L.~Truskinovsky, {\og On the critical nature of plastic flow: One
  and two dimensional models\fg}, \emph{Int. J. Eng. Sci.} (2012).

\bibitem{Salman2012-zg}
O.~U. Salman, A.~Finel, R.~Delville, D.~Schryvers, {\og The role of phase
  compatibility in martensite\fg}, \emph{J. Appl. Phys.} \textbf{111} (2012),
  \cdrnumero 10, p.~103517.

\bibitem{Salman2011-ij}
O.~U. Salman, L.~Truskinovsky, {\og Minimal integer automaton behind crystal
  plasticity\fg}, \emph{Phys. Rev. Lett.} \textbf{106} (2011), \cdrnumero 17,
  p.~175503.

\bibitem{Salman2021-sb}
O.~U. Salman, I.~R. Ionescu, {\og Tempering the mechanical response of {FCC}
  micro-pillars: an Eulerian plasticity approach\fg}, \emph{Mech. Res. Commun.}
  (2021), p.~103665.

\bibitem{Salman2019-cg}
O.~U. Salman, B.~Muite, A.~Finel, {\og Origin of stabilization of macrotwin
  boundaries in martensites\fg}, \emph{Eur. Phys. J. B} \textbf{92} (2019),
  \cdrnumero 1, p.~20.

\bibitem{Salvalaglio2020-eb}
M.~Salvalaglio, L.~Angheluta, Z.-F. Huang, A.~Voigt, K.~R. Elder,
  J.~Vi{\~n}als, {\og A coarse-grained phase-field crystal model of plastic
  motion\fg}, \emph{J. Mech. Phys. Solids} \textbf{137} (2020), p.~103856.

\bibitem{Salvalaglio2019-kt}
M.~Salvalaglio, A.~Voigt, K.~R. Elder, {\og Closing the gap between
  atomic-scale lattice deformations and continuum elasticity\fg}, \emph{npj
  Computational Materials} \textbf{5} (2019), \cdrnumero 1, p.~48.

\bibitem{Sanderson2016-ht}
C.~Sanderson, R.~Curtin, {\og Armadillo: a template-based C++ library for
  linear algebra\fg}, \emph{J. Open Source Softw.} \textbf{1} (2016),
  \cdrnumero 2, p.~26.

\bibitem{Sanderson2018-rt}
\bysame , {\og A {User-Friendly} Hybrid Sparse Matrix Class in C++\fg}, in
  \emph{Mathematical Software -- {ICMS} 2018}, Springer International
  Publishing, 2018, p.~422-430.

\bibitem{Sandfeld_undated-nm}
S.~Sandfeld, M.~Zaiser, {\og Pattern formation in a minimal model of continuum
  dislocation plasticity\fg}, \emph{Modell. Simul. Mater. Sci. Eng.}, p.~1-18.

\bibitem{Schaedler2011-gn}
T.~A. Schaedler, A.~J. Jacobsen, A.~Torrents, A.~E. Sorensen, J.~Lian, J.~R.
  Greer, L.~Valdevit, W.~B. Carter, {\og Ultralight metallic microlattices\fg},
  \emph{Science} \textbf{334} (2011), \cdrnumero 6058 (en), p.~962-965.

\bibitem{Sethna2017-tm}
J.~P. Sethna, M.~K. Bierbaum, K.~A. Dahmen, C.~P. Goodrich, J.~R. Greer, L.~X.
  Hayden, J.~P. Kent-Dobias, E.~D. Lee, D.~B. Liarte, X.~Ni, K.~N. Quinn,
  A.~Raju, D.~Z. Rocklin, A.~Shekhawat, S.~Zapperi, {\og Deformation of
  Crystals: Connections with Statistical Physics\fg}, \emph{Annu. Rev. Mater.
  Res.} \textbf{47} (2017), \cdrnumero 1, p.~217-246.

\bibitem{Shan2008-nh}
Z.~W. Shan, R.~K. Mishra, S.~A. Syed~Asif, O.~L. Warren, A.~M. Minor, {\og
  Mechanical annealing and source-limited deformation in submicrometre-diameter
  Ni crystals\fg}, \emph{Nat. Mater.} \textbf{7} (2008), \cdrnumero 2 (en),
  p.~115-119.

\bibitem{Sharma2018-iw}
A.~Sharma, J.~Hickman, N.~Gazit, E.~Rabkin, Y.~Mishin, {\og Nickel
  nanoparticles set a new record of strength\fg}, \emph{Nat. Commun.}
  \textbf{9} (2018), \cdrnumero 1 (en), p.~4102.

\bibitem{Shchyglo2012-nz}
O.~Shchyglo, U.~Salman, A.~Finel, {\og Martensitic phase transformations in
  {Ni--Ti-based} shape memory alloys: The Landau theory\fg}, \emph{Acta Mater.}
  \textbf{60} (2012), \cdrnumero 19, p.~6784-6792.

\bibitem{Shenoy1999-gh}
V.~B. Shenoy, R.~Miller, E.~b. Tadmor, D.~Rodney, R.~Phillips, M.~Ortiz, {\og
  An adaptive finite element approach to atomic-scale mechanics---the
  quasicontinuum method\fg}, \emph{J. Mech. Phys. Solids} \textbf{47} (1999),
  \cdrnumero 3, p.~611-642.

\bibitem{Skaugen2018-rf}
A.~Skaugen, L.~Angheluta, J.~Vi{\~n}als, {\og Separation of Elastic and Plastic
  Timescales in a Phase Field Crystal Model\fg}, \emph{Phys. Rev. Lett.}
  \textbf{121} (2018), \cdrnumero 25 (en), p.~255501.

\bibitem{Sorkin2014-jy}
V.~Sorkin, R.~S. Elliott, E.~B. Tadmor, {\og A local quasicontinuum method for
  {3D} multilattice crystalline materials: Application to shape-memory
  alloys\fg}, \emph{Modell. Simul. Mater. Sci. Eng.} \textbf{22} (2014),
  \cdrnumero 5 (en), p.~055001.

\bibitem{Sornette2012-kh}
D.~Sornette, G.~Ouillon, {\og Dragon-kings: Mechanisms, statistical methods and
  empirical evidence\fg}, \emph{Eur. Phys. J. Spec. Top.} \textbf{205} (2012),
  \cdrnumero 1, p.~1-26.

\bibitem{Sparks2019-ie}
G.~Sparks, Y.~Cui, G.~Po, Q.~Rizzardi, J.~Marian, R.~Maa{\ss}, {\og Avalanche
  statistics and the intermittent-to-smooth transition in microplasticity\fg},
  \emph{Phys. Rev. Materials} \textbf{3} (2019), \cdrnumero 8, p.~080601.

\bibitem{srolovitz1986dislocation}
D.~Srolovitz, P.~Lomdahl, {\og Dislocation dynamics in the 2-d
  Frenkel-Kontorova model\fg}, \emph{Physica D: Nonlinear Phenomena}
  \textbf{23} (1986), \cdrnumero 1-3, p.~402-412.

\bibitem{Starkey2020-qb}
K.~Starkey, G.~Winther, A.~El-Azab, {\og Theoretical development of continuum
  dislocation dynamics for finite-deformation crystal plasticity at the
  mesoscale\fg}, \emph{J. Mech. Phys. Solids} \textbf{139} (2020), p.~103926.

\bibitem{Steinmann2006-qw}
P.~Steinmann, A.~Elizondo, R.~Sunyk, {\og Studies of validity of the
  {Cauchy--Born} rule by direct comparison of continuum and atomistic
  modelling\fg}, \emph{Modell. Simul. Mater. Sci. Eng.} \textbf{15} (2006),
  \cdrnumero 1 (en), p.~S271.

\bibitem{Tadmor1996-qi}
E.~B. Tadmor, M.~Ortiz, R.~Phillips, {\og Quasicontinuum analysis of defects in
  solids\fg}, \emph{Philos. Mag. A} \textbf{73} (1996), \cdrnumero 6,
  p.~1529-1563.

\bibitem{Tadmor2011-mo}
E.~B. Tadmor, R.~E. Miller, \emph{Modeling Materials: Continuum, Atomistic and
  Multiscale Techniques}, Cambridge University Press, 2011 (en).

\bibitem{Takeuchi1975-vw}
T.~Takeuchi, {\og Work Hardening of Copper Single Crystals with Multiple Glide
  Orientations\fg}, \emph{Transactions of the Japan Institute of Metals}
  \textbf{16} (1975), \cdrnumero 10, p.~629-640.

\bibitem{Truskinovsky2005-gx}
L.~Truskinovsky, A.~Vainchtein, {\og Quasicontinuum modelling of short-wave
  instabilities in crystal lattices\fg}, \emph{Philos. Mag.} \textbf{85}
  (2005), \cdrnumero 33-35, p.~4055-4065.

\bibitem{Truskinovsky2004-xp}
L.~Truskinovsky, A.~Vainchtein, {\og The origin of nucleation peak in
  transformational plasticity\fg}, \emph{J. Mech. Phys. Solids} \textbf{52}
  (2004), \cdrnumero 6, p.~1421-1446.

\bibitem{Uchic2004-ax}
M.~D. Uchic, D.~M. Dimiduk, J.~N. Florando, W.~D. Nix, {\og Sample dimensions
  influence strength and crystal plasticity\fg}, \emph{Science} \textbf{305}
  (2004), \cdrnumero 5686 (en), p.~986-989.

\bibitem{Uchic2009-rj}
M.~D. Uchic, P.~A. Shade, D.~M. Dimiduk, {\og Micro-compression testing of fcc
  metals: A selected overview of experiments and simulations\fg}, \emph{JOM}
  \textbf{61} (2009), \cdrnumero 3, p.~36-41.

\bibitem{Uchic2009-jl}
\bysame , {\og Plasticity of {Micrometer-Scale} Single Crystals in
  Compression\fg}, \emph{Annu. Rev. Mater. Res.} \textbf{39} (2009), \cdrnumero
  1, p.~361-386.

\bibitem{Valdenaire2016-yp}
P.-L. Valdenaire, Y.~Le~Bouar, B.~Appolaire, A.~Finel, {\og Density-based
  crystal plasticity: From the discrete to the continuum\fg}, \emph{Phys. Rev.
  B Condens. Matter} \textbf{93} (2016), \cdrnumero 21, p.~214111.

\bibitem{Van_Hoang2019-ob}
V.~Van~Hoang, N.~H. Giang, {\og Compression-induced square-triangle solid-solid
  phase transition in {2D} simple monatomic system\fg}, \emph{Physica E}
  \textbf{113} (2019), p.~35-42.

\bibitem{Van_Vliet2003-yg}
K.~J. Van~Vliet, J.~Li, T.~Zhu, S.~Yip, S.~Suresh, {\og Quantifying the early
  stages of plasticity through nanoscale experiments and simulations\fg},
  \emph{Phys. Rev. B} \textbf{67} (2003), \cdrnumero 10, p.~104105.

\bibitem{Varadhan2006-xb}
S.~N. Varadhan, A.~J. Beaudoin, A.~Acharya, C.~Fressengeas, {\og Dislocation
  transport using an explicit Galerkin/least-squares formulation\fg},
  \emph{Modell. Simul. Mater. Sci. Eng.} \textbf{14} (2006), \cdrnumero 7,
  p.~1245.

\bibitem{Vattre2014-be}
A.~Vattr{\'e}, B.~Devincre, F.~Feyel, R.~Gatti, S.~Groh, O.~Jamond, A.~Roos,
  {\og Modelling crystal plasticity by {3D} dislocation dynamics and the finite
  element method: The {Discrete-Continuous} Model revisited\fg}, \emph{J. Mech.
  Phys. Solids} \textbf{63} (2014), p.~491-505.

\bibitem{Wang2014-sb}
J.~Wang, I.~J. Beyerlein, C.~N. Tom{\'e}, {\og Reactions of lattice
  dislocations with grain boundaries in Mg: Implications on the micro scale
  from atomic-scale calculations\fg}, \emph{Int. J. Plast.} \textbf{56} (2014),
  p.~156-172.

\bibitem{Wang2018-fk}
J.~Wang, Y.~Wang, W.~Cai, J.~Li, Z.~Zhang, S.~X. Mao, {\og Discrete shear band
  plasticity through dislocation activities in body-centered cubic tungsten
  nanowires\fg}, \emph{Sci. Rep.} \textbf{8} (2018), \cdrnumero 1 (en),
  p.~4574.

\bibitem{wang2012pristine}
Z.-J. Wang, Z.-W. Shan, J.~Li, J.~Sun, E.~Ma, {\og Pristine-to-pristine regime
  of plastic deformation in submicron-sized single crystal gold particles\fg},
  \emph{Acta Mater.} \textbf{60} (2012), \cdrnumero 3, p.~1368-1377.

\bibitem{Weinan2007-hd}
{Weinan}, P.~Ming, {\og Cauchy--born rule and the stability of crystalline
  solids: Static problems\fg}, \emph{Arch. Ration. Mech. Anal.} \textbf{183}
  (2007), \cdrnumero 2 (en), p.~241-297.

\bibitem{Weinberger2008-qw}
C.~R. Weinberger, W.~Cai, {\og Surface-controlled dislocation multiplication in
  metal micropillars\fg}, \emph{Proc. Natl. Acad. Sci. U. S. A.} \textbf{105}
  (2008), \cdrnumero 38 (en), p.~14304-14307.

\bibitem{Weiss2019-yl}
J.~Weiss, W.~Ben~Rhouma, S.~Deschanel, L.~Truskinovsky, {\og Plastic
  intermittency during cyclic loading: From dislocation patterning to
  microcrack initiation\fg}, \emph{Phys. Rev. Materials} \textbf{3} (2019),
  \cdrnumero 2, p.~023603.

\bibitem{Weiss2015-eh}
J.~Weiss, W.~B. Rhouma, T.~Richeton, S.~Dechanel, F.~Louchet, L.~Truskinovsky,
  {\og From mild to wild fluctuations in crystal plasticity\fg}, \emph{Phys.
  Rev. Lett.} \textbf{114} (2015), \cdrnumero 10, p.~105504.

\bibitem{Xia2015-vo}
S.~Xia, A.~El-Azab, {\og Computational modelling of mesoscale dislocation
  patterning and plastic deformation of single crystals\fg}, \emph{Modell.
  Simul. Mater. Sci. Eng.} \textbf{23} (2015), \cdrnumero 5 (en), p.~055009.

\bibitem{Xu2019-ex}
S.~Xu, J.~R. Mianroodi, A.~Hunter, I.~J. Beyerlein, B.~Svendsen, {\og
  Phase-field-based calculations of the disregistry fields of static extended
  dislocations in {FCC} metals\fg}, \emph{Philos. Mag.} \textbf{99} (2019),
  \cdrnumero 11, p.~1400-1428.

\bibitem{Yu2017-mg}
W.~Yu, Z.~Wang, S.~Shen, {\og Edge dislocations interacting with a {$\Sigma$11}
  symmetrical grain boundary in copper upon mixed loading: A quasicontinuum
  method study\fg}, \emph{Comput. Mater. Sci.} \textbf{137} (2017), p.~162-170.

\bibitem{Zaiser2013-ff}
M.~Zaiser, {\og Statistical aspects of microplasticity: experiments, discrete
  dislocation simulations and stochastic continuum models\fg}, \emph{J. Mech.
  Behav. Mater.} \textbf{22} (2013), \cdrnumero 3-4.

\bibitem{Zaiser2019-qm}
M.~Zaiser, P.~Moretti, H.~Chu, {\og Stochastic crystal plasticity models with
  internal variables: Application to slip channel formation in irradiated
  metals\fg}, \emph{Adv. Eng. Mater.} (2019), \cdrnumero adem.201901208 (en),
  p.~1901208.

\bibitem{Zepeda-Ruiz2020-cl}
L.~A. Zepeda-Ruiz, A.~Stukowski, T.~Oppelstrup, N.~Bertin, N.~R. Barton,
  R.~Freitas, V.~V. Bulatov, {\og Atomistic insights into metal hardening\fg},
  \emph{Nat. Mater.} (2020) (en).

\bibitem{Zepeda-Ruiz2017-gu}
L.~A. Zepeda-Ruiz, A.~Stukowski, T.~Oppelstrup, V.~V. Bulatov, {\og Probing the
  limits of metal plasticity with molecular dynamics simulations\fg},
  \emph{Nature} \textbf{550} (2017), \cdrnumero 7677 (en), p.~492-495.

\bibitem{Zhang2017-mb}
K.~Zhang, Y.~Feng, F.~Wang, Z.~Yang, J.~Wang, {\og Two dimensional hexagonal
  boron nitride ({2D-hBN)}: synthesis, properties and applications\fg},
  \emph{J. Mater. Chem.} \textbf{5} (2017), \cdrnumero 46 (en), p.~11992-12022.

\bibitem{Zhang2007-hb}
M.~Zhang, J.~Zhang, D.~L. McDowell, {\og Microstructure-based crystal
  plasticity modeling of cyclic deformation of {Ti--6Al--4V}\fg}, \emph{Int. J.
  Plast.} \textbf{23} (2007), \cdrnumero 8, p.~1328-1348.

\bibitem{Zhang2020-ax}
P.~Zhang, O.~U. Salman, J.~Weiss, L.~Truskinovsky, {\og Variety of scaling
  behaviors in nanocrystalline plasticity\fg}, \emph{Phys Rev E} \textbf{102}
  (2020), \cdrnumero 2-1 (en), p.~023006.

\bibitem{Zhang2017-cg}
P.~Zhang, O.~U. Salman, J.-Y. Zhang, G.~Liu, J.~Weiss, L.~Truskinovsky, J.~Sun,
  {\og Taming intermittent plasticity at small scales\fg}, \emph{Acta Mater.}
  \textbf{128} (2017), p.~351-364.

\bibitem{Zheng2010-ag}
H.~Zheng, A.~Cao, C.~R. Weinberger, J.~Y. Huang, K.~Du, J.~Wang, Y.~Ma, Y.~Xia,
  S.~X. Mao, {\og Discrete plasticity in sub-10-nm-sized gold crystals\fg},
  \emph{Nat. Commun.} \textbf{1} (2010), \cdrnumero 1, p.~144.

\bibitem{Zheng2018-zn}
S.~Zheng, D.~Zheng, Y.~Ni, L.~He, {\og Improved phase field model of
  dislocation intersections\fg}, \emph{npj Computational Materials} \textbf{4}
  (2018), \cdrnumero 1 (en), p.~20.

\bibitem{Zhong2008-pj}
Y.~Zhong, T.~Zhu, {\og Simulating nanoindentation and predicting dislocation
  nucleation using interatomic potential finite element method\fg},
  \emph{Comput. Methods Appl. Mech. Eng.} \textbf{197} (2008), \cdrnumero 41,
  p.~3174-3181.

\bibitem{Zhu2004-wr}
T.~Zhu, J.~Li, K.~J.~Van~Vliet, S.~Ogata, S.~Yip, S.~Suresh, {\og Predictive
  modeling of nanoindentation-induced homogeneous dislocation nucleation in
  copper\fg}, \emph{J. Mech. Phys. Solids} \textbf{52} (2004), \cdrnumero 3,
  p.~691-724.

\bibitem{Ziegenhain2010-tu}
G.~Ziegenhain, H.~M. Urbassek, A.~Hartmaier, {\og Influence of crystal
  anisotropy on elastic deformation and onset of plasticity in nanoindentation:
  A simulational study\fg}, \emph{J. Appl. Phys.} \textbf{107} (2010),
  \cdrnumero 6, p.~061807.

\bibitem{Zimmerman2001-cz}
J.~A. Zimmerman, C.~L. Kelchner, P.~A. Klein, J.~C. Hamilton, S.~M. Foiles,
  {\og Surface step effects on nanoindentation\fg}, \emph{Phys. Rev. Lett.}
  \textbf{87} (2001), \cdrnumero 16 (en), p.~165507.

\bibitem{Zimmerman2009-my}
J.~A. Zimmerman, D.~J. Bammann, H.~Gao, {\og Deformation gradients for
  continuum mechanical analysis of atomistic simulations\fg}, \emph{Int. J.
  Solids Struct.} \textbf{46} (2009), \cdrnumero 2, p.~238-253.

\end{thebibliography}
\def\bysame{\leavevmode ---------\thinspace}
\makeatletter\if@francais\providecommand{\og}{<<~}\providecommand{\fg}{~>>}
\else\gdef\og{``}\gdef\fg{''}\fi\makeatother
\def\cdrandname{\&}
\providecommand\cdrnumero{no.~}
\providecommand{\cdredsname}{eds.}
\providecommand{\cdredname}{ed.}
\providecommand{\cdrchapname}{chap.}
\providecommand{\cdrmastersthesisname}{Memoir}
\providecommand{\cdrphdthesisname}{PhD Thesis}

\end{document}